# A Responsive Framework for Research Portals Data using Semantic Web Technology

**Muhammad Zohaib Computer Science Department Government College University Faisalabad zohaib@gcuf.edu.pk**

## ABSTRACT

As the amount of data on the World Wide Web continues to grow exponentially, access to semantically structured information remains limited. The Semantic Web has emerged as a solution to enhance the machine-readability of data, making it significantly more accessible and interpretable. Various techniques, such as web scraping and mapping, have been employed by different websites to provide semantic access. Web scraping involves the extraction of valuable information from diverse data sources, such as the World Wide Web, utilizing powerful string manipulation operations.In the research field, researchers face the challenge of collecting relevant data from multiple sources, which requires substantial time and effort. This research aims to address this issue by designing a framework for the semantic organization of research portal data. The framework focuses on the extraction of information from two specific research portals, namely Microsoft Academic and IEEE Xplore. Its primary objective is to gather diverse research-related data from these targeted sources. By implementing this framework, researchers can streamline the process of collecting valuable information for their work, saving time and effort. The semantic organization of research portal data offers enhanced accessibility and interpretability, facilitating more effective and efficient knowledge discovery. This research contributes to the advancement of research data management and promotes the utilization of semantic web technologies in the academic community.

# 1. Introduction

## 1.1 Overview

Computer is one of the greatest inventions made by the human. With the passage of time modern communication have given rise through internet. Internet is a huge source of information. Is the era of science computer devices are one of the greatest inventions ever made by the human. Modern communications have also going to raise using internet with the passage of time. It is known that WWW (World Wide Web) over a domain is great source of knowledge. On the internet huge amount of data as well as information is placed over internet in many formats and forms. Unfortunately, this large amount of data major part is unstructured. Data over the WWW is increasing with the passage of time. IT professional, researchers, common internet users are playing their role to increase amount of data. Because of all above reasons it is very hard to search useful, relevant, and right information. Researchers are working on to make this unstructured data in the form of structured and portable files.

In 1960's, the basic concept of web semantic was actually purposed by the researcher named Tim-Berner Lee on the basis of that concepts , one can made a framework with the capability of retrieving relevant materials in structured form. Web semantic provides the facility of finding, reusing, combining the relevant information. (Berners-Lee *et al.,* 2001).

Now a day, trend of queries are commonly based on their keywords. The idea and model of semantic based search was introduced to solve limitations of relevant keywords that make search possible. Extracting relevant information is actually art of searching documents as well as relational data bases have information from their documents over server. Web semantics are build on MRI (Machine Readable Information) along with flexible RDFs (Resource Description Frameworks) approaches that represents the information. Some frameworks of web semantic have the capability to share and reused the data.

Web scrapping is one of the basic technologies to collect, reuse and to make structured data over small scale and large scale. Pervious study shows that many researchers purposed their web scrapper architectures to solve web semantic issues. All of the purposed models of web scrapping were gathering their data from different web resources according to their algorithms. All these algorithms aim to collect data from large amount of unorganized data sources. Different algorithms work differently but show same results (Bakers et al., 2003).

Any web semantic framework can be developed to interact with many research web portals at the same time. Every research portal has different approaches to represent the data and query generation. On the basis of these queries their response with relevant material. Now a day many researcher have interest to collect data of different ontology's like software engineering, artificial intelligence, system engineering and bio-informatics. All the ontology models are developed by using OWL (Ontology Web Language) this is a standard language to make web semantic models over globe. RDF (Resource Description Framework) was defined as exchange of data without any markup attributes.

JSON is a way of exchanging data without using markup attributes in HTML.The expansion of the Semantic Web standards is the World Wide Web. The standard of the Semantic Web established the typical data and interlinked the standards on the web.Mostly essentially the Resource Description Framework (RDF). The idea of the Semantic Network Model was proposed in the mid 1960s by the researcher Allan M. Collins and psychologist Elizabeth F. Loftus as a structure to semantically organized information. At the point when connected in the setting of the current web, it broadens the system of hyperlinked comprehensible website pages by embeddings machine-discernable metadata about pages and how they are identified with one another. This automatically computerized operators to get to the Web and perform more undertakings for the benefit of clients. The expression Semantic Web was instituted by Tim Berners-Lee, the creator of the World Wide Web (WWW) and chief of the World Wide Web Consortium (W3C). Author directs the improvement of proposed standards of Semantic Website. He declared the semantic we that the" data on the websites can be managed indirectly or directly by computers".

A significant numbers of the advancement are proposed by the World Wide Web Consortium (W3C). They were existed or situated under W3C umbrella. These are utilized as a part of different texts especially which managing data that includes a constrained and characterized by limited area and where sharing information is a very typical need, for example, researcher's research or information and business communication among different organizations. What's more, different advancements with comparative objectives have developed, for example, micro formats.

Berners-Lee originally the goal of Semantic Web as:I have a vision about the web that machines can do everything and able to capture and control information on the websites, the content links and transaction and communication between human or machines. This can only be done by the semantic web and semantic web can do this in a better manner. But when it does and yet to be appearing, the daily trade mechanisms, government and our every day routines will be controlled by computers and conversation to these Computers. The intellectual agent's humanadvertize for ages will finally materialized.

Presently the main focal point of a World Wide Web Consortium(W3C) groups, A Semantic Web idea purposed by Tim Berners-Lee, Architect of the World Wide Web (WWW). The WWW also altered the technique of communication in which we communicate, in such a way we can do our own business and get proper information, semantic web is the next door of web evaluation, Berners-Lee described Semantic Web "a web of data that can be processed indirectly and directly by computers."

The Semantic Web is viewed as an integrator crosswise over many contents, data applications and frameworks. It has applications distributed in publishing and blogging, and numerous different zones. As per the World Wide Web Consortium, "The Semantic Web gives a typical system and structure that permits information to be shared and reused crosswise over application, organization, and group limits". That term was composed by Tim Berners-Lee for a web of data that are arranged by personal computers. While its commentators have scrutinized its possibility, defenders contend that applications in industry, science and human sciences exploration have effectively demonstrated the legitimacy of the concept.

Scientific American article 2001 by Berner-lee, Lassila and Hendler depicted a normal development of the current Web to a Semantic Web. Berners-Lee in 2006 and their associates expressed that: This basic thought is remains to a great extent undiscovered and largely hidden. In 2013, semantic web are more than four million Web spaces contained a Semantic Web markup.

The Semantic web idea was proposed by inventor Tim burners-Lee, Tim Burner is also called the inventor of the World Wide Web (WWW) and different sites like URL etc. At the present time the semantic web is well know and popular and very familiar in peoples of the researcher. In the early stage of 1960 the idea of semantic web was proposed and kept by three different researchers. With the discovery of the semantic web the information is more manageable, readable, and more accurate, easily accessible and organized. This information is also useful for the researchers because the information is more relevant to the topic of the research. Fundamentally before the existence of semantic web relevant data is not easily accessible, reasonable and difficult to understand for the computers and human and simply information are not processed and readable by the computer.

The utilization of semantic web is expanding step by step, not just vast and major sites like Google are utilization semantic web way to deal with collaborate a huge number of sites yet numerous little sites are additionally utilizing semantic web to deal with gather information inside they could call their own destinations. While gathering information from sites on little scale or vast scale, scrapping is a best standout method amongst the most helpful procedure to gather information. There are numerous scrappers available presented and depend on semantic web and encourage their clients to diverse in many types of area. Semantic Web promote out current web to a web of information. This is accomplished due to a limited extent, by data in data, Ontology, intellectual Agents.

ResourceDescription Framework and Ontology Web Languageare two buildings blocks on the Semantic Web;Different algorithms' techniques are applied by this scrapper to gather unarranged and unstructured needed information form a large scale. At whatever point we can get to information on the web as per our needs not very many sites are accessible on the web are utilizing semantic web innovations. Also the information on the web are going toward greater excessively greatest in each portion shortest time. Consequently information on the web is presently considering "enormous information" which is a most recent term in the software engineering. As not very many information are organized and arranged semantically on web for getting to execute different undertakings for information resembling discovering, again reuse or brushing to achievea few different valuable data so there are many number of issues are interfacing while we discovering related information on websites, an excess of unimportant information is additionally consolidated likewise with our obliged information.

There is a huge amount of information are placed on internet in many different forms and formats. Majority of information is unorganized and unstructured it is almost impossible to organized information in proper format. Only the addition is not only from professionals but an average internet user is also playing his/her role to increase data. Because of these kinds of involvements or nonprofessionals it is difficult task or collects the right, relevant and useful information from that huge amount of data. So semantic web is helpful to collect relevant amount of data and resolve these kinds of problems.

From the previous couple of years site is changing to semantically sorted out networks to encourage their clients. There are numerous web scrapper in view of semantic web as it encourages to get focused on and significant information from huge measure of information. Be that as it may, there is not any scrapper which can gather information from more than one entrance of research websites. As the quantity of documents, paper, and publication and research articles from over the world is expanding, numerous portal of research are create to collaborate among these publication there is an in number requirement for a system to create connection among these research gateways. System will be expected to created architecture to communicate many different research portals at a time. Latest web frameworkwill design to be suggested for retrieve the data from this structure. To add such structure or framework new web ontology will be recommended for using the Web Ontology Language (OWL), which is a universal standard. The Ontology is measured as an essential segment of a semantic web and it is alsoillustration of information or we can say that it is an arranged of thoughts in a particular region. The usage these Ontology to create the data in more valuable association however the data is assembled by using web scrapping. Web Ontology Language is used for framework, exchanging and encoding the Ontology

There is a huge problem while collecting data from different types of formats, human have the capacity to get information from different types of data sources because human can use and read different, formats but the problem with machines is that they can read, write or use single type of format or data type. Semantic web is designed to collect different kind of data and process it there for the use of semantic web is increasing rapidly and gradually, especially vital portals and search engine are mostly based on semantic web because of its relevancy and accuracy.

As it is explained that a very limited amount of information on World Wide Web (WWW) is semantically organized, the exertion of collecting relevant, reliable and useful information is increasing day by day. Some time relevant data is collected but the reliability of the data is doubtful or the data is not useful for any purpose, in simple words the data is relevant but not more than garbage. The researchers from almost every field are also facing these kinds of problems.

It is mostly observed that researchers collect data from different kind of journals. Now a day's many research portals are providing the facility to interact more than one research journal like Microsoft academics. But there are also many research portals that are providing different journals for their users. So in this case it is very important for a researcher to cover whole industry to collect more relevant and useful data. But the problem is the same as explained above that these portals are not using same kind of format so framework must be proposed by using semantic web that can collect data from different research portals and combine it in one format.

There are many scrappers available in market based on web semantic performing different kind of tasks but still there is a strong need of web scrapper based on semantic web to collect data from different research portals to facilitate researchers because there is not such a scrapper available in market. Research portals are collections of multiple journals and many of them are using web scrapper based on semantic web to generate queries and extract relevant data from different journals within their research portal. But the number of scientific journals and research portals are increasing day by day as well as the number of publications is also increasing. As the number of publications is increasing new areas of researches are also introducing that creates ambiguity and creates hurdles to collect relevant and accurate data while data is also getting huge amount. So, to make possiblefor the scientist a architectural model will be developed to attract and interact peoples with different research portal. These portals have multi formats and data types so the key role of scrapper will be to extract data from different data types and semantic web approaches will facilitate to do so. The user will generate queries and the scrapper based on semantic web will collect the data according to those queries and algorithm. So in less time researcher can collect data from many portals and semantic web techniques will ensure that data is relevant and useful.

Web Ontology Language is utilized for outline of design configuration, trading and exchanging of web ontology's. Ontology Web Language in view of Extensible Markup Language and RDF and proper semantic for semantic websites.RDF is outlined and planned as information about other data; it is utilized as system for theoretical demonstrating or strategy for calculated displaying of that specific data which is executed in web assets.

Resource description framework (RDF) utilizes a mixture of documentations, information organizations as well assentence structureat the same time asExtensible Markup Language (Xml)lies situated principles to instruct reports for the configuration which is intelligible for computer and human understandable (Lauser*et al.* 2006).

Villamor designed a scraping model on basis of three different level scraping service, syntactic scraping and semantic scraping model. Scraping service presented skills to collect the information from websites at high level by providing an interface to intelligent agents and generic applications. A semantic RDF model is defined in semantic scraping model. It provides a declarative approach in scrapping process by dividing HTML into fragments. While the syntactic scraping, provides implementation of the defined model for specific technology using the fragments.

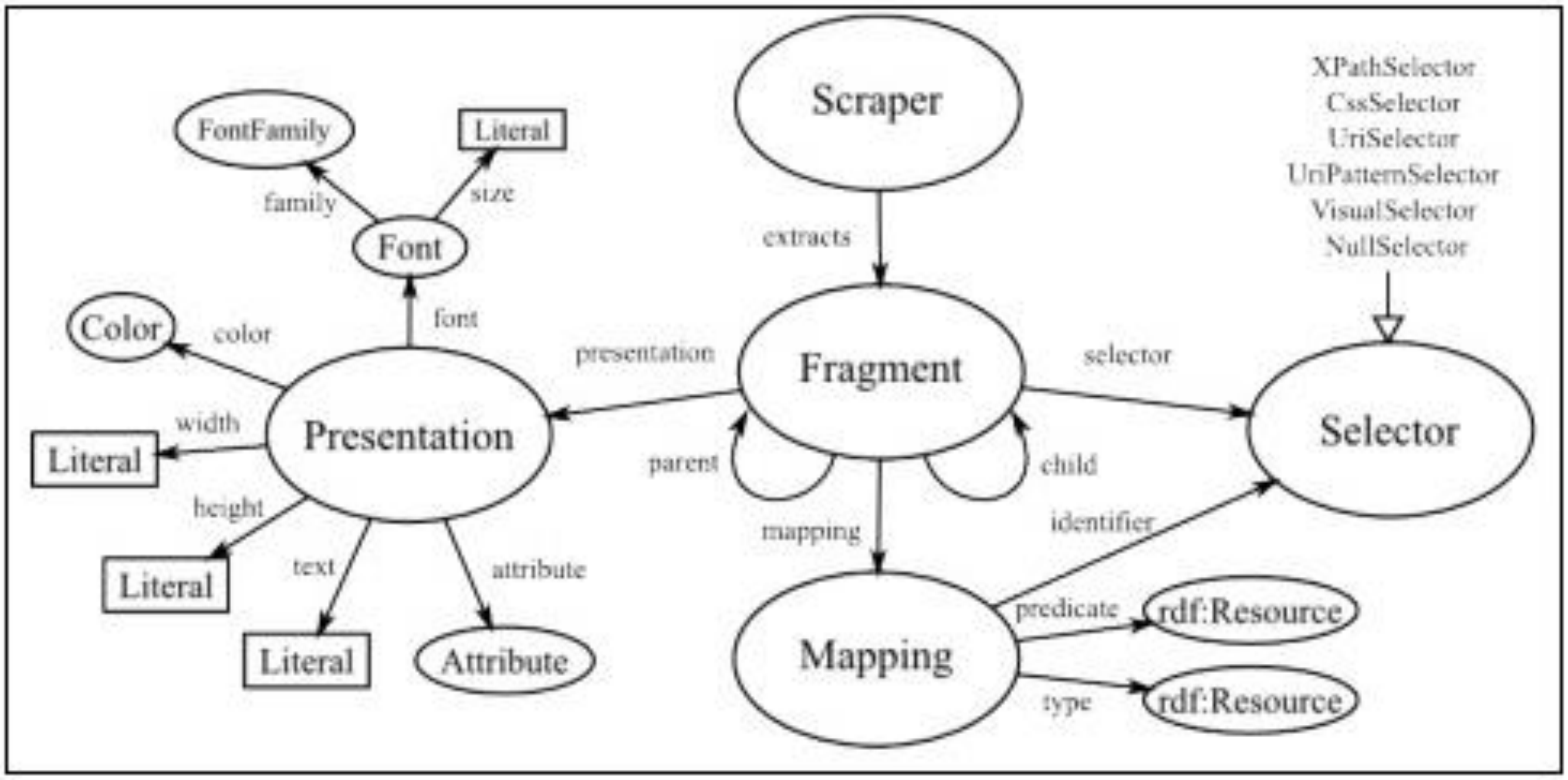

**Figure 1.1:** Semantic scraping RDF model

Madhudescribed a brief survey of semantic web search engines like Yahoo, Google, and Bing are used to retrieve information. As indicated by that study, the principle point and goal of these internet searchers the search engine is used to retrieved significant and important information. Semantic web is assuming an essential part to accomplish this research. The aim of the semantic search engine is in the starting point. In these web search tools the data is depicted utilizing RDF. The second most imperative piece of a semantic web is Ontology. OWL and Resource Descriptive Framework architecture are utilized to characterize the web Ontology Language. These two models are prescribed by World Wide Web Consortium (Madhu*et al.*2011).

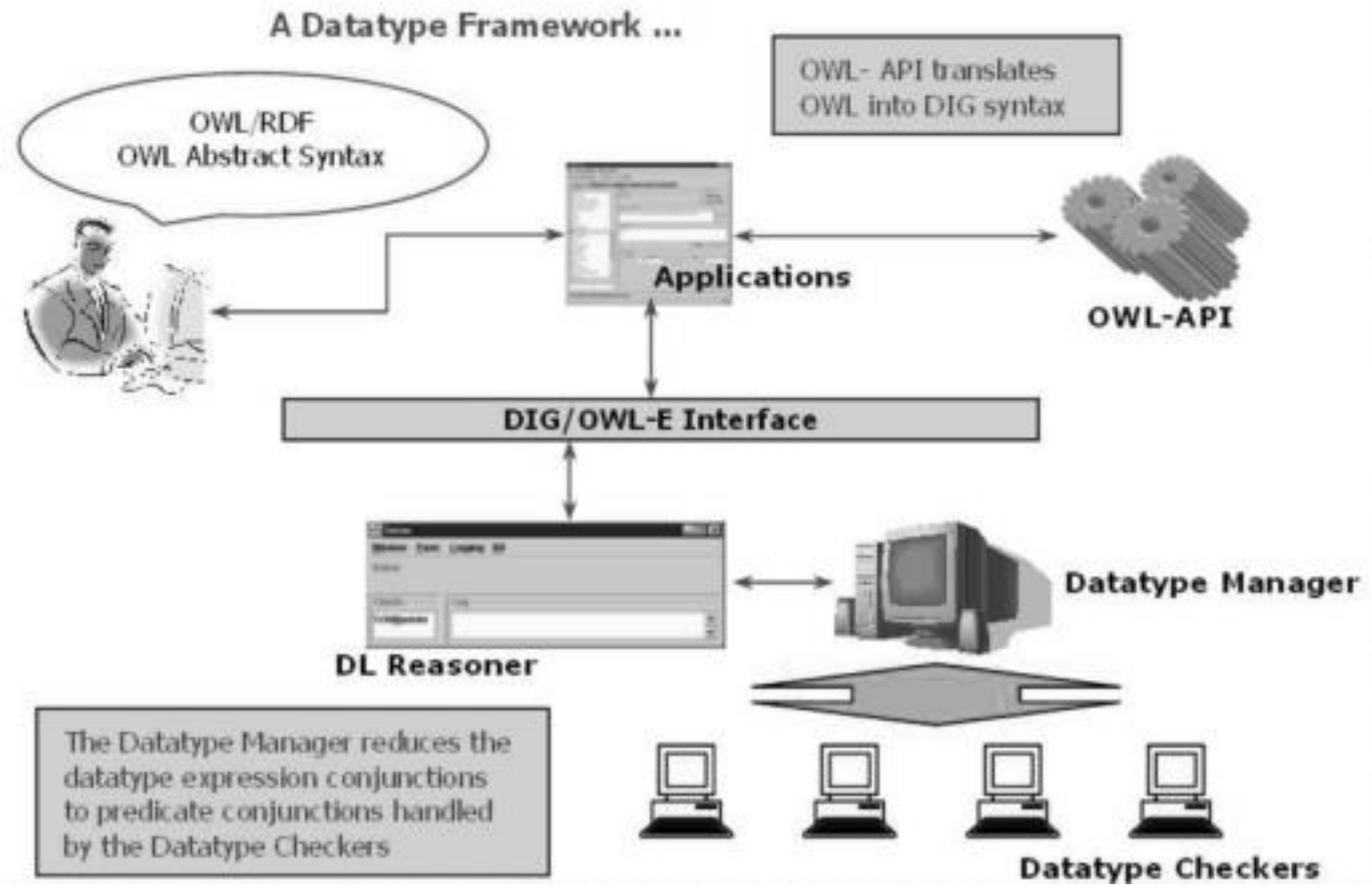

**Figure 1.2:**Flexible Ontology Reasoning Architecture

The working of the framework will be quite simple. Researcher will generate simple query containing the keywords related to his/her research. The search will be query based; scrapper will get the keyword and match those keywords in targeted research portal to get the relevant data according to algorithm and ontology. Some useful string operations will be used to collect information from this bulky amount of data for extracting useful data.

The main purpose of the framework is to decrease to work load of researcher and facilitate him/her. It will also helpful to increase the capacity of researchers and interaction between the emerging fields of research. It will also ensure that researcher can easily cover whole area of his/her interest.

Numerous sites are giving the semantic component to get to information on the web base on retrieval and scratching strategies. Scratching system serves to concentrate retrieved information from the web as indicated by our needs in the support of Ontology. For specialists this is a period and exertion expending procedure to gather precise and helpful information for their exploration by interfacing with diverse sort of sources. This is essentially composed as a responsive system of framework for exploration entries information utilizing semantic web innovation. The Semantic Web takes and gives the arrangement further. It includes distributed in languages particularly intended for information, Resource Description Framework (RDF), Web Ontology Language (WOL), and Extensible Markup Language (XML). Hyper Text Markup Language (HTML) stated reports and the connections between them.

## 1.2 Semantic Web Architecture

The fundamental structural planning of semantic web contains Identifiers (Uniform Resource Identifiers) and character code as Unicode. Over this layer is the Syntax layer, characterizing the syntactical relationship and the base here is XML. Over this layer is the Data Interchange layer with RDF characterizing. Above it the query handling part is taken care of by SPARQL and the scientific categorizations is determined by RDFS. The Ontology is represented by OWL and principles by RIF/SWRL. Above it is the unifying and the evidence layer. All the previously stated layers were scrambled utilizing Cryptology over these is the trust layer.

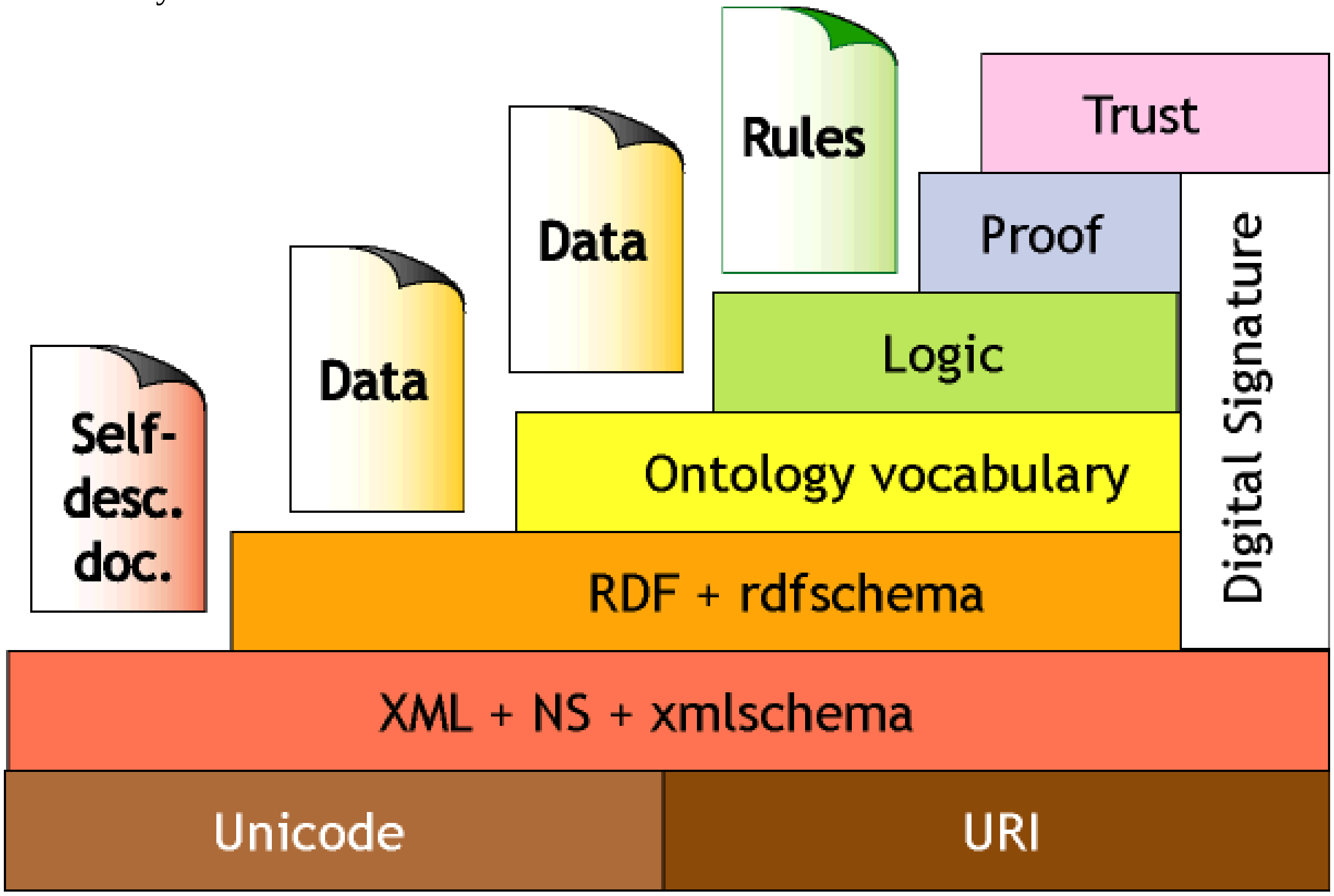

**Figure 1.3**:A Layered Approach of Semantic Web

## 1.2.1 Logic:

Logic layer makes a characterized organization and standard regulations for the internet searchers which are dependable to create the final data in semantic web. This layer which is situated up to the Ontology layer is utilized to clarify the reasonable words in machine level. In Ontology layer, researcher's machine can identify with the semantic web base ideas. Along these lines, to enhance the force of semantic processing of the search engines, the logic layer is used.

## 1.2.2 Proof:

Regulations verification implies that the machines must lay on them recovering the data in large amount measure of the data. These regulations are truth be told scripts and the software programs.

## 1.2.3 Trust:

One of the objectives of Semantic Web is to achieve trust. As in Semantic Web the pursuit operation happens utilizing the web crawlers, the clients must have the feeling of dependability about the security, improvement and modification and organized, recovering the data. This layer is vital in internet shopping from security and trust perspective

### 1.2.4Unicode and Uniform Resource Identifier (URI):

This layer shows the text and the sending method of them to the web environment. Also, it is possible to use this layer and make all web environment resources accessible and explainable. The HTML pages are the documents which are implemented in this layer. URI is used to identify the concepts in Semantic Web. In fact it is a kind of URI to identify the resources in web. Unicode is used to support the multilayered language.

### 1.2.5 Extensible Markup Language (XML):

This layer is a standard language for arranging the data using tagging and like HTML; it is script language to some extant goes through the way. But it differs in this way that the information on it is saved in syntactic form and is easily accessible or to be linked. This specification makes different elements of the web pages attributed different and favored specifications.

### 1.2.6Resource Description Framework (RDF) and RDF Schema (RDFS):

RDF(Resource Description Framework) is a XML (Extensible Markup language) based language which is created to explain the concepts and create the documents in SW. And using a set of mathematic and semantic relations, it is possible to form the relation between the data in a logic way which is accessible directly. RDF documents are able to explain the words in a way that it is understandable to the machine. Now, the different languages are used for presenting the contents of the web pages like HTML and XML.

This advancement of technologies is sort out to represent descriptions that supplement substance of Web documents reports. Thus, content may show the descriptive information put away to in web and stored in Web-accessible databases, markup inside of reports especially in HTML or XHML are scattered with XML, all the more frequently, absolutely in XML (with layout or signs put away independently).These description make possible for the content administrators to add meaning to the content to the computers for readable. That describe construction of knowledge we cover concerning that material. this such a way, computers be able to transform learning itself, rather than content, , utilizing procedures like human deductive thinking and derivation, in this way acquiring more critical results and helps personal computer to execute robotized information, computerized information collection and research.

The data of Semantic web is also become the part of the web and to be managed separately of domain, platform or application. We identify the difference of World Wide Web (www) today. This carries unlimited information in the form of documents. These documents can be searched by using computers. But these documents can be read first by humans and interpreted before any helpful information can be extrapolated. Computer gives you all type of information and always present with you but it does not understand what type of data to display and which information is to display and better for us and related to specified conditions. On the other hand, semantic web is the way in which relevant data and documents are tooshown in web.

With the goal that computers can accumulate handle and change information in helpful ways.

In case, you want to generate query of different articles on different research portal. First, you have knowledge about it. Then you can choose your better search engine. The results generated by the query are irrelevant or mixed and not helpful unfortunately. We are unable to find the accurate information. After reading different linked pages and topic reading we are able to find correct information about World Wide Web articles.

Semantic web is the only way in which we can extract the relevant results of your search. Semantic web enable the environment in which operators of semantic web is utilized for search different research articles in the web. In which different types of web services are used. At this the result will be related to the search. Semantic web agent can search and display topics related research on the network.

## 1.3HISTORY:

### 1.3.1Web 1.0:

As indicated by Tim Berners-Lee the first execution and illustration of the web, which can be differentiate by static web instead of updated blogs and networking tools. Representing the Web 1.0, could be considered as the "read-only web." At the end, the early web permitted clients and end user to read and search more information. Web 1.0 started with the arrival of the WWW to people in general in 1991, the users are not familiar or popular in such a way of content contribution and communication. Web 1.0 introduced to the first step toward the World Wide Web (www). In the general the Web before the "blasting of the Dot-com rise" in 2001, peoples are familiar and it is also the turning point towards web and internet. Which are totally comprised of Web pages that are linked with hyperlinks,only the owner or admin of the site can publish the search articles.

### 1.3.2 WEB 1.0 DESIGN ELEMENT

**Some common outline of the design components of the web 1.0 website:**

- Static pages are used rather than dynamic client generated contents.
- The utilization of framesets.
- HTML augmentations, for example, the <blink> and <marquee> labels presented amid the first browser war.
- Online guestbook.
- GIF button, ordinarily 88x31 pixels in size advancing websites browser and different items.
- HTML form structures sent through by means of email. The client will fill a form, and after filing form client clicking the submit button and submit their email. Customer would send to an email containing the detailed form.

### 1.3.3 Web 1.0 Example:

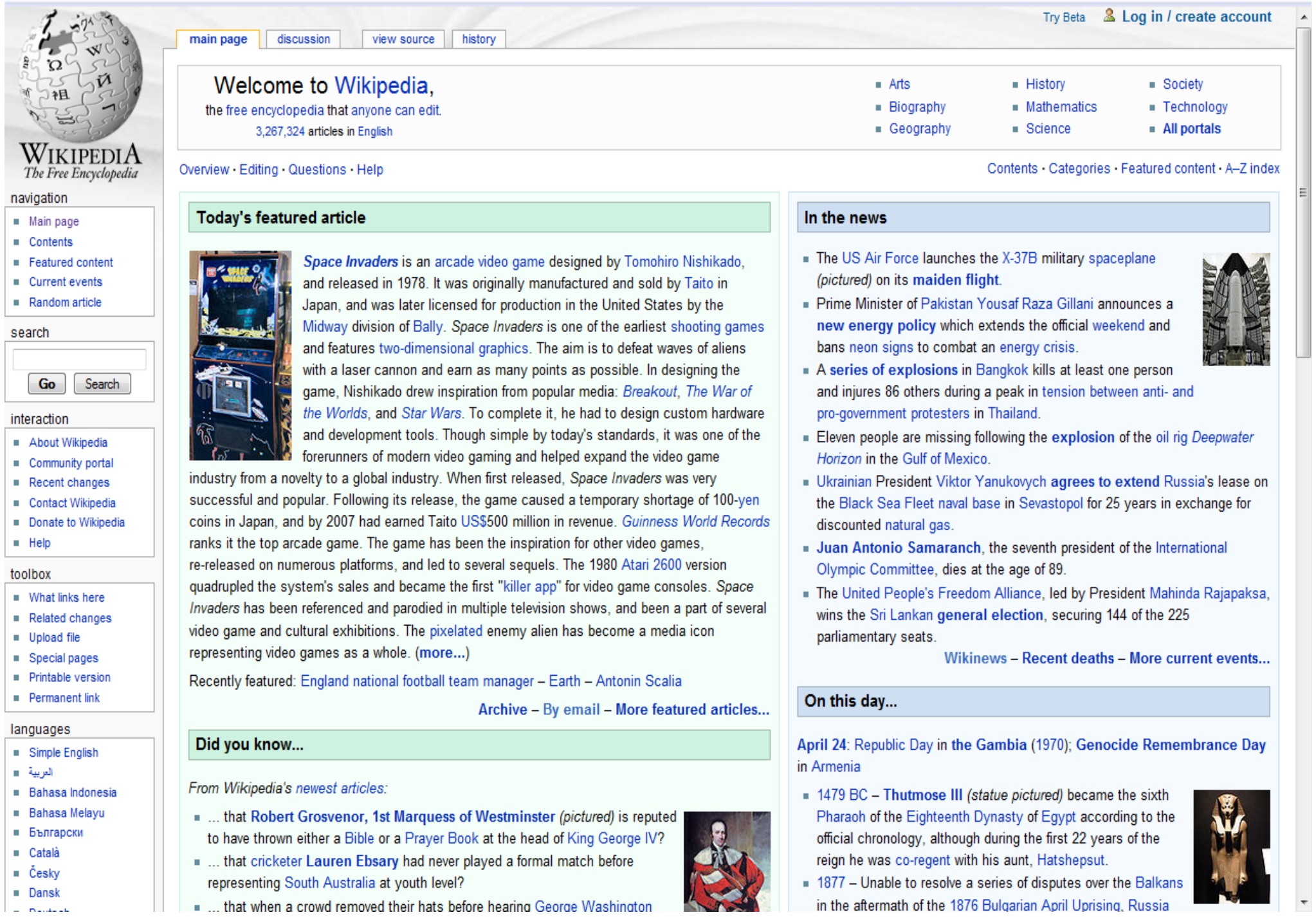

**Figure 1.4 Web 1.0 Example**

- Wikipedia is best suited example of the web 1.0 on the website permits the client to just view pages, however the client collaboration is least and the webpage is fundamentally static.

### 1.3.4 Web 2.0 :

The expression "Web 2.0" is generally and regularly connected with web applications that encourage intelligent data sharing, interoperability, and client association and configuration effort on the World Wide Web. Illustration of Web 2.0 incorporate online groups, facilitatedadministrations, web applications, long range interpersonal, person to person communication, feature sharing destinations, wikis, and different websites. Web 2.0 allow and enablethe customers to connected with many othercustomer or modifyweb site content, relatively than non-intelligent sites somewhere customers are limited to the uncomplicated review or review of information is given to it.Despite the fact that the term proposes another variant of the World Wide Web, technical specification are not be updated, but relatively to growing changes in the ways clients and developer use the Web.

## 1.4 Characteristicsof Web 2.0:

Web2.0 sites permit clients to accomplish more than simply to extract data. They can expand on the intuitive of Web 1.0to provide network as platform give processing, permitting clients to run programming applications completely through a program. Clients can possess and claims the information on a Web 2.0 site, anddoingsmanage information. These websites destinations may have architecture of participation interest that encourages customer to build the estimation of the application as they utilize it.

According to Best the characteristics of Web 2.0 are: rich customer support, customer assets, component content, and data about data, web models and flexibility. Moreover the characteristics are, for example, frankness, opportunity and total information by technique for customer hobby, can in like manner be seen as fundamental key attributes of Web 2.0

## 1.4.1Web 2.0 Examples:

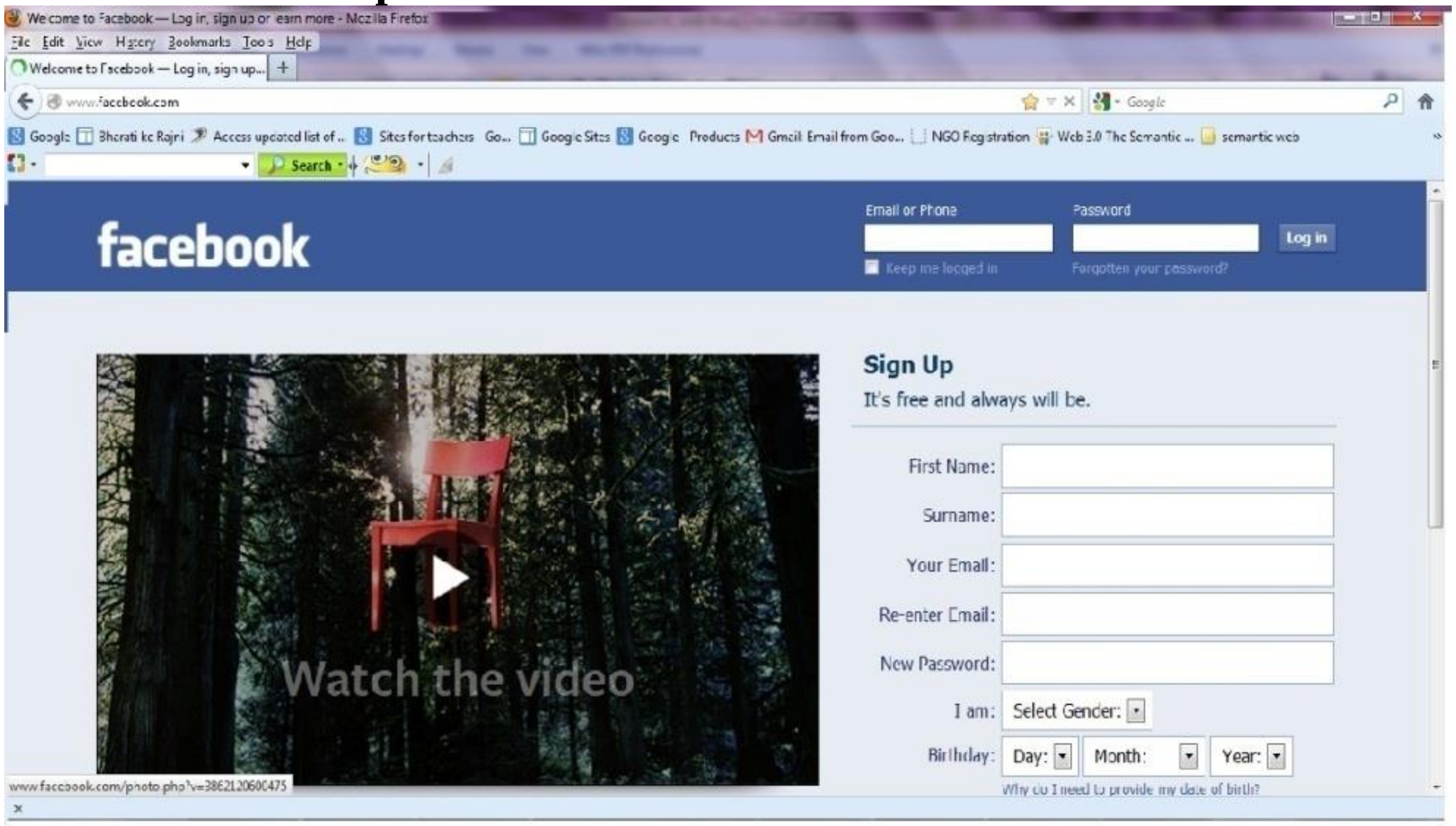

**Figure1.5 Web 2.0 Examples**

Face book is a common and prominent networking site and it is popular example of web 2.0. Face book website allows the user to share information, communicate each other all over the world, upload the pictures and videos type the message and share information all over the world etc.

## 1.5Web 3.0- A Basic Introduction

The best part of the semantic web is that machine processed the data easily all over the world. You imagine that it is the best way of World Wide Web showing the linked data all over the world. Tim Berners-Lee proposed idea of semantic web the inventor of hypertext markup language and World Wide Web etc. Committed group of people are working to show improvement in World Wide Web. These committed people improve the performance of different languages and tools, developed the standards of the organization. The growing rate of the semantic web is also bright and semantic web is also very popular. Information hidden is files of HTML is useful in some contexts, the problem is that the major portion of the information is placed on the web that is in different form like HTML , this information is impossible to be utilized in large amount because such type of system is not available to organized the published data World Wide. We need to build such type of systems which gives the information easily globally for example we can easily extract any information of local sports and plane time. Different kind of data are available and showed in different sites but data are in the form of html. It is difficult to use such form of information in some contexts that one might to do so. Semantic web shows in two forms. Semantic web shows the data in simple format, extract data from different sites. It also shows recording and connection of databases and moves to an endless set of databases and relation of data to real world projects.

## 1.6 The Vision of Semantic Web

## Today’s Web

The World Wide Web (WWW) has transformed the techniques in which different peoples can communicate one and another through web and increased their business through communication on the internet. It placed at the heart of an uprising which is right now changing the created world towards a learning economy, and all the more extensively talking, to an information society. This advancement has additionally changed the way we consider PCs. Initially they were utilized for computation numerical problems. As of now their prevalent utilization is data handling, normal applications being information bases, content preparing, and recreations. At present there is a move of centre towards the perspective of PCs as passage focuses to the data gateways. A large portion of now a day’s currently Web contents is appropriate for person utilization. Indeed, still content produced naturally from data bases is generally displayed not including first basic essential data found in information bases. Mill employments of the Web today include people looking for data, seeking and contacting different people, checking on the indexes of online stores and filling so as to request items out structures, and review grown-up material to run mill. These exercises are not especially very much upheld by programming tools. Apart from the presence of connections which set up associations between reports, the principle important, without a doubt basic, sort of apparatuses are web engines. Decisive word based web indexes, for example Google and AltaVista and yahoo are the basic principle device utilizingwebsite today. Itis clearly understandable that the Web do not have huge amount of achievement it was,not possible for the web crawlers. Yet there not major issues connected with their utilization.Here we list the fundamental ones. High review, low accuracy, Even if the principle applicable pages are recovered, they are of little utilize if another 28,758 gently pertinent or immaterial records were likewise recovered.

• **Low or no recall**:

Routinely it shows that we can’t retrieve any answer for our interest that we requested, vital or more important significant pages are not recovered and retrieved. In such a case low recall is a less successive issue with flow internet searchers, it does happen. This is frequently because of the third issue.

• **Results highly sensitive to vocabulary:**

Often we need to utilize semantically comparative essential words to retrieve the outcomes in which according to our expectation, demand and need to wish; the important archives utilized many techniques to the original, unique query. Which is conduct insufficient, since it is only possible by semantically comparable queries should give you the same comparable results.

• **Results are single Web pages:**

In the event that we require data that is spread over different reports, then we must generates many queries to retrieve and gather the related documents about generated query, and after that we should manually separate the fractional data and set up it together.

## 1.7Motivation

A Framework for Semantic Organization of Research Portal Data focuses to facilitate research work. Information technology is facilitating many different fields but also increase day by day of data is creating many problems and ambiguities. It is very difficult to collect relevant, significant and authentic data from huge amount of information available on internet. There are many types of research portals available for playing their roles to facilitate research work; the project is going to open a new field to facilitate research work.

## 1.8Problem Statement

There is a lot of need to gather information which is more relevant and precise, efficient, well organized amount of data collected from many research journal and portals. To solve such kind of problems we need to develop a framework which is capable to understand the functionality of different research portals to gather information and manage it semantically.

## 1.9Objective:

The goal of this exploration research work is to build up an extended new approach which will make the strategy simple to plan general data retrieval framework for semantic organization to meet the requirement of user.

## 1.10 Scope of Research

The research is open to introduce a very large field in research topic areas. There are many latest research journals are available, introduced in market and lot of working as well in market. So, large no of research journals and portals can be share their information at one place in future.

# 2. Review of Literature

Horrocks and Schneider (2003) explained that the Semantic Web was relative major dependent on a formal importance for the constructs of its language. For language of the semantic website to work well together their formal implications must utilize a typical view. Else it won't be conceivable to accommodate records written in many languages. The proposal of representation fundamental Resource Description framework (RDF) and RDFS is especially difficult in manner, as it has a few different unordinary perspectives in both syntactic and semantic. It is additionally not more essential to necessary for people to characterize the language structure and manufacture utilized for each kind of data exchange. This is taken care of by HTML, XML and related standards. Burner's layer cake is also displayed in figure initially introduced in XML. The graph delineates and shows that a Semantic Web Framework in which language of expanding force is layered on each other at top. Unhappily the connections between neighboring layers are not determined, either regarding sentence structure or semantics.

Celik and Elci (2006) describes writing about something new or preferable regarding to explore in the terms of semantic web services that deal with client in a best manner to fulfill their deeds. This system may be use ontology web which can be describes services and its languages (OWL-S). There may be a lack of semantic parts which is more prominent in quantities of web administrations in the web fields. Analysis of these services in the search operations may become problems or errors regarding to current web terminologies and their techniques. Search agent that can also use to find links related to the Semantic Search Agent (SSA), it may utilized to locate links which are rely on semantic web services as fulfill the requirements of clients requests.SSA Performs a two types of functions which can be defined as their names enhancing. Enhancing is a term or technique which treat synonym. The second one is matching the Semantic web software l/o and with the link of client l/O. We are using two types of properties for the increase or improvement in semantic for the input terms of client .we told you that kinds or properties are synonym. We may apply different type of properties that can be employed as demand and needs of the interest of its controlling effects, when we may greater the number of its properties the performance can be better in the way of improving.

Dehors*et al*. (2006) stated that Reusing existing web assets for e-learning is an extremely encouraging and exceptionally advanced thought in the examination field of website based education of online training, particularly for wise or versatile frameworks where the expense of composing is huge. Systems instruments are as yet missing and achievability at sensible expense is a pending issue. In paper we suggest a unique method to deal and reuse learning assets in the view of information building, designing and Semantic Web advancements. He looks the web, uses learning item stores or contact partners for accessible pedagogical material. A consequence of his pursuit he chooses a slide demonstrate that, to the extent he can tell, contains essential ideas of the course. Keeping in mind the end goal to help understudies stayed with paper gifts and encountering inconveniences discovering data, they choose learning substance ought to be made open through an keen web framework. A presentation of the pedagogical substance taking into account express semantics ought to encourage understudies in perusing and comprehension the course.

Namgoong*et al.* (2006) stated that web and its information are persuading the chance to be exhaustless different scientist's favors in these gets some information about information and resources. A Semantic web association are establishing web contents and showing to the computer machine that can be clear up the semantic web. The Semantic web was finds course that provide for the clients to locate a best web associations to the clients. They could be a spotlights over semantic to discover the consistencies and among its point by point position to contain particular results. In the context it additionally figuring may be upgrade by matching algorithm to build its outcomes progression as per requirement of a client and persuading in web associations as well as it could be less.

It is additionally prerequisite of viable administrations with numerous sorts of inputs and yields in that on the grounds that when there is no coordinating information accessible in web administrations. In this manner we are expecting a framework that can be give solid impression discovery administrations to a system base robots and clever specialists. That can be applying in semantic coordinating and it can be attached calculation for the development revelation administrations.

The algorithm can register and usage semantic between the customers need and its promotion that can be requirements. The utility that can be expected worth that mean the publicizing is a same with the solicitation in a custom. There for an aggregate assemble of web administrations can be give capable a customer valid way web administrations when there is no coordinating web administrations or present web administrations. With the goal that we can expect that it is a valuable and semantic degree that is compelling to web administrations to a same semantic and its recommended structure. As with respect to future work we can broaden a mean full arranging systems and strategies for that work process in element shape so it will make wise operators to do make and change is work process taking into account environment and conditions.

Pereira and freire (2006) This article describes the increment of tools related to semantic web since last three years but these semantic tools has very strong limitations. The common limitation of these tools is that these are designed to work with a very few or limited layers of architecture related to semantic web. So the market strongly needs to develop a tool which can integrate different layers to utilize maximum number of layers in minimum time frame. Architecture is proposed in the research paper for that specific limitation named "SWedt". The tool focus on fully standardized layers and their functionalities as well as just standardized layers to integrate and it allows the user to create semantic web documents easily.

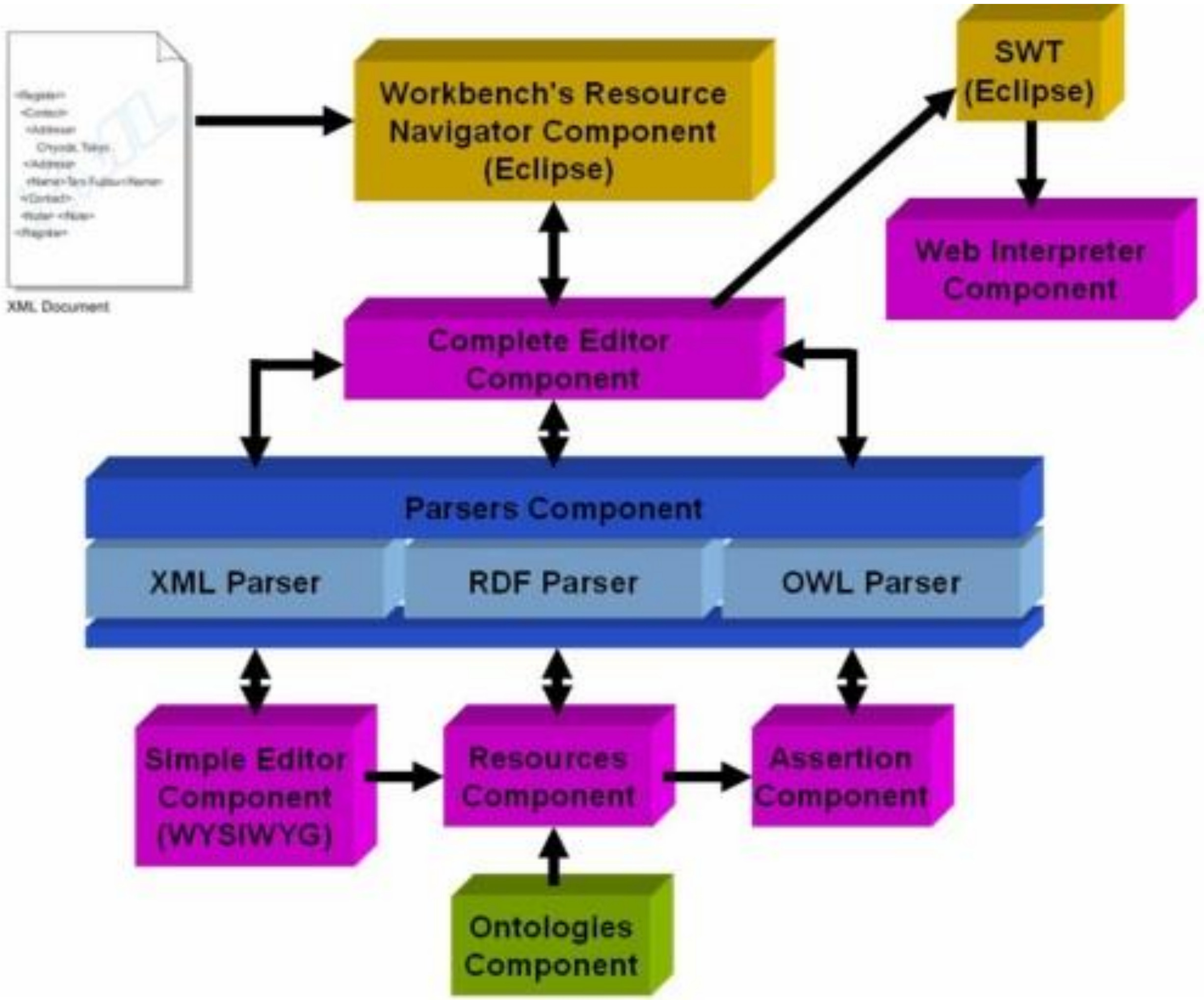

**Figure 2.1:**SWedt Architecture

Bell *et al*. (2007) described that most of Semantic architecture are in the form of ontologies, that are used by many web services aims to find and utilized methods that gives us sustainable in recreating models that use their self again and again. Actually semantic web in the form of ontology, used by web inter-operation and can be obtained by using component analysis that recognizes the essential subcomponents at the level of unreliable granularity. These approaches commonly connected with software and many other web services that are able to design and retrieved in order to make data understandable. These types of approaches can be simulated in form of models that focus on the cross domain. In this search it was claimed that large no of web semantic models provides the opportunity of reusing data with cross domain functionalities. The reuse of a copy components models are use for joint working in an organization that can be restricted the same semantic problems and with theory issues. The current situation and any action or conduct that can be turned into a statement of fact in web components that are powerful in statics in nature reducing the fundamental semantic mainly; for the most parts of web components that are nature deficiency the main semantic Material or masonry used to support a structure that is requirement to explore the new format techniques in semantic web. Semantic model is the form of ontology that can measure and provide an approach to support act of process model to reuse, semantic explanation is the achieving when we can establish a identity for the requirements and verity levels of ontology. The ontology is engineering framework and the discovery how such that ontology can be created. This is mentions' and subsequently that are used for a model reuses, the ontology is the consistent services in hospital and the national blood services, the ontology engineering is the framework that invented. However the uncovered domain is the semantic and adopting the Intruding or tending the connection between the participants. Regardless specific CSPS is the work that has wide connection for the simulation community.

Bo *et al.* (2007) described that semantic web is able to compute and calculate the algorithm of data. Data is illustrated that as a result of to relate critical issues and its characteristic to elaborate the information to topological machine working semantic web is developed on the techniques of computations in base of its data and subject to source. There for two creators should be developed to relate semantic web dataresults show that data and information as well as references will be effective as a result of search query in semantic web. Web semantic is capable of compiling or computation.In semantic web we need to tell you about the issues of information and its quality and its applications contradiction and its excess .in these terms we will present an ontology that must be based semantic trust that can be computed as its methods of information regarding to this method can say that can be exist in our real society. We can generalize these as link of semantic web by each and any node on the calculating the trust to neighbor nodes.

Li *et al.* (2007) declared the semantic web as a breakthrough in data processing and named as "big bang". The most important and useful tool for collecting data from different sources of data is search engines. The machines, these days in semantic web, care only about the location and the type of data. Therefore a lot of loopholes still exist in search engines and the most popular and advance search engines are still unable to satisfy user. The article proposed a search engine based on relations and implemented in virtual environment. In that virtual environment of semantic web tested the architecture and major algorithm related to relation-based search engine. As the search engine is based on relations, these relations and concepts are defined using RDF. The search engine retrieves data by following these concepts and relations.

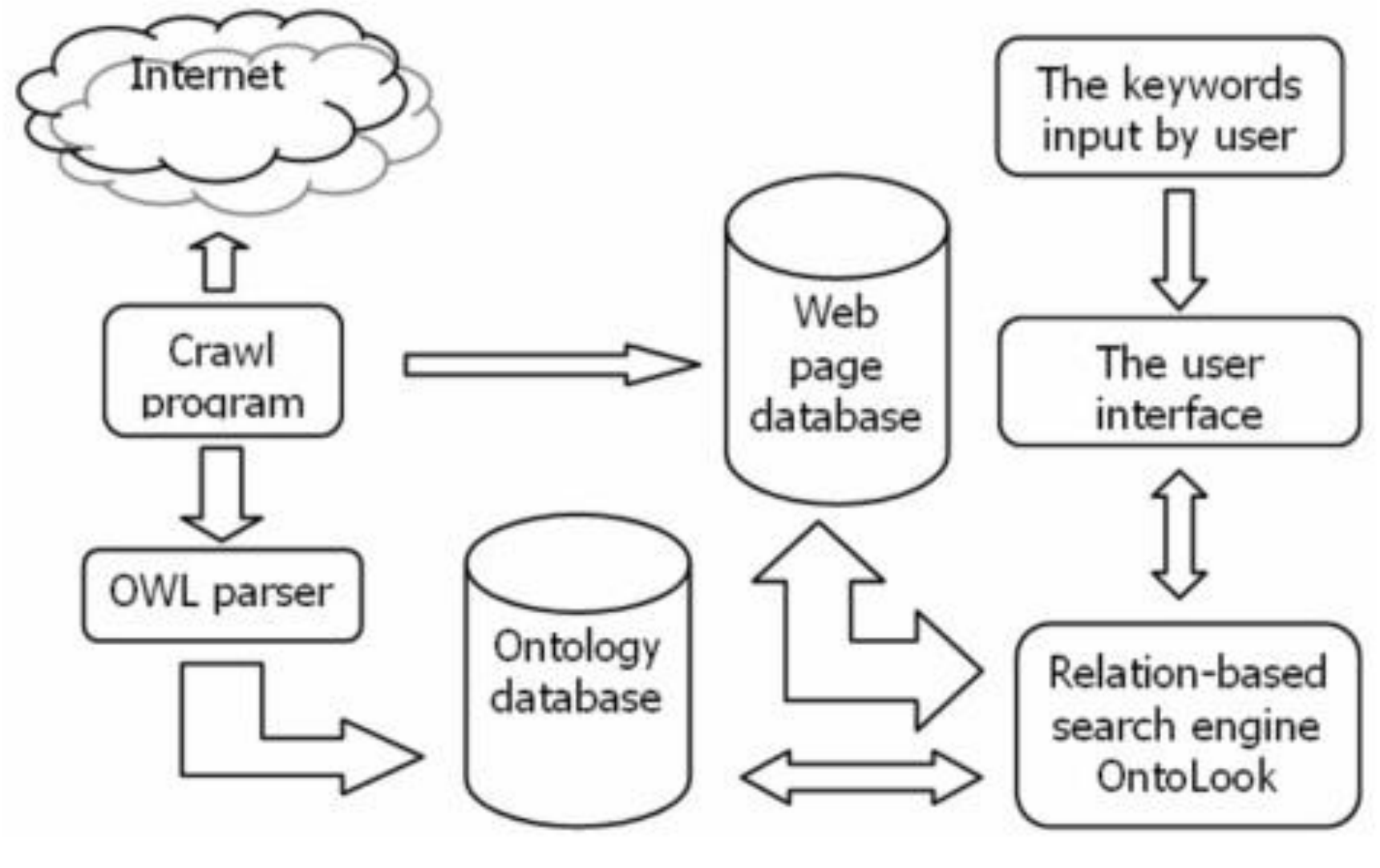

**Figure 2.2** Ontology Database

Pan (2007) make an investigation for KDD (Knowledge Driven Data) in era of web semantics.That paper was published in IEEE and was entitled “A Flexible Ontology Reasoning Architecture for the Semantic Web”.This article was explaining that any model based on KDD can utilized the features of semantic and it enhances the capability to extract meaning full required data. The author of this paper purposed a reasoning framework which was flexible in perspectives of finding the recreation of the samples as well as web ontology languages (WOL). The discussed features of semantic model was providing the facility to the client in terms of defining is own data types. In this model a client was able to change its required data type along with adding new one. This web semantic model was very useful to customize the required data according to client desires and requirements. These two features of framework provide the facility to the user to define their own predicates and data types as well as user can change its data type.

Farshad*et al.* (2008) described that web semantic terminologies are becoming more and more popular due to its effectiveness on different fields of research. In the ear of science and technology, software agents in multi agent system (MAS) aim to find related data from different many resources. These frameworks produced flexibility, adoptability to get different kind of composition tools and technology. XML and JSON format of data can be moved from one node to another node easily. And browsers can understand their format. In this research author proposed architecture based on web scrapping and claimed that this model was fit to their requirements and was able to deploy on WWW (World Wide Web).

Taswel(2008) stated that initial stages of development web semantics are easy to work in flexible environment. Taswel stated that semantic web is in remain state of development .it was not yet achieve the targets enclosed by its creators. The web browsers domain name systems are able to recognized published web portals in the original form. More over unlabeled portals can be purposed for basic infrastructure to gain Meta data of web portals with in a framework that will be able to serve as a bridge among semantics and original portals. The name of this domain system can be DNS can be browsers and it can be benefiter and motivated to the people and clients to register to publishing their web sites as the cost of original web. This is an analogous resource label system. The benefits and working together in semantic network that will motivate the people to register resources able labels in semantic web it is the Domain Ontology Oriented to redevelop resources. Portal and doors are spouse to be a as a system for these type of registrations label tag and others details are in description. They are analogous to IRIS and it’s domain in registration process.This kind of door and portals are design to be helpful semantic web.

Gang (2009) explained that existing web administrations advances are just syntactic level. So we can have still semantic crevices that cloud is cross space asset advancement or disclosure differing in character or substance asset question. Interpretationsstarting with one place to then onto the next at the level of semantic. So, that semantic web is looking to completely demonstrate the enormous or more prominent data asset of the web as information that PCs can clarify significance consequently. With the goal that this paper can clarify how semantic web advancements included qualities in a building semantic geospatial web administrations utilizing rest that can be based outline approaches. Such as a computer or computer software, that can be integrated into or used with another device like GIs applications and standards.

Hitzlerand Harmelen(2010) explained that Semantic web clarified large number of developed strategy to recover shared data. At the point when an operators like a web administration, often the database finishes a message, the specialists has numerous implicit questions as a top priority for the most part. Standard exploration on this subject confronts genuine difficulties, which compels us to question set up lines of examination and to reconsider the hidden methodologies.Message created by the specialists contains large number of significance. A sensible semantic web must have the capacity to characterize the significance in extraordinary way. Semantic web advancement begins from the extremely essential idea (learning representation and thinking) towards certain objectives. The objective of semantic web is to have the capacity to make frameworks and association, these institutes and frameworks would be to cooperate in all ranges of organizations. As per the article in nowadays semantic web is utilizing lighter semantic methodologies and will be skilled to utilize advance semantic methodologies in 10 years and is ready to control huge information and intensely include in zone of content reading and machine interpretation.

Ma and Xie (2010) described that clarified that Automatically getting their structure is the condition that must exist or be set up before something can happen or be considered to markup language and which is best programmed sharing consolidating of discrete components or substances to frame a aware entire programmed way in which such parts are joined or related of semantic web administrations. So we can say firstly to break the Owl an unpredictable entire so shaped for composite procedure in light of philosophy reasons which can be understand the programmed displaying of semantic web administrations with Petri and set forward for thought an innovation methodology of dynamic and programmed consolidating of unmistakable parts or components to frame an entire semantic web administrations. Presently we are presented the technique for pass on a thought or impression of; portray to depict programmed for semantic web administrations procedure models to incorporate programmed and made up of unmistakable segments procedure models. With the goal that work purposes a specialized focuses for the element and it can be the demonstration of assembling or making up by joining parts of semantic web.

Jayatilaka and Wimalarathe (2011) stated that the research describes that the roots of semantic web is ontology and these ontology have key factor in semantic web that's why is becoming the future of all web technologies. But the ontologytakes much time to construct and the designer of the ontology needs important domain knowledge because most of the ontologyis designed manually. The main hurdle in the fast growth of semantic web is the same reason. The paper is focused on the problems occur during extraction of knowledge from huge number of web sources for development of Ontology. Research has a new introduction of fresh source for semantic web that is web usage patterns in ontology learning. The web usage patterns have two parts of its methodology; one is web content mining to collect the useful and relevant information and the other one is web usage mining to observe the behavior of web user. By combining these two parts much useful and more realistic conceptual relationships can be get. These concepts can also be used for search engine optimization purposes to rank the most relevant and useful website higher than other websites.

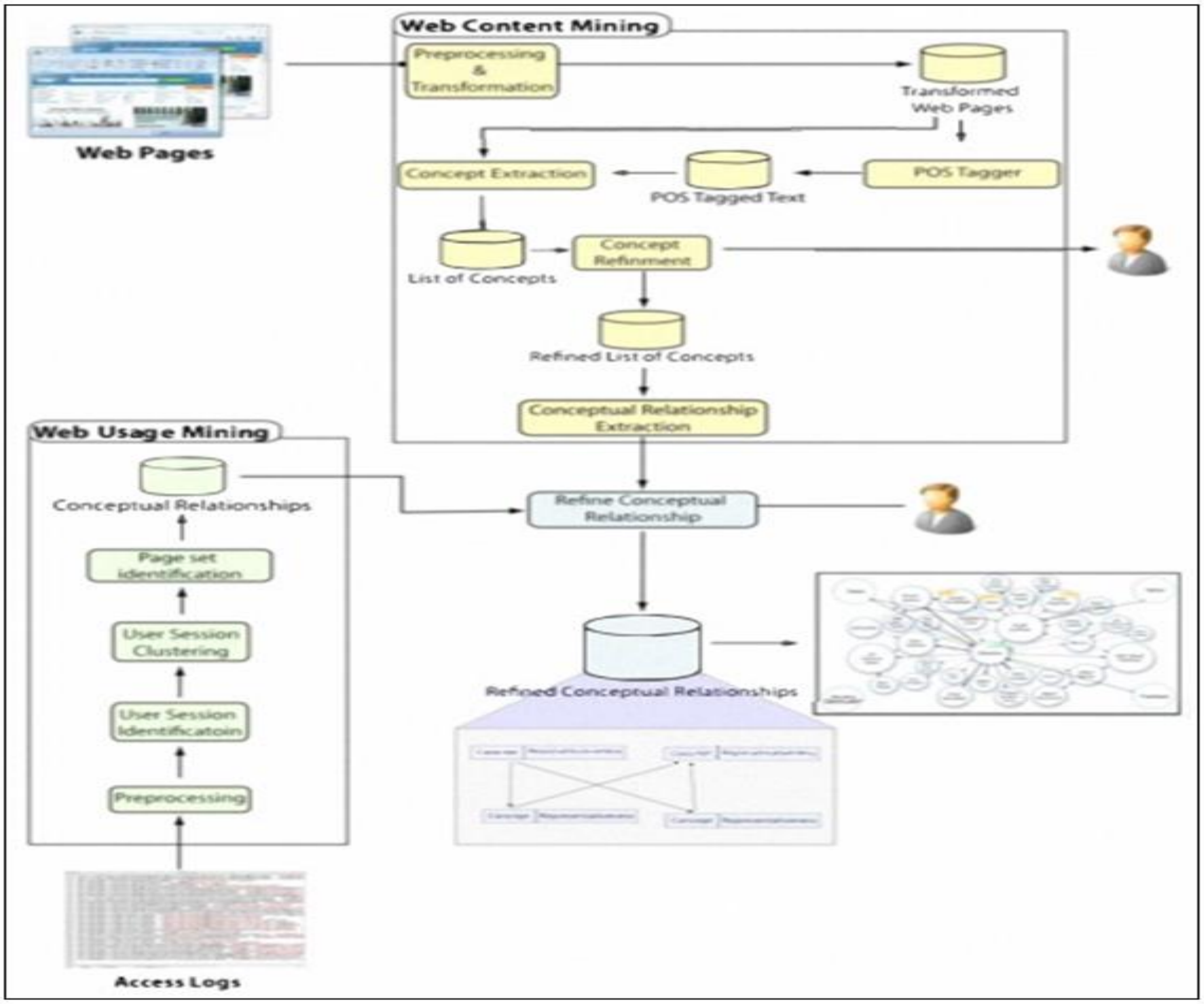

**Figure 2.3:** Knowledge Extraction for Semantic Web Model

Liu *et al.* (2011) explained that in application based on large data base, the web pages are depended continuously. The web data must be integrated on a wide scale to use these resources professionally. To achieve the goal of deep data integration the key factor is to use deep data efficiently so the primary work is to discover deep data sources. The paper proposes a mechanism to discover a method for deep web data entry. The first phase of this process is to establish ontology by forwarding a deep web entry. The second step is to form a crawler that will judge the web forms by crawling in all web resources one by one. Some of the web pages will extend their attributes and if the value of attribute is greater than the value defined in ontology the page will be download. It will complete the process to find the high quality deep web pages.

Sanjay and Rizvi (2011) explained that Retrieve helpful data from the web is the hugest issue of sympathy toward the recognition of semantic web. This may be accomplished by a few methods among which Web utilized Mining, Web Scrapping and Semantic Annotation plays an essential part. Web mining empowers to figure out the applicable results from the web and is utilized to remove significant data placed behind the server patterns discovery. Web utilization mining is a sort of web mining which mines the data of access routes behavior of clients going to the web locales. . It can be accomplish by numerous routes with Semantic Annotation, Web scrapping assume a vital part. Web scratching, is likewise method, is an advancement of separating useful data One of the another method is web scraping is a procedure of retrieval valuable data from HTML pages which may be executed utilizing a language of scripting also called (PSP) prolog server Pages depend on prolog . The third one is a Semantic explanation is a system which made it achievable to include semantics and a formal formation to text based documents. An essential feature in semantic data retrieval which may be conducted by tool which is called KIM (Knowledge Information Administration)

Villamor*et al.* (2011) stated that Regardless of the expansion of Semantic Web Services, just a small amount of data in the Internet gives a semantic access. Late activities, for example, the developing and rising of Linked Data Web are giving semantic access to use information by porting accessible assets to the semantic to utilized different technologies, e.g. database scratching and semantic mapping. This article proposes a non specific structure for web scratching in view of semantic advancements. This structure is organized in three levels: scratching administrations, semantic scratching model and syntactic scratching. Characterize a system for web scratching for the extraction of RDF diagrams that speak to substance in HTML archives. We have utilized this model to manufacture a scrubber uses RDF based extractors to choose sections and information from web records also, manufactured RDF diagrams out of unstructured data.

Xie*et al.* (2011) published an article entitled “Ontology based semantic web service clustering”. According to the research it is very important operation to discover web services in architecture of web service. The research focuses to classify the web services before it perform the operation of web service discovery that is based on ontology defined by OWL-S Language. The research uses some sort of functions belongs to web service and identify the similarities amount these functions. A mathematical modeling language named Petri net is used to explain the process of web services. The similarities among these web services are defined by an exact concept of semantic resemblance of domain ontology. The concepts of semantic similarities have different aspects like concentration of concepts, connection between two concepts, and pathway between the connected concepts and the antisense of these connections.

Xu et *al.* (2011)explained that the article portrays that semantic web ontology is not able to comprehend or right the wrong data set in semantic web. The article proposes an achievable arrangement in type of semantic web. It is an expansion to semantic online on learning to bolster the functional execution of semantic web. Essentially is the primary centering purpose of the application which is taking into account ontology and it unequivocally need to review consistently. So it is vital to receive such a system which permits enough space to ensure enough access effectively. The procedure would be done on the bases of characterized model of semantic web unpleasant ontology utilizing a standout amongst the most widely recognized media of capacity, the proportion database. The paper additionally depicts the capacity source seriously and the ontology is anything but difficult to store in social database taking into account plan

Alafif and Sasi(2012) explained Semantic web index is the new era of traditional web crawler that extracts exact and significant data from the web. Search client query utilizing Semantic Web Documents (SWDs) to answer the query that are found in ontology’s record. The range of query was greater than one range in domain. The semantic web crawlers, for example Swoogle, and Watson don't distinguish the ranges of client's question while extracting the related result of the query. Extracted list items are not in a single range of the client keyword. Domain and Range Identifier (DRI) components are already placed in existing web crawler to solve the problem. DRI components are utilized ontology that shows Semantic web Documents (SWDs) to recognize and categorized the domain and range to validate in domain.

Budan and Chen (2012) stated that BPEL is a famous standard in SOA (Service Oriented Architecture) venture for indicating business process. An engineer needs to utilize particular programming tools to enhance develop process structure and determine Web administrations input output of the process. The time has come devoting to indicate each business process from low-level Web administrations. The author suggested semantic expanded BPEL model for SOA process era, and actualizes the model in useful BPEL (Business Process Execution Language) IDE (Integrated Development Environment).Arrangements to consolidate more resources in resource store to make conceivable adaptive process. The structural planning comprise of an Ontology framework for ongoing connections among distinctive reusable resources. The layer upgrades multi-layer displaying components.

Choi and WooSeop(2012) stated that information was increasing on the web due to social media and related attractive services of the web.Information on the web has been expanded by online networking, media and extended day by day new technologies. To use these colossal information incorporate sensor information, semantic sensor web maintenance stage is required. What's more, it can give the setting beneficial semantic web service. The fact that sensor produced information is so powerful; it can't generally upgrade and restore the information. Hence, the conveyed semantic sensor web structural was planning to partition preparing of setting context data and service oriented data.

Gherabi*et al.* (2012) stated that information on the Web was increasing in social databases. Subsequently, it is more critical to create the relationship between social databases Ontology for store information on web. In such cases, we get another development system to relate the information store in interactive or inter-related databases the Semantic web. A system has created, which move effectively RDB (Relational Database) into OWL (Ontology Web Language) report. The outcomes are essential to demonstrating that the proposed system is reasonable and effective.These agents are responsible to increase or decrease the quality of data. Both the consumer and the provider of the web pages are involved in these quality attributes.  The quality of Data is named as information quality of the web pages. The paper describes some concept of security and reliability of web semantic based on some ideas of dependable semantic web which was proposed by the same author.

Li and Yang (2012) explained there a great dealing of web activities that are present in web surrounding. Websites and their web services can be enhanced and categorize that right explore innovations so that web services is allow users to have many types of choices so that is very difficult to semantic evolution. The response rate of this web service is very low. so that it is the method which is approved to be a possible and analytical and agreeable via an instance experiments, it can be compact that bring less semantic computation and promotion reaction rate by classification of some Unrelated or inapplicable to the matter in issue of web services .on a more notes that is experience of user that can be improved and better. So now in this paper we are presenting a group of the same or similar elements gathered or occurring closely index that is based on user need and demands by records the hit rate of the services to choose best option in cluster index by using the same clustering method. The cluster type of the manifestation so, we can't compute semantics, so that close it is great to improve the ability of semantic query. Still there are some issues that are needed to improve or which can be better our future. Like these types when we getting cluster more effectively with the ability to do best using of information and we can talk in to deeds. Now we will keep it to improve and develop the cluster index well in future.

Malik and Rizvi (2012) published as we know that for satisfying the sensible web, Ontology is one of the most significant backbones of semantic web. Its importance may not be denied for understanding the mission of semantic web to include semantics or machine reasonable information to the current intelligible web in order to encourage a more effective information extracted and reuse on shareable premise on web. It implies that Ontology empowers to formation and conceptualizes the common information of a specific domain. However the major testing issue is the enormous decentralized nature of web which does not allow having a solitary extensive Ontology however left with the likelihood of having a few littler ontology which are hard to actualize and organized. Presently, Ontology has a few discriminating viewpoints where Ontology configuration and improvement is one of the major issues which may be best acknowledged with the assistance of a contextual investigation.

Paliwal*et al.* (2012) explained that it is unreal like it can, admittedly, be hard to distinguish between being idealistic and being unrealistic. As new services are introduced they are connected with semantic descriptions. In this description many of them already connected with semantics clearly developed or formulated. These lattices are lay on the nature or quality or character affiliation parts of web administrations examples .this our methodology may be use to semantic relationship positioning for setting up semantic importance and a hyper faction example is utilization for gathering web administration parameters into significance full association. There terms are re unions by semantic relevance and that must be the mechanical advantage to discover rank web services. We read in numerous books or books about this sort of web methodologies for web administrations revelation that may be re union semantic and measurable affiliation measurements. Semantic measurements are in view of semantic period of Ontology. These measurements taking into account the affiliation parts of web administrations for this situation when their info or yields are additionally concern. This is our Ontology that get-together semantic relationship.These kinds of organizations may be gather by semantic significance are then mechanical term to discover and rank web services. These kinds of terms which use pattern of process or may be contract prominent of semantic associations it can be proved authentic source or more effective web services.

Pretzsch*et al.* (2012) explained that data is published on large scale on internet and most of these sites are unable to arrange data in proper format. These types of webs are the large way of informative valid data. Now a day this type of data is not available by professional publishers, but there is a fact that everybody who is using these data and generates information can be satisfied. a very big part of that types of content are located in that web forums. At that stage it is very easy to share knowledge which can be on our access. So, that the size making a very difficult to do discussions on that relevant topic. Automatic system is a filter and rely the points of information, unluckily a state of satisfaction is present in a human-readable designs and layouts and that cannot be planned which can be process able by its automatic system. So that is very important to separate the content from the web discussion and the layout design by doing information and the process of any business. This is the paper which shows FedEx, that is system for automatic data and its selection .the selected data from any forum that is match able it to unified data scheme. The paper presents the process of rebuilt a unified data schema for a big set of forums. We are further discussing how we can find process to obtain despite resistance factors with in posts such as a writer author body text and publication data from all types' forums follow the structure of their discussing boards.FODEX is the system that can make a test the relevant or appropriate of those concepts are achieved good results on some factors and searching forums posts from page. that is our work now to direct or draw toward a common center points that show a wrong terms and results .it is our plan to unite factors classifications with a rule that is based on approaches ,these extraction is Composed or appearing by refined block segments levels to token level. These results are going to be presented in future publications.

Sujan*et al*. (2012) declared that Semantic web build the information of the web machine justifiable and ontology is the center of semantic web innovation and enhancement. Ontology catches the area learning in a non specific way and gives a usually agreed comprehension of a region. It helps the internet searcher to figure out the significant and experimentally right data. It also helps the researcher to find the appropriate and tentatively the correct information. Majority of the data on internet and unstructured to the point it is difficult for the computer to understand and these data only understand by the humans. Yet the measure of information is so tremendous that they must be handled by effectively by machines. The semantic webs frameworks and its structural engineering makes the information machine justifiable that will help the internet searcher to do significant, vast search. It upgrades the quality of the data benefits and investigates data administrations.

Yu *et al.* (2012) stated that According to the author, the development of semantic web is increasing day by day therefore there is a strong need to evaluate the quality of semantic web. Currently available tools to evaluate semantic web are also based on some specific applications or same technology of semantic web which is not enough to evaluate the general features of a semantic web. The critical points are collected from these characteristics of a semantic web to evaluate the balance. After summarization of these critical points the key factors of balance evaluation are presented. The research is helpful to open a new angle to analyze the semantic web. The author further explained that balance is the key in life and everywhere it's important in all areas from the art and it's detailed. Search work mainly when we want to make an application that is used in our daily routine. Recently growing rate increasing day by day in Semantic web applications are displaying. So that there are a few types of numberresearch on the computation and bench scratch in semantic website technology itself and can be special application types, when we are computing a general thing of semantic web applications that can't be sufficient.We are trying to analyze general factors for the computation of the vast range of semantic web and its applications for the function of future working out. So, now we are mainly focus on the demonstration of the major features of computation and its progression factor .Our analysis survey tell us the recent semantic web application future development of semantic web evaluation. So, we are able to give recognized matrices for the act counting of stability computations and regarding too simple factors and evolution of typical semantic websites application that can be depend on them to count its performance and efficiency.

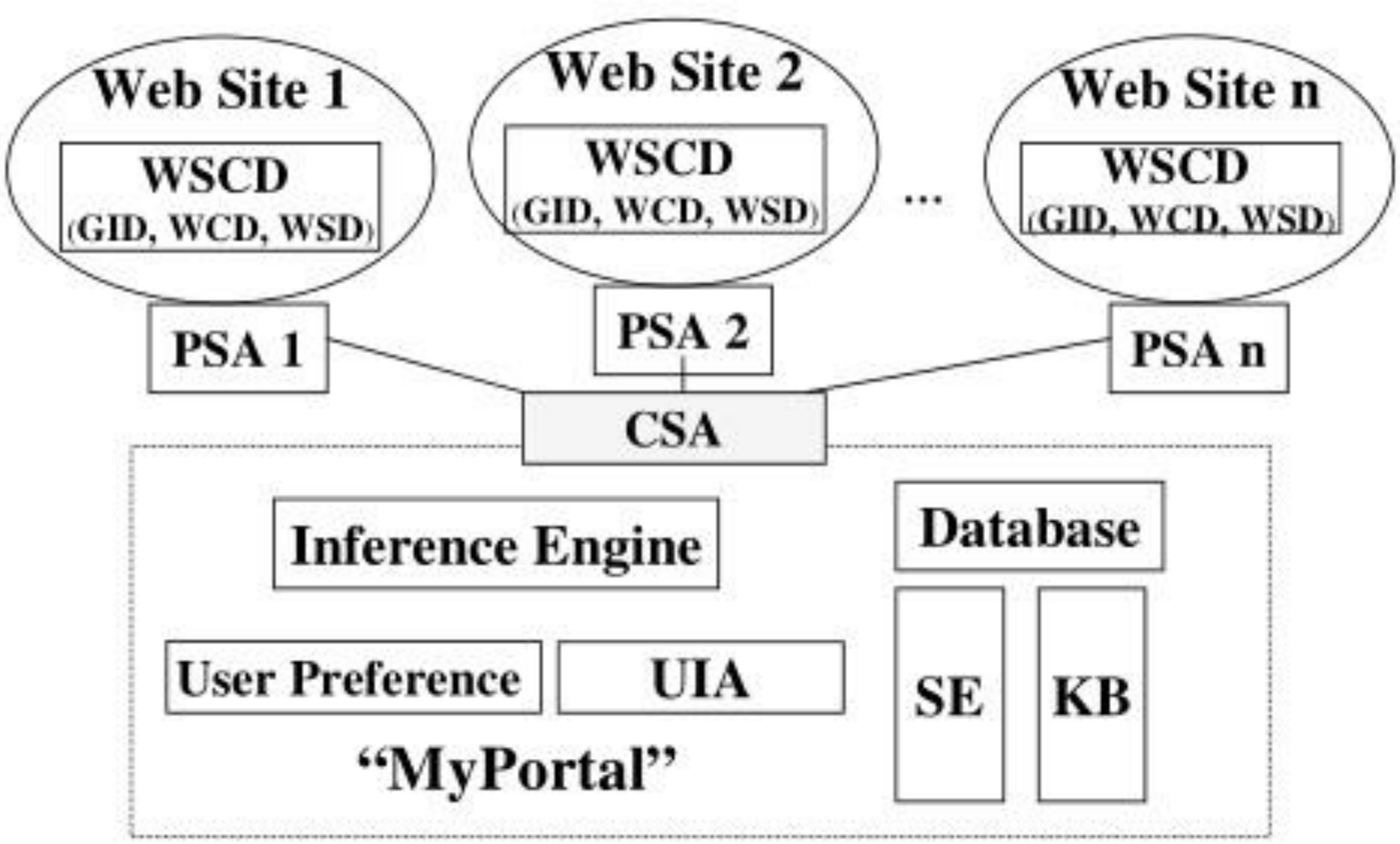

**Figure 2.4:** Semantic Web Application Model

Aroma and Kurain (2013) stated that knowledge base ontology of semantic web have a great deal which perform different operation on the semantic web for its enhancement.Basically, it is backbone of the ontology based function systems which are to be invoked simply. As we know that for satisfying the sensible web, Ontology is one of the most significant backbones of semantic web. Its importance may not be denied for understanding the mission of semantic web to include semantics or machine reasonable information to the current intelligible web in order to encourage a more effective information extracted and reuse on shareable premise on web. It implies that Ontology empowers to formation and conceptualizes the common information of a specific domain. However the major testing issue is the enormous decentralized nature of web which does not allow having a solitary extensive Ontology however left with the likelihood of having a few littler ontology which are hard to actualize and organized. Presently, Ontology has a few discriminating viewpoints where Ontology configuration and improvement is one of the major issues which may be best acknowledged with the assistance of a contextual investigation. The paper describes a technique named novel search to achieve this goal. The web scheme using these days is only collecting relevant data while to collect useful data semantic web is a proper solution. Semantic web helps to expand the database support to achieve machine readability that helps to understand the exact requirement of information. The paper use semantic web by implementing mapping techniques and main purpose of the paper is to open new door for new semantic web algorithm by using novel technique for classification and semantic mapping to categorize the concepts. It's a growth of web resource that tend to be need greater that is search scheme for the information which is in a process. so in any single user that can play a part for a new information that can be added to the web every day. This can be big supply that can be divided area in any origin which can be added. Without a being nothing more than what is specified as relation. So we can say that this type of scheme must be applied for the bringing out a related result on the query web data. Current web scheme can be bring out only some related pages for the results. So in that type of order to rewrite the semantic matched results to make as perfect or effective as possible results ranked can be apply over the matched results, it can give more reliable Having a bearing on or connection with the matter to be ranked high.

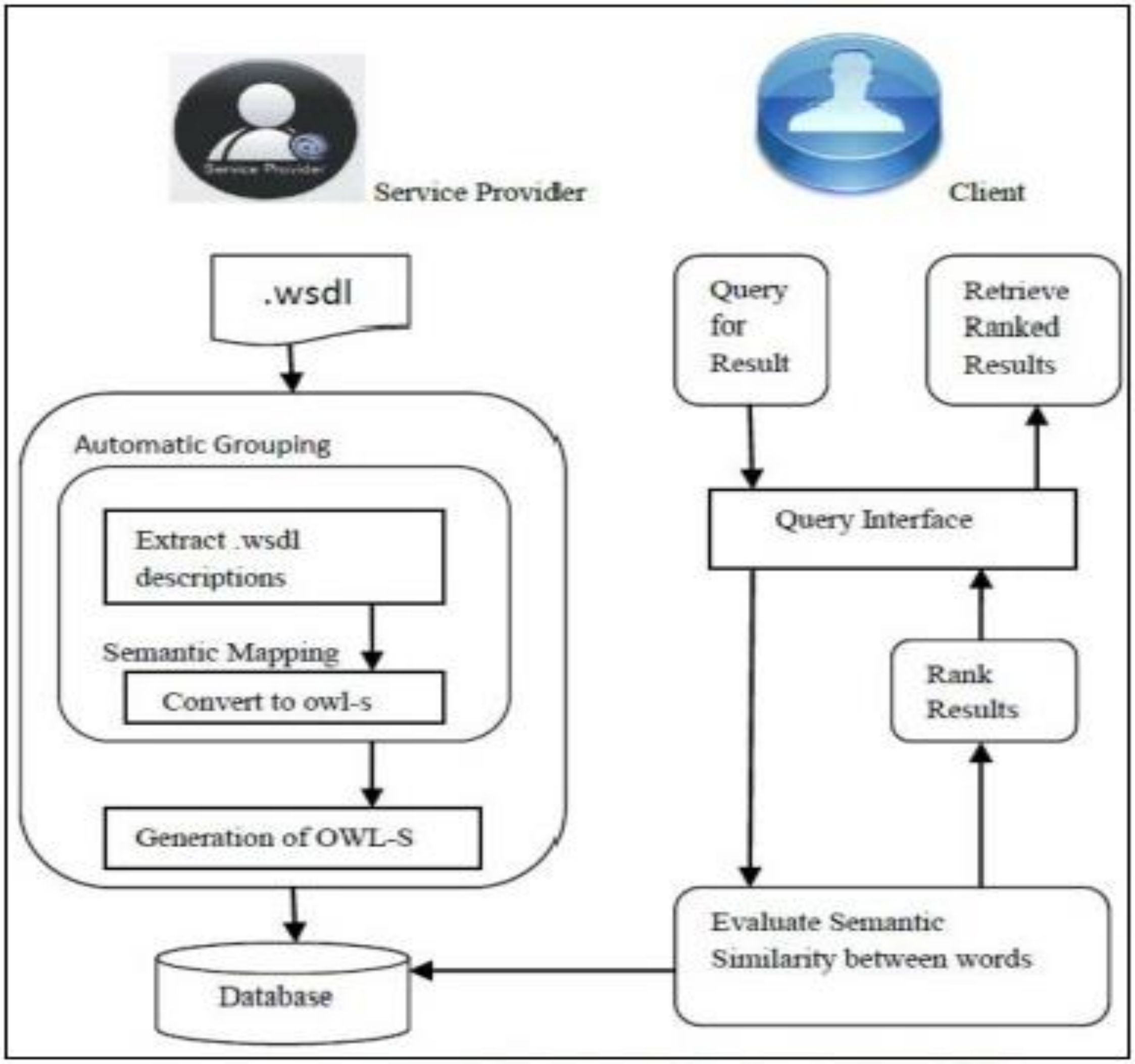

**Figure 2.5:** Deep Web Research Discovery Model

Barbin and Maleki (2013) stated that with the progression of time advanced correspondence innovations and systems have likewise settled an extraordinary arrangement. Web increased day by day due to modern technology. Huge source of information are available on the web. It is impossible to retrieve and find correct information from such type of large data. The data on the internet is placed globally. Client's retrieved the required data for their demand by utilizing the online web services on the internet. Information Technology (IT) trend has also increased.This information Technology has established the world like global village. Semantically a web is the only way to control the redundancy of the data it also improve the searching techniques and structure the correct information. In this research article described that technology is needed to optimize the information and creating relationship in the exiting information and the World Wide Web (WWW). The Semantic web is the growth of the existing websites through content mining and arrangement of combination of accurate and continues information is formed. Semantic web also creates the method for the machine to handle the information.

Huang *et al.* (2013) explained that the fast increase of web possessionsfor need have enhanced Search techniques for data recovery. Each and every client gives a piece of new data to be added to the web consistently.The huge amount of information complete is of many different .There is a need to change look procedures to gather applicable and suitable and right data. Semantic web serves to comprehend the precise prerequisite of data. So that the related methodologies on the current semantic mapping plan that are recorded and shows principle key components. The primary proposed of this revelation or advancement calculation that can be consolidates the strategy for semantic as a rule measure between words to be connected with supplied inquiries. So insight on data to re compose for accomplished more solid applicable and proficient data.

Kumar and Dwivedi (2013) described that two different technologies were projected in a few couple of year to carry, complete task or web content automatically. These advancements are agent technology and Semantic Web. During operators performed by these agent the specialists is to perform their particular goal with no interruption of outer source Semantic Web gives the semantics of information to machine . Agent based semantic web applications improvements are difficult for such peoples which lies outside the research groups. His operator’s technology uses the semantically coordinated data. There are a few minds boggling issues that the problems or task are Increase and expanding day today. These large amounts of unstructured data increasing rapidly and data are control by the computers. In this way, new clients might be trained for the preparation to utilize PCs later on. Agent bases semantic web technologies can deal with the issue in simple ways and less preparing time is required to the clients.

Shaikh and Kolhe (2013) described that a present web there are billions of site pages and contain data which is suitable to be comprehended by people. Be that as it may, this data is less helpful for programming instruments on the grounds that this data is semi organized and it doesn't contain semantics. This adds difficulties to the computerized handling of data. In the event of organized data like XML or RDF, robotized handling of content furthermore, looking data turns out to be simple for machines utilizing question languages like SPARQL, yet composing such inquiry is very little agreeable to all clients.

Shirgave and Kulkarni (2013) described that Dangerous and snappy development of the World Wide Web has brought about multifaceted Websites, requesting upgraded client aptitudes and modern tools to help the Web client to locate the required data. Finding needs data on the Web has turned into a basic element of ordinary individual, instructive, and business life. Therefore, there is an interest for more modern devices to help the client to explore a Web websites furthermore, to locate the fancied data. The clients must be furnished with data and administrations particular to their needs, instead of an undifferentiated mass of data.

Vasile*et al.* (2013) described that the research describes that there are new media sources introduced in past few years like face book, twitter and blogs. These media sources are facilitating users to share their information and knowledge. But the problem is that the data is much higher in quantity as compare to other media sources and not authentic. The research focus to develop a web crawler that is specializing for these kinds of media called social media. The crawler also focuses on the scalability because of the wideness of targeted area. The crawler operates single node as well as hundreds or more to make it useful for different kind of situations. The architecture of crawler, its components and their relations are given below.

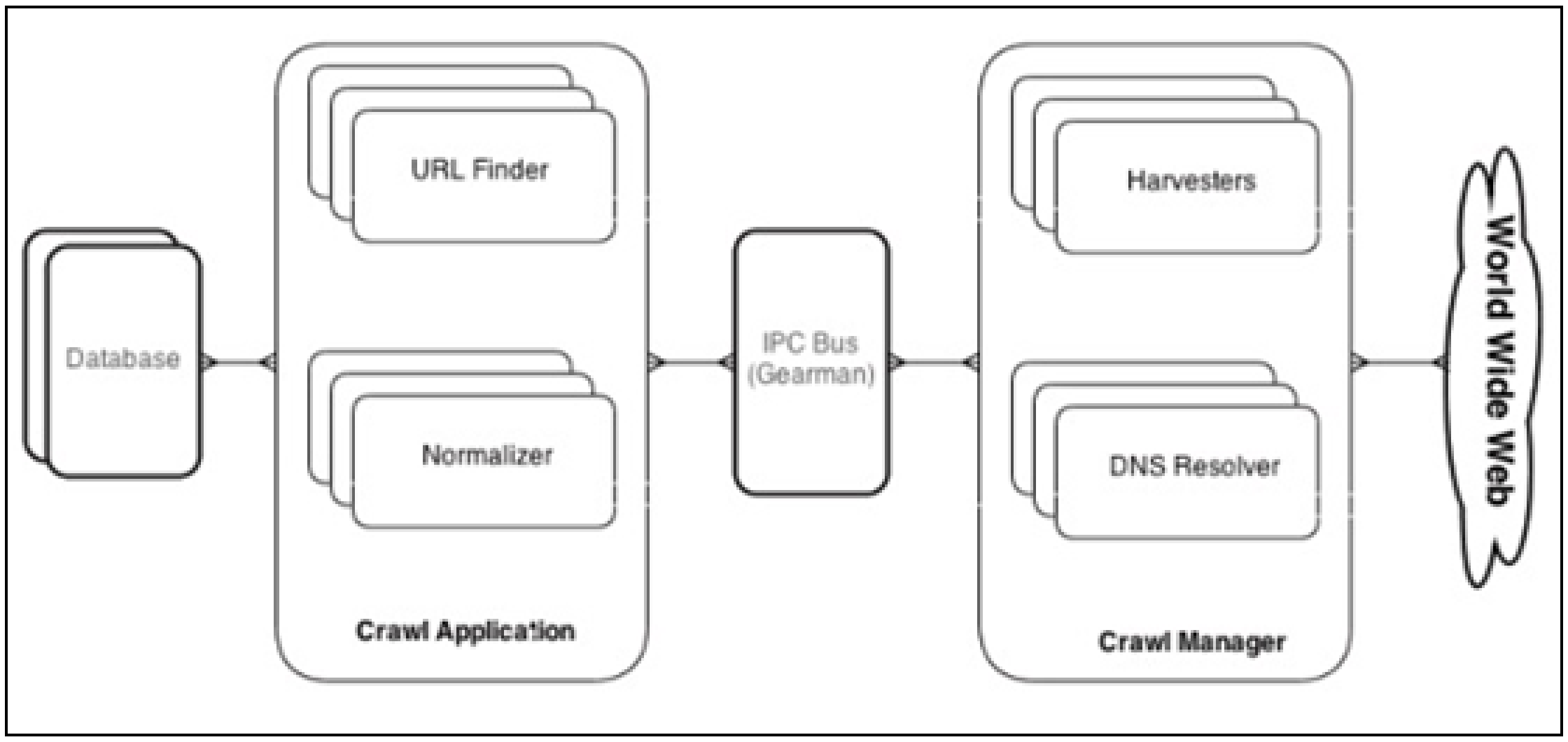

**Figure 2.6:** Specialized High Performance Web Crawler.

Al-Bukhitan*et al.* (2014) described that the vision of semantic Web is to have a Web of information rather than Web of archives in a shape or frame that can be handled by machines. This vision could be accomplished in the current Web utilizing semantic annotation. Because of exponential growth, development and tremendous size of the Web sources, there is a need to have a quick and programmed semantic annotation of Web reports. The instrument takes a URL of Web record and the relating comparing then creates an outside annotation of the Web report utilizing Asset Description Framework (RDF) language. Semantic Web technology give diverse intends to the data for machine handling. The most widely recognized organization is Resource Description framework (RDF) for the separated data and Web Ontology Language (OWL) for ontological representation of ideas, their relationship and semantic standards that could be connected on the information. The vast and greater part of Web semantic annotation apparatuses just bolster Latin languages and difficult to adjust for Arabic languages.

Azad and Abhishek (2014) stated that Semantic entropy can help scientists to choose how to choose and work with words, as well as which keywords to work on. Numerous scientists are attempting to redesign the after effect of web mining by promoting semantic structure of the web so one can get the applicable and effective data from the web, yet effective and applicable, relevance data extricating from the Web still confronts a major test. Semantic Synaptic web mining present method data mining which interlinks with website of information to distinctive information having low entropy so one can discover the most precise, significant and exact data on the web. The achievement of World Wide Web can't stay away from the commitment of words in the web content and association between networks content in light of the fact that it is essential for us, Which and how to work with keyword and their inter relationship with the contents of the websites. This point will improve the content of website.

Thangaraj and Sujatha (2014) described that the present web information retrieval framework recovers significant and related data depend on the essential keywords words which are insufficient for measures and gathers large amount of information. It gives inadequate abilities to catch the ideas of the client which the client’s needs and the connection between the keywords. These restrictions lead and give direction to the thought of the client reasonable inquiry which incorporates ideas and implications. This study manages the Semantic Based Information extracted System for a semantic web look and gave an enhanced calculation to recover the data in an effective and manageable manners.

# 3. Material and Methods

## 3.1 Overview:

A Framework for Semantic Organization of information retrieval that will base on Application Programming Interface (API) focuses to facilitate users using different devices with structured and tabular data that is portable. Information technology is facilitating many different fields but also increase of data is creating many ambiguities. It is very hard to collect relevant and authentic data from huge amount of data available on internet. There are many convenient applications that playing their roles to facilitate users in a specific domain, the research is going to open a new field to facilitate users in semantic organization.

Information for semantic web uses conceptual representation of content beyond plane keywords such as knowledge bases. This architecture will handle the concept representation of content. Quires extractions match the semantic relevant results. The plane key words entered by the user is expended at converted into the Symantec query by the different ways such as by matching, by using the links by using the thesaurus. When the semantic query will match the semantically indexed web content should match with two parts keywords and concepts of the semantic query. The www allow user to share the information and data from the large data repositories all over the world the user will also generate the query relation to his/his topic. The framework will extract the data from targeted resources, organize it symmetrically and provides to the user in tabular form in structured format such as XML and JSON form.

Here an API Based conceptual Architectural framework is going to develop for extracting data from different research portals having multiple journals and huge amount of research articles. At the start we are two research portals for the collection of data, first one is IEEE Explore and second one is Microsoft Academics.. The application will provide a graphical user interface and web service to the research to collect the queries from user and provide the desired output of generated query in requested format. The user will generate the query or will send a request to API related to his/her research work, the framework will extract the data from both of targeted sources (Microsoft Academics and IEEE Explore) organize it semantically and provide it to user in JSON,XML and tabular form. Here's a pseudo code summary that can be used to implement Semantic based web scrapper.

1- Firstly, view the source of the web source web page and compare the source page and our output web page.
2- Extract the source below and put in into XML,JSON Document object, regular expression is used to do so.
3- Convert initial XML schema to the custom schema XML,JSON format. XmlOML.xslt to convert initial XML schema to custom schema XML Format.
4- Apply XSLT fill “htmlOML.xslt” to transform the output XML into the HTML Format.
5- Implement the javascript to show the corresponding content dynamically by getting the element match with keywords provided by user according to defined ontology.
6- Show desired data in a window by binding the data retrieved through step 5.

These steps describes the complete procedure to collect data, process data and get the required material, through which we can collect more relevant data from two different portals. The quantity of portals can be increase more to collect more data at one. We have to design some Ontology for difference categories of research. These categories are divided into two properties; first one is data property and second is object property.

Primary data will be collected from target Web portals Framework will be established for data mining from three different, heterogeneous research portals or websites like IEEE Explore, Science Direct and Microsoft Academic. A Graphical User Interface will be provided, User will generate query regarding his/her research and framework will provide semantically organized data from these research portals.Requirements will be gather and analyze regarding to IEEE Explorer, Microsoft academic and Science Direct. Main focus will be on the structure of these portals. These portals have huge amount of research repository. Here portals will be studied with respect to their user interfaces and the information that they contain. Elicited requirements will be analyzed. A comparison of three of research portals will be made. Focus will be on identifying that what kind of material is needed to collect data from these three portals. Also working of these portals will be analyzed in detail.

A detail analysis and research related to algorithms used by these portals will be done. Either they use any specific algorithm or not? Based on the analyzed requirements, a new ontology will be designed and developed in this phase.A web scraping or data extraction technique will be design to collect data from research portals.Moreover Graphical User Interface will be designed in this phase for semantic representation of extracted data. Finallydesigning and implementation of testing plan (Mohammadian., 2007).

## 3.2 Architecture of Framework

The basic architecture of Framework is as followed, The SUD will generate a query to web services server, and the web services server will generate the query on both research portals (IEEE Explore and Microsoft Academics) and collect the relevant data from these two portals, after that these results will provided to SUD by web services server. The working of framework in IEEE Explore and Microsoft Academics will be discussed further in detail.

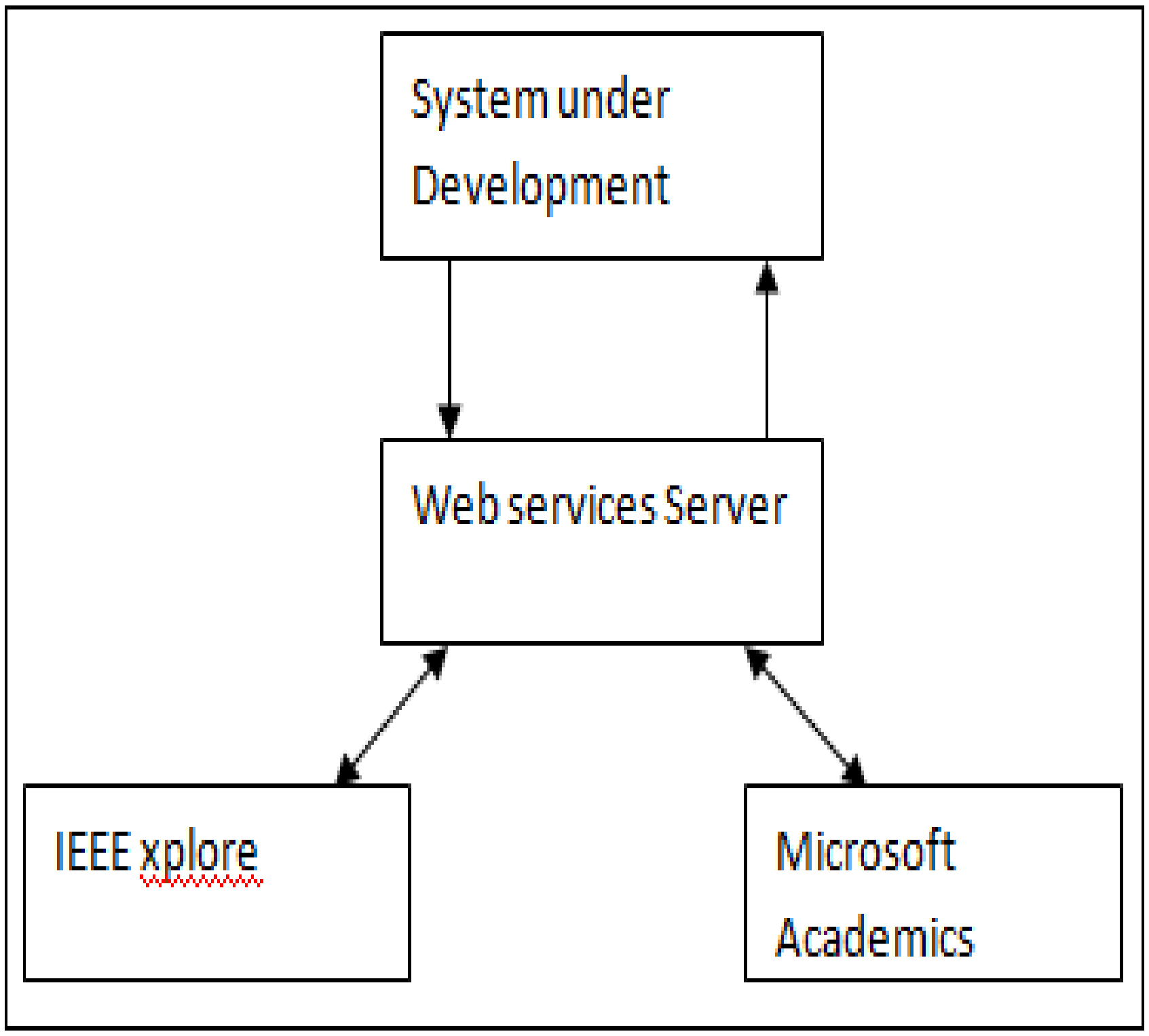

**Figure 3.1:**Main architecture framework

## 3.3Architecture of Microsoft Academics

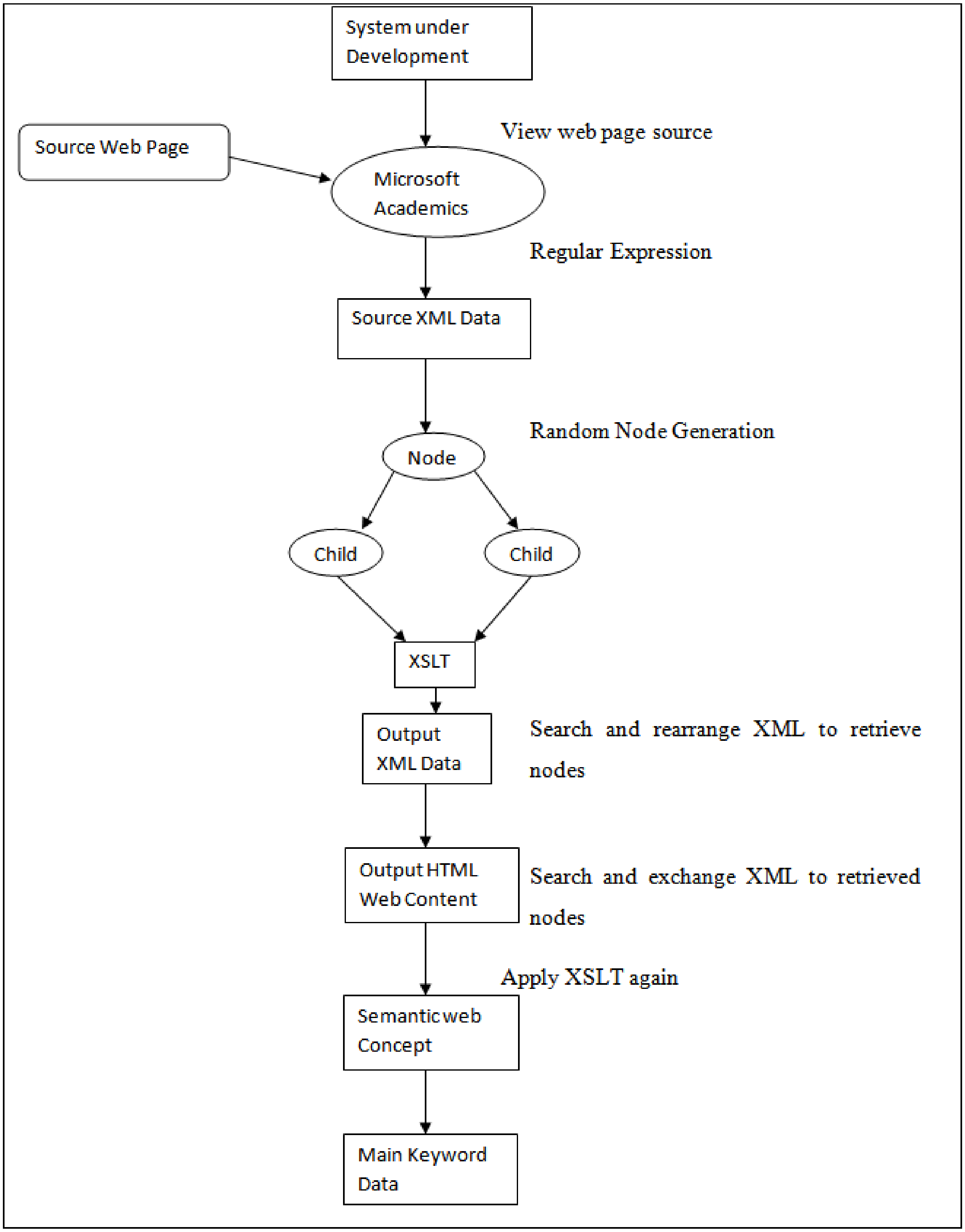

**Figure 3.2**Architecture of Microsoft Academics

## .4 Architecture of IEEE Explore

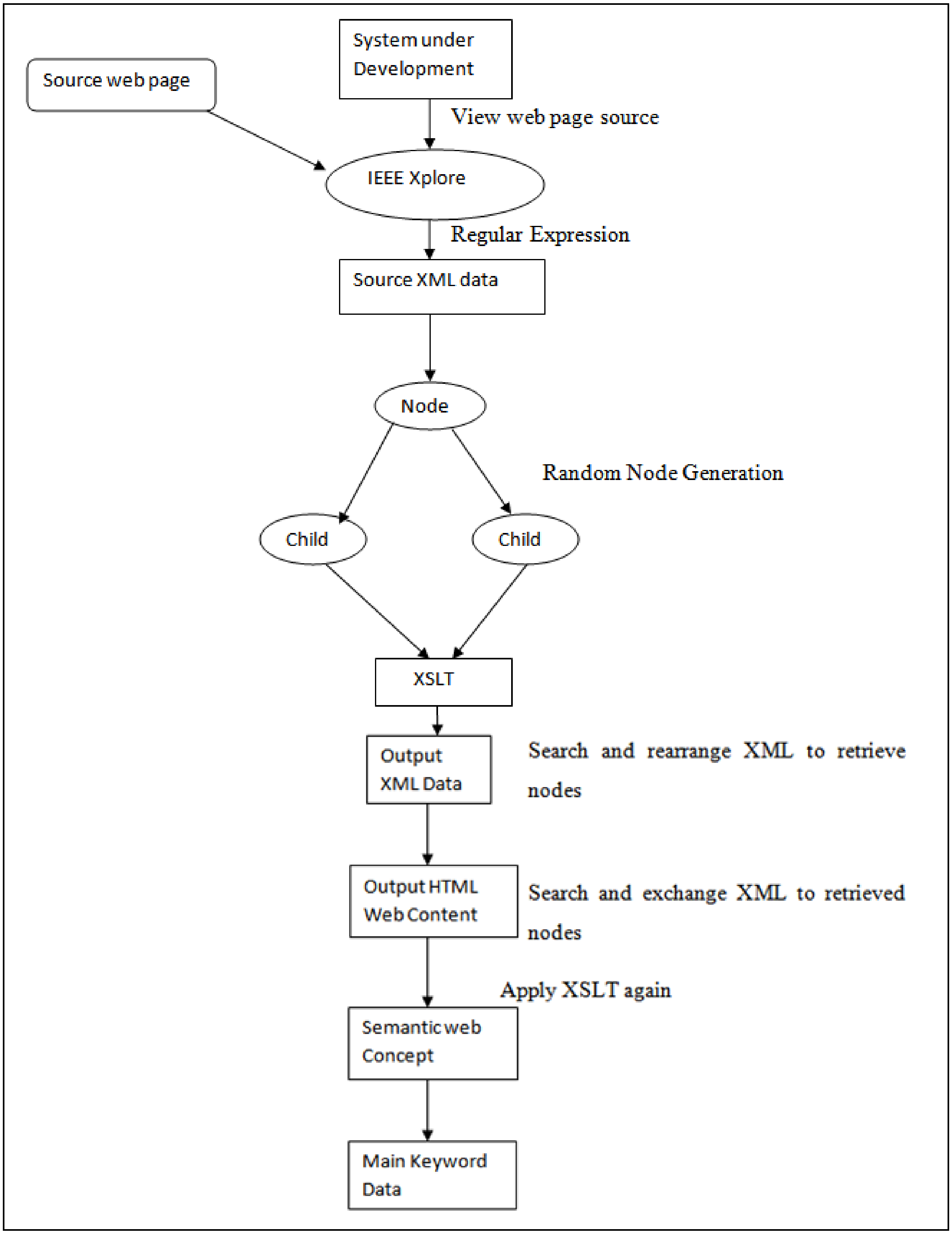

**Figure 3.3:**Architecture of IEEE Explore

## 3.5 Web Ontology’s

Ontology is considered to be the basic unit of semantic web, this web Ontology is used to represent knowledge. It is also defined as the set of concepts in particular domain. These, Ontology’s are used to represent and organize the data collected by web scrapper in more efficient and productive format. Ontology is very easy to view the relations among all classes, sub classes and super classes. For example software defined networking is a sub class of networking and networking is also a super class of different classes. Ontology formally stated number of items which signify to imperative ideas, for example, classes of articles and the connections between them. To think about theoretical data over two learning bases on the web, a system must have an approach to find basic implications and the answer for this is to gather data at a typical spot called Ontology. Incorporating Ontology is separated with three stages: metaphysics catch, cosmology coding and conceivable mix with existing ontology. An Ontology life cycle includes ventures like detail, conceptualization, formalization, mix, execution and preservation. For the framework it would be easy to understand the relevancy of generated query with other keywords that would help us to generate more relevant, targeted and useful data(Snášel, .2007).

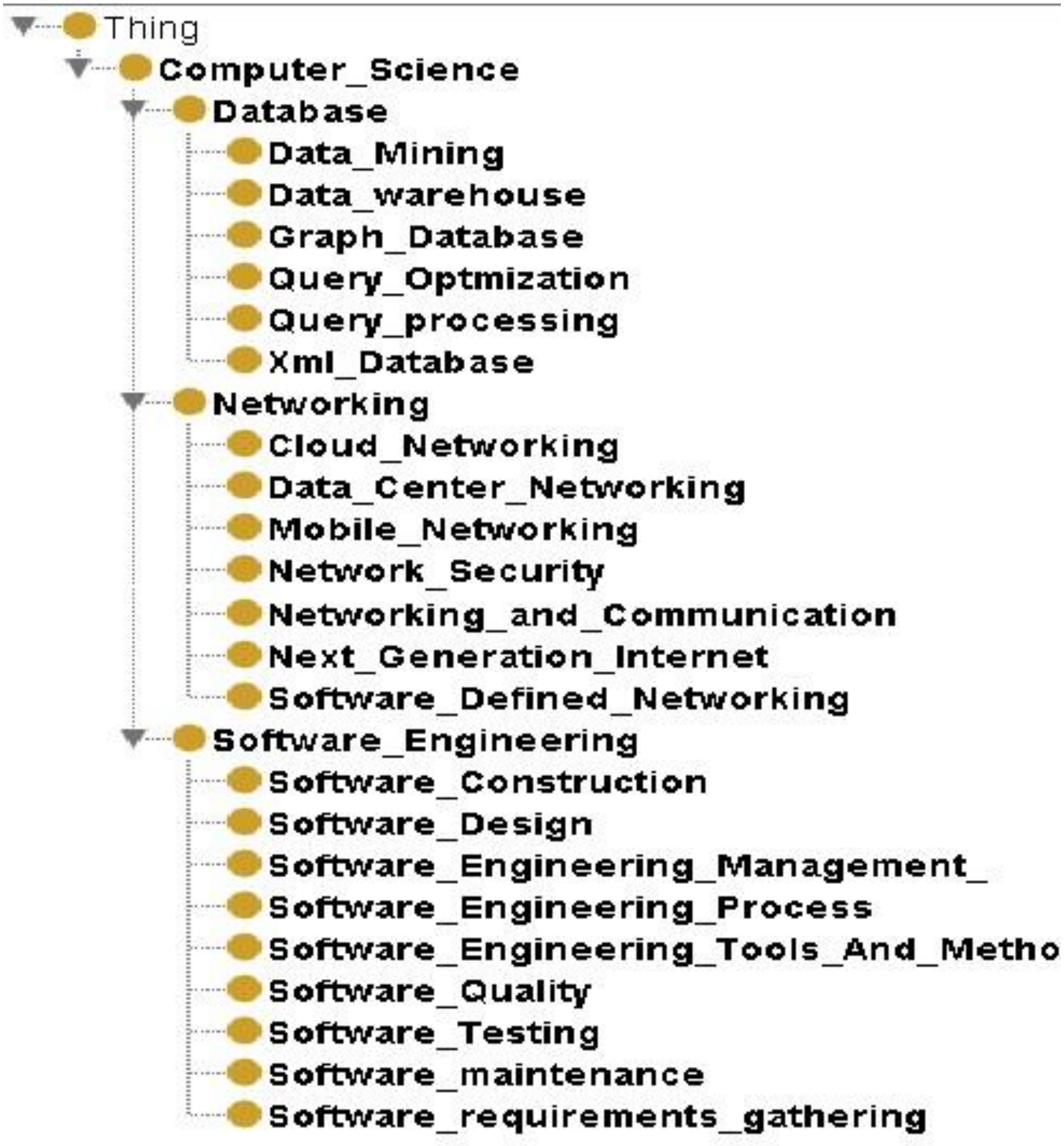

**Figure 3.4:**Ontology of Computer Science

In ontology given above, Computer science is main class that contains Database, Networking and Software engineering as its subclasses. Data mining, Data warehouse, Graph database, Query optimization, Query process and XML Database contains the super class Database which is a subclass of Computer science. It represents the connection among subclasses and also super classes of computer science. Same as networking is a subclass of computer science and super class of cloud networking, data center networking, mobile networking, network security, network communication, next generation internet, software defined networking. Now next software construction, software design, software engineering management, software engineering tools and methods, software engineering process, software quality, software testing, software maintenance, software requirement gathering are subclasses of software engineering which contains the same super class Computer science.

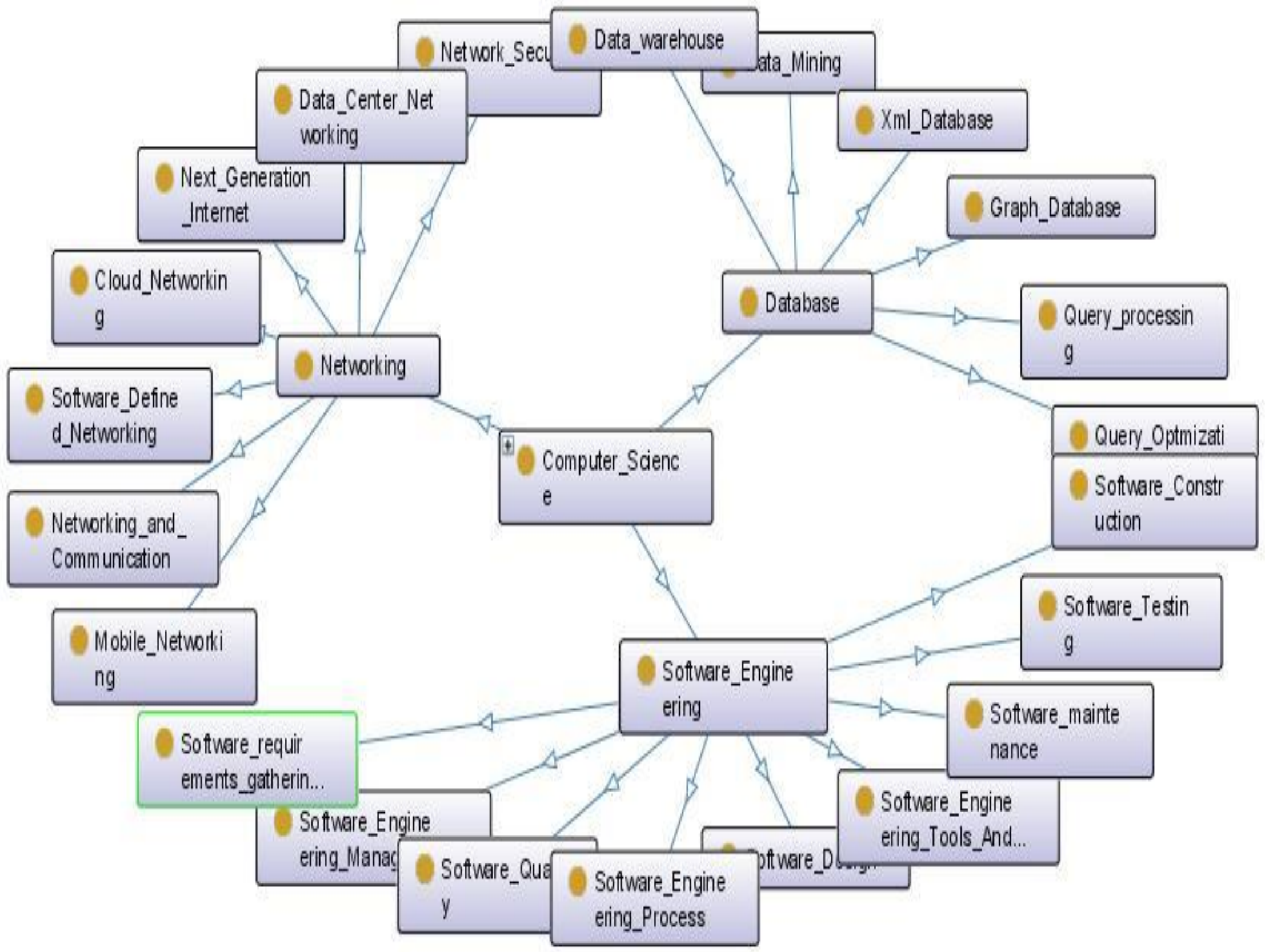

**Figure 3.5:** Graphical view of Computer Science Ontology

The picture given above is graphical representation of discussed ontology. Here it is very easy to view the relations among all classes, sub classes and super classes. For example software defined networking is a sub class of networking and networking is also a super class of different classes. For the framework it would be easy to understand the relevancy of generated query with other keywords that would help us to generate more relevant, targeted and useful data.

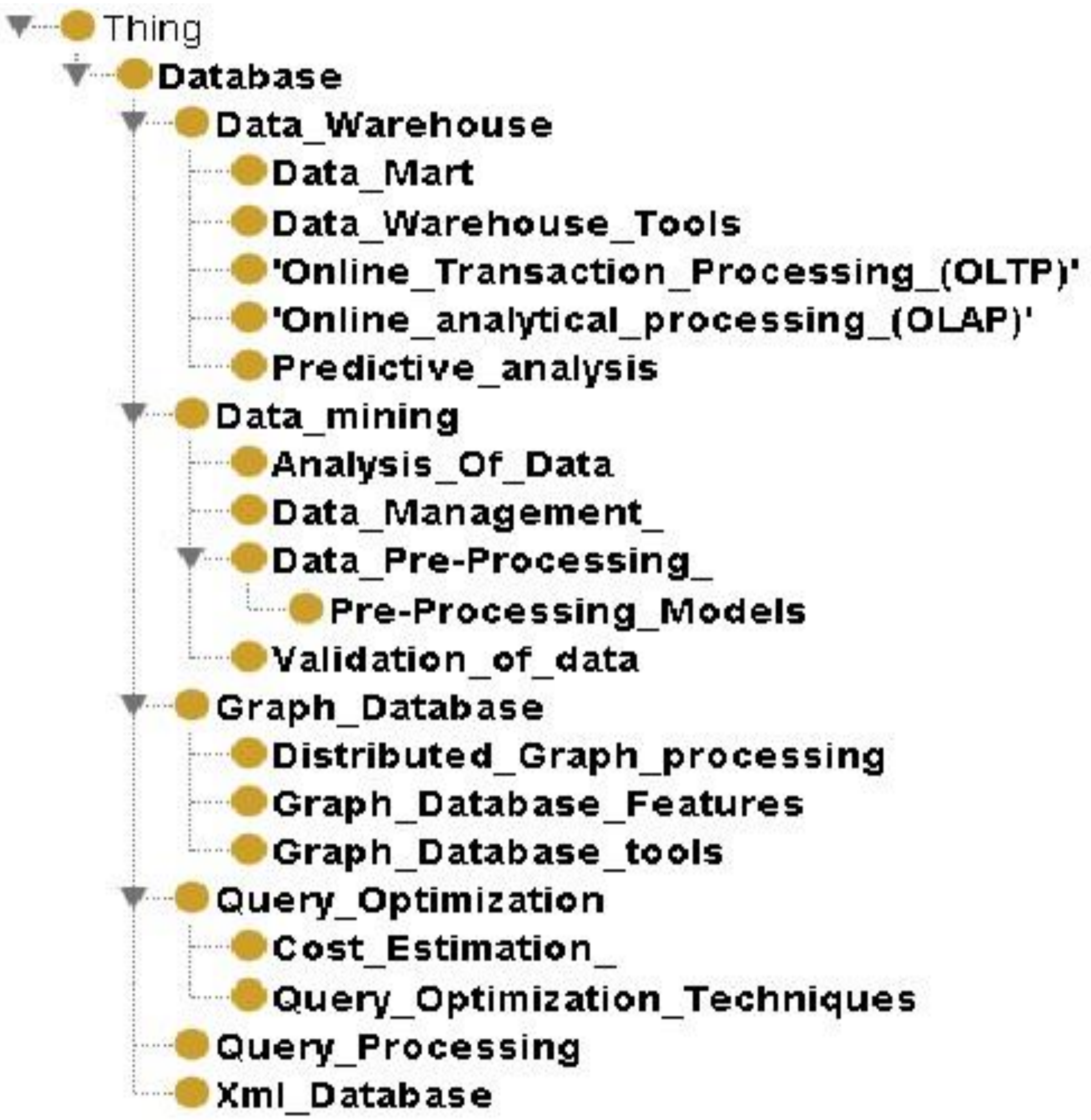

**Figure 3.6:** Database Ontology

The ontology given above is representing the detailed relationships of class database and its sub classes. Here database contains sub classes of data warehouse, data mining, graph database query optimization, query processing and xml database. These all subclasses are also super classes of many other classes like data warehouse is super class of data mart, data warehouse tools, OLTP, OLAP and predictive analysis. Data mining is super class of analysis of data, data management, data pre-processing and validation of data. While data pre-processing is also super class of pre-processing models. Ontology can easily explain the relation among between pre-process models, data pre-processing and data mining. Further in ontology it is explained that graph database that is a subclass of database is also a super class of distributed graph processing, graph database features and graph database tools, same as cost estimation and query optimization techniques are subclasses of query optimization that is also a sub class of database. Query processing and xml database are also subclasses of database but they are also acting as individuals.

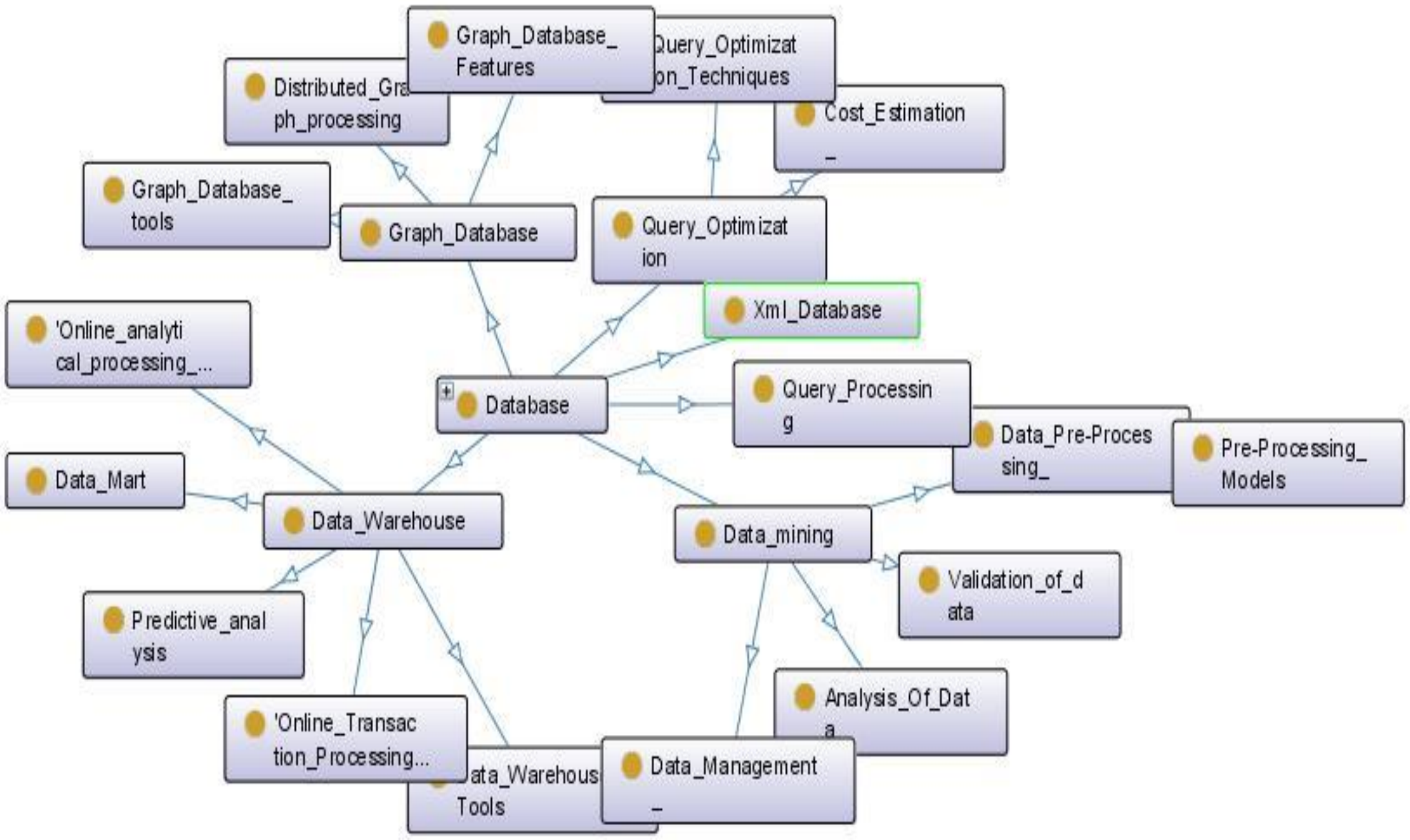

**Figure 3.7:** Graphical View of Database Ontology

The picture given above is graphical representation of discussed ontology base on database which is a subclass of computer science. Ontology provides a very clear view of the relations among all classes, sub classes and super classes. For example data management is a sub class of data mining and data mining is sub class of databases which belongs to the super class defined in previous ontology.

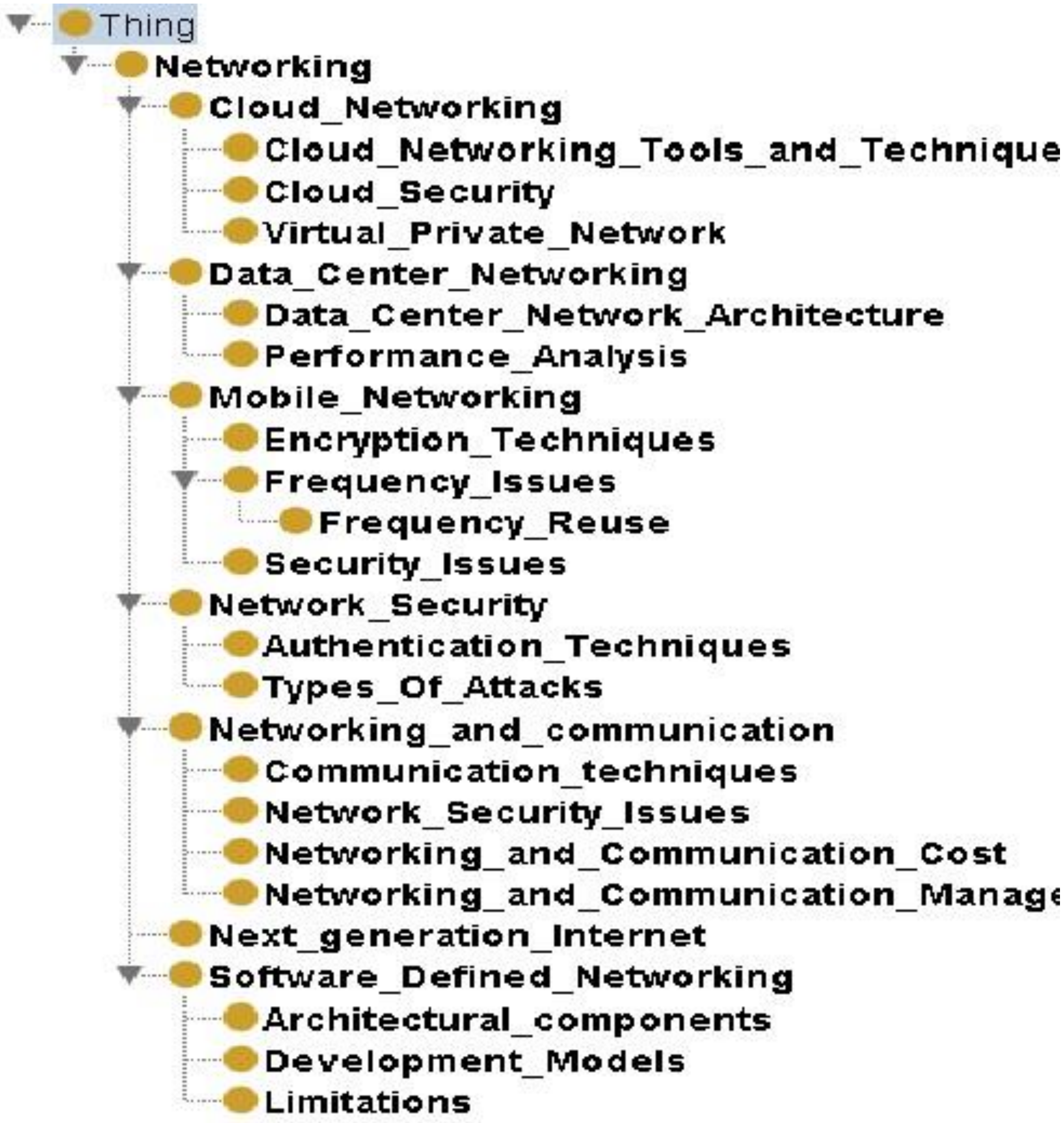

**Figure 3.8:** Networking Ontology

The ontology given above is representing the detailed relationships of class networking and its sub classes. Here networking contains sub classes of cloud networking, data center networking, mobile networking, network security, networking and communication software defined networking and next generation networking. These all subclasses are also super classes of many other classes like mobile networking is super class of cloud networking tools and techniques, cloud security and virtual private networking (VPN). Data center networking contains subclasses of data center networking architecture and performance analysis. Some subclasses of networking like mobile networking have sub classes those are also super classes. For example frequency is subclass of mobile networking and super class of frequency reuse. Mobile networking has also sub classes like encryption techniques and security issues. Further in network security we have authentication techniques and types of attacks as its sub classes. Network and communication that is a sub class of networking is also super class of communication techniques, network security issues, network and communication cost, network and communication management. While next generation internet that is a subclass of networking do not contain its own subclass while software designed networking have architectural components, development models and limitations as its sub class.

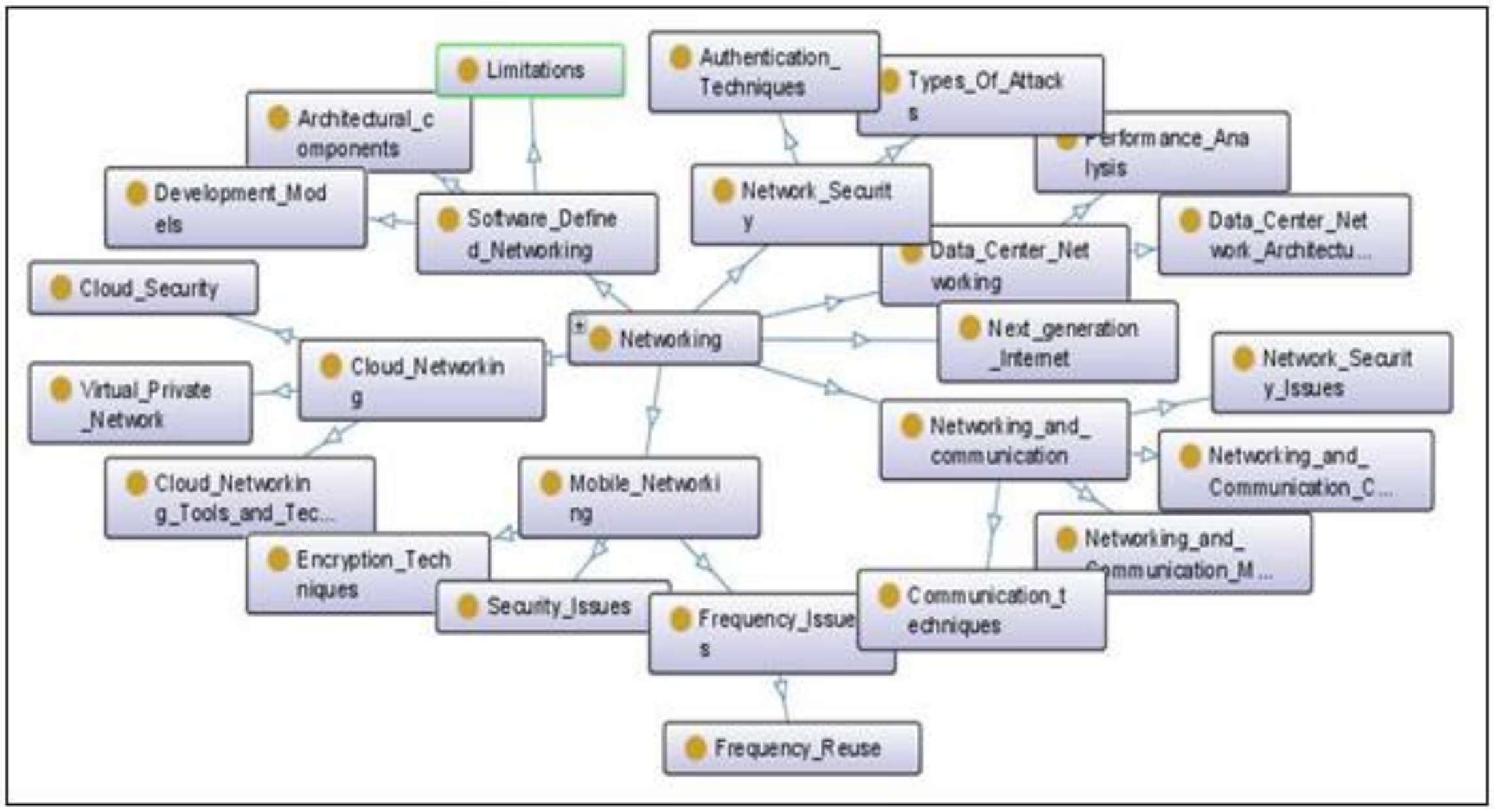

**Figure 3.9:** Graphical view of Networking Ontology

The figure given above is the graphical view of networking ontology showing the relationship among its sub classes and sub classes of these sub classes having networking as super class. For example we can observe that frequency reuse is a class having super class frequency issues that is a sub class of mobile networking that contains its super class networking.

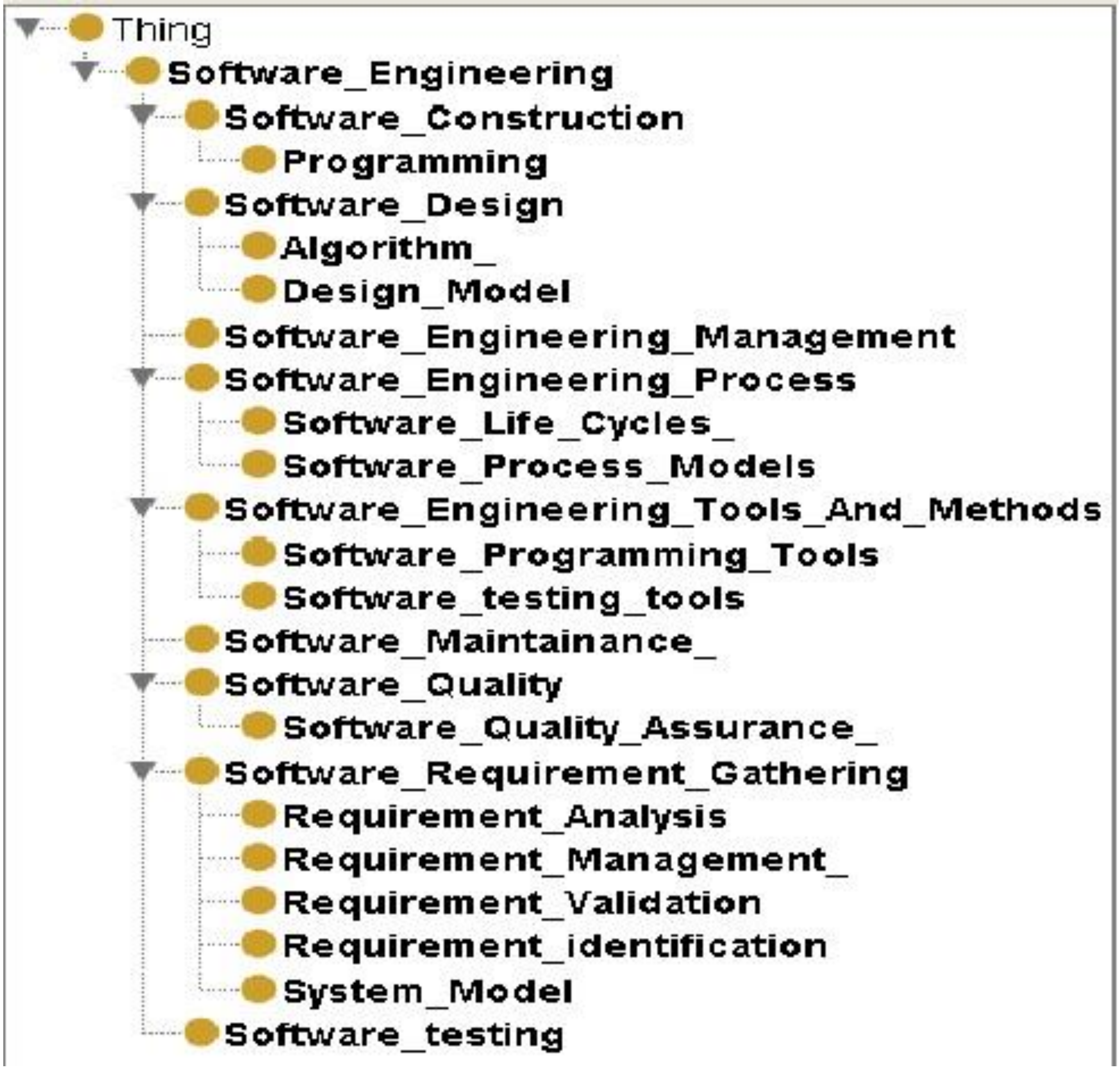

**Figure 3.10:** Software Engineering Ontology

The ontology given above is representing the detailed relationships of class software engineering and its sub classes. Software engineering is sub class of Computer science in ontology one while it is super class of Software construction, Software design, Software engineering management, Software engineering process, software engineering tools and methods, software quality, software requirement gathering and software testing. While software construction is super class of programming and software design is super class of algorithm and design model. Software engineering management does not contain any sub class. Software programming tools and software testing tools are the sub classes of software engineering tools and method. Software maintenance and software testing also do not contain any sub class. Software quality assurance is subclass of software quality while software requirement gathering is super class of requirement analysis, requirement management, requirement validation, requirement identification and system model.

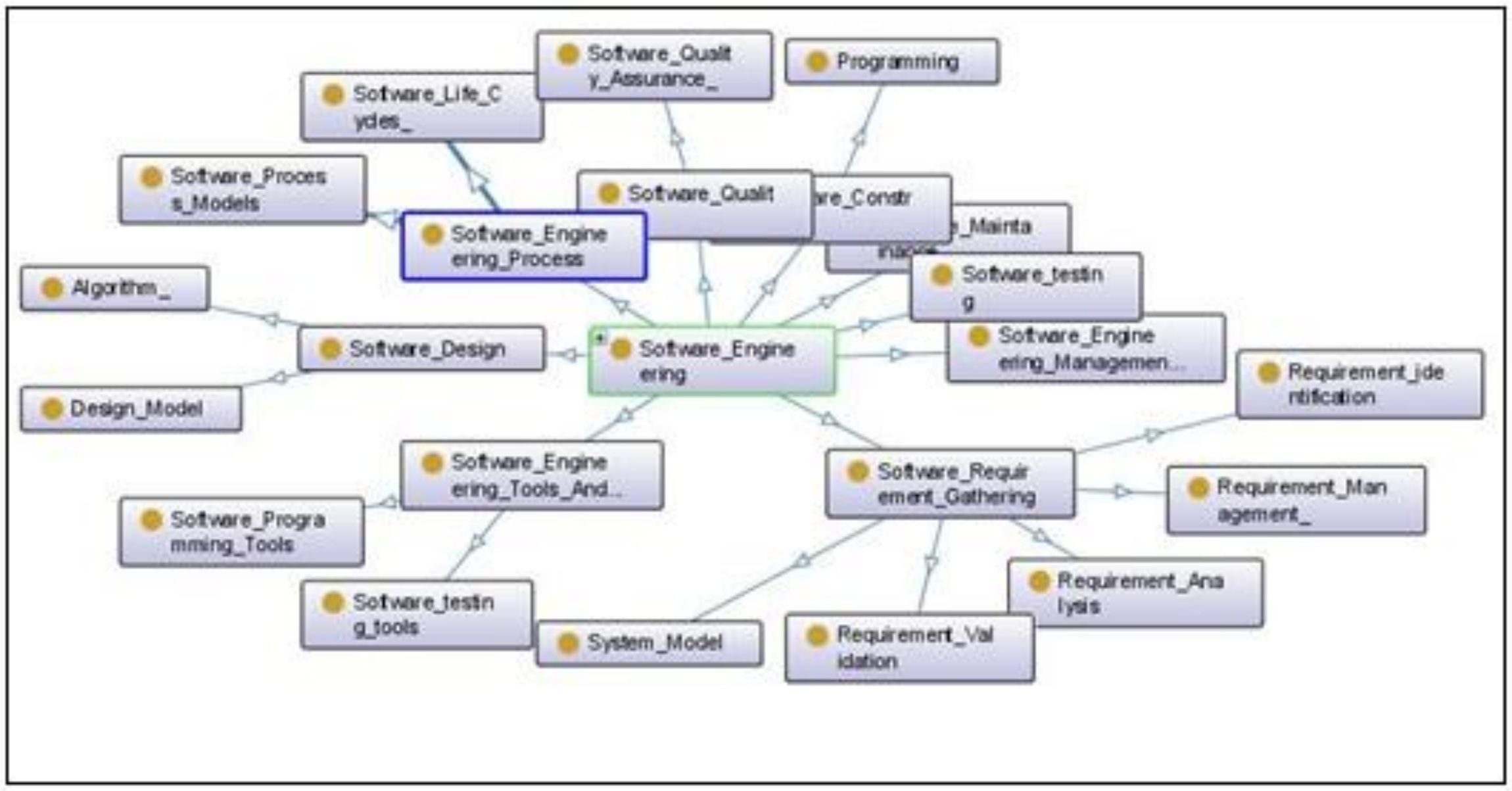

**Figure 3.11: Graphical View of Software Engineering**

The figure above is the graphical representation of Software engineering ontology showing the relationship among its sub classes and sub classes of these sub classes having Software engineering as super class. For example we can observe that design model is a class having super class Software design that is a sub class of Software engineering.

# 4. RESULTS AND DISCUSSION

Since we alter the query and different results are produced here some queries are generated and results are retrieved are given below. The results shown by the query “Computer of science” are given below. These results are retrieved from both research portal Microsoft Academics an IEEE Explore.

| Title | Abstract |
|---|---|
| Computer-Aided Geometric Design: A bibliography with Keywords and Classified Index | Computer-aided geometric design is a subject which involves the representation, specification, design, manipulation, display, and analysis of free-form curves and surfaces. This eclectic field draws on techniques and principles from numerical analysis, approximation theory, computer graphics, interactive computer systems, mechanical and geometric design, and many other areas. Nevertheless, computer-aided geometric design is an endeavor with a character, emphasis, and purpose of its own, combining techniques and principles from other fields with unique approaches that are often somewhat enigmatic and confusing to the novice. Many things sound and seem familiar, yet the directions and motivations are distinct to this discipline. This decade is proving to be an era of intense interest in computer-aided geometric design, an activity attracting people from many sectors including computing and manufacturing. As the discipline more firmly establishes valid principles and standard techniques, it will undoubtedly be of increasing importance in computer-aided manufacturing and a wide range of other applications. |
| Women in Computer Science or Management Information Systems Courses: A Comparative Analysis | This chapter contains sections titled: Method, Materials, Discussion, Conclusions, Acknowledgments |
| Computer Simulation Models of Human Behavior: A History of an Intellectual Technology | The history of the growth and development of the technology of computer simulation is reflected in an analysis of 2034 sight-read and classified simulation studies of human behavior published before 1971. The limiting goal of the work was an exhaustive bibliography of these simulation studies. The empirical studies referenced are classified into four major model categories for analysis: 1) individuals, 2) individuals who interact, 3) individuals who aggregate, and 4) individuals who aggregate and interact. Each of these studies is also classified into one of eight types, according to the empirical relationship between the model and reality. Additional classifications are employed to describe methodological studies. The analysis includes estimates of the completeness of the bibliography and of the reliability of the classification scheme, as well as the distributions of studies by category and type. |
| Review of computer vision education | Computer vision is becoming a mainstream subject of study in computer science and engineering. With the rapid explosion of multimedia and the extensive use of video and image-based communications over the World Wide Web, there is a strong demand for educating students to become knowledgeable in computer imaging and vision. The purpose of this paper is to review the status of computer vision education today. |
| Computer Science | This chapter contains sections titled:<br>Unexplained Differences<br>Status of Women in Professional Life<br>Changing Representation of Women in Computer Science<br>Growth of Computer Science as a Discipline<br>Leadership: Women at Higher Levels<br>Summary<br>Some Possible Explanations<br>Future Research Questions<br>Strategies for Change<br>Closing Thoughts<br>Acknowledgments<br>References |
| The Rise of Computer Science | This chapter contains sections titled: The Humble Programmer, Comptologist, Turingeer, or Applied Epistomologist?, Computer Bureaus and Computing Laboratories, Trading Zones, Is Computer Science Science?, Fundamental Algorithms, “Cute Programming Tricks”, Science as Professional Identity |
| The Oregon Report Computer Science and Computer Engineering Education in the 80's | The computer's pervaiveness in the 1980's will demand even more professional talent, broader computer education, and perhaps even licensing of the computer professional. Industry, education, and professional groups must cooperate to meet these challenges. |
| Electronics Technology and Computer Science, 1940-1975: A Coevolution | This paper explores the relation ship between two disciplines: electrical engineering and computer science, over the past 40 years. The author argues that it was the technology of electronics - the exploitation of the properties of free electrons - that finally permitted Babbage's concepts of automatic computing machines to be practically realized. Electrical Engineering (EE) activities thus "took over" and dominated the work of those involved with computing. Once that had been done (around the mid-1950s), the reverse takeover happened: the science of computing then "took over" the discipline of EE, in the sense that its theory of digital switches and separation of hardware and software offered EE a guide to designing and building ever more complex circuits. |
| Computer Graphics, Interactive Techniques, and Image Processing 1970-1975: A Bibliography | Computer graphics, interactive techniques, and image processing are among the developments in the constantly evolving computer science field that impact the potential user ever more rapidly. This bibliography attempts to compile all articles, books, conference papers, and technical reports about computer graphics and man-machine interaction that have been published in English from 1970 to 1975. Because the literature pertaining to computer graphics and man-machine interaction is immense, this bibliography will no doubt be incomplete. Suggestions and contributions for future supplements to the bibliography should be sent to the compiler. |

The result shown by the query “Reverse Engineering “is given below.

| | |
|---|---|
| Research on Reverse Engineering Technology of Complex Product | This paper expatiates on the basic concept about reverse engineering. It puts forward a new working mode which takes innovation as the core. According to 3D reconstruction of complex product as the goal, it analyzes the working process of reverse engineering. The mode and process constructs CAD solid model of complex product under universal CAD software environment and lays a good foundation for product innovation design based on prototype. This paper points out that feature recognization is the main research content and key link. It decides the working quality and working speed of reverse engineering. It is pointed out in this paper that there is a particular parameter system in reverse engineering. It is composed of original design parameters, objective prototype parameters and reconstruction parameters. After discussing the concepts of all kinds of parameters and analyzing the errors in reverse engineering, it is indicated that the parameters evaluated in reverse engineering today are actually reconstruction parameters. However the original design parameters must be evaluated and the product is manufactured by the original design parameters in most cases in mechanical manufacturing field. |
| Reverse engineering of software life cycle data in certification projects | Some applicants, developers, and commercial-off-the-shelf (COTS) software vendors have proposed reverse engineering as an approach for satisfying RTCS/DO-178B objectives for airborne software. RTCA/DO-178B, Software Considerations in Airborne Systems and Equipment Certification, serves as the means of compliance for most airborne software in civil aircraft. DO-178B defines reverse engineering as "the method of extracting software design information from the source code" and provides guidance particular to reverse engineering, when it is used to upgrade a development baseline. For purposes of this paper, reverse engineering is an approach for creating software life cycle data that did not originally exist, cannot be found, is not adequate, or is not available to a developer in order to meet applicable DO-178B objectives. Reverse engineering is not just the generation of data - rather it is a process to assure that the data is correct, the software functionality is understood and well documented and the software functions as intended and required by the system. Reverse engineering is not, as some software developers propose, just an effort to generate the software life cycle data without intent to build in quality and the resulting design assurance. This article explores reverse engineering in airborne software projects, by explaining a definition for the certification domain, describing the motivation for its use, and documenting the certification concerns. Two actual cases of reverse engineering are also described to illustrate the certification concerns in real projects. |
| Engineering design for software: on defining the software engineering profession | Since the mid-1980s, software engineering has been accepted as a formal field of study in academia. Software engineering education is maturing from specialized courses in computer science, to numerous Master's programs, and more recently to the advent of undergraduate as well as PhD programs. What is new today is the widespread impetus from many fronts to consider software development as engineering profession. The notion of whether software development is engineering can be answered in a number of ways. In this paper, the authors look at generally accepted definitions of engineering and show their correspondence or applicability to software development. They demonstrate through a detailed analysis how prominent features that cut across all engineering disciplines are found in software engineering as well. They conclude with a discussion of the educational implications |
| Software Reverse Engineering to Requirements | The aim of reverse engineering is to draw out many kinds of information from existing software and using this information for system renovation and program understanding. Based on traditional practice, reverse engineering and requirements engineering are two separate processes in software round trip engineering. In this paper, we argue that it is necessary to recover requirements from the reverse engineered outcome of legacy system and by integrating this outcome in the requirements phase of software life cycle, it is possible to have a better requirements elicitation, and clear understanding of what is redundant, what must be retained and what can be re-used. So we have presented a revised model of traditional re-engineering process and also described the rationality of the proposed model. In the paper we have also discussed briefly about software reverse engineering, requirement engineering and their basic practices and activities. |
| Domain-retargetable reverse engineering | A user programmable approach to reverse engineering is described. The approach uses a scripting language that enables users to write their own routines for these activities, making the system domain-retargetable. The environment supported by this programmable approach subsumes existing reverse engineering systems by being able to simulate facets of each one and provides a smooth transition from semi-automatic to automatic reverse engineering |
| An Information Security Engineering Paradigm for Overcoming Information Security Crisis | The information security crisis should be overcame by means of information security engineering paradigm. However, definition, approach and paradigm on security engineering are not clear yet. In this paper we survey on definitions on security engineering, and propose a new definition and paradigm. Approaches and research topics on security engineering, to overcome the security crisis, modeled and described. Results of paper are useful for establishing consensus on security engineering in community of information security and cryptography |
| Incorporating green engineering in the chemical engineering curriculum | Introducing green engineering concepts to undergraduate students is recognized as increasingly important by industry and the general populace. Implementing green engineering principles at the start of the design process can lead to substantial environmental benefits and cost savings in the pursuit of more sustainable processes and products. The most common method to introduce environmental engineering is through a senior/graduate level elective course on environmental engineering; however, green engineering concepts can be incorporated in core engineering courses. In 1998 the Environmental Protection Agency initiated a program to develop a text book on green engineering, disseminate these materials, and assist professors in using these materials through national and regional workshops. The textbook, "Green Engineering: Environmentally Conscious Design of Chemical Processes," by David Allen and David Shonnard, is designed for a senior/graduate chemical engineering course. Through this program, material from this text can be incorporated throughout the curriculum. Teaching aids have been developed that include: presentation materials, lecture notes, example and homework problems, case studies, and experiments. These are tailored to fit general freshmen and sophomore engineering, as well as core chemical engineering courses. |
| Lessons learned in data reverse engineering | Reverse engineering of data has been performed in one form or another for over twenty-five years (1976-2001 approx.). The author describe the lessons learned in data reverse engineering (DRE) as contributed in a survey of data reverse engineers. Interesting is the fact that some of the lessons learned tell us how we are doing in the process of initial database design as well as how difficult the DRE process really is. It is hoped that from these lessons learned, we can assist in the suggestion of the next steps that are needed in the DRE area and promote discussion among the DRE community |
| A comparison of four reverse engineering tools | Reverse engineering tools support software engineers in the process of analyzing and understanding complex software systems during maintenance activities. The functionality of such tools varies from editing and browsing capabilities to the generation of textual and graphical reports. There are several commercial reverse engineering tools on the market providing different capabilities and supporting specific source code languages. We evaluated four reverse engineering tools that analyze C source code: Refine/C, Imagix4D, Sniff+, and Rigi. We investigated the capabilities of these tools by applying them to a commercial embedded software system as a case study. We identified benefits and shortcomings of these four tools and assessed their applicability for embedded software systems, their usability, and their extensibility |

The result shown by the query “Remote Sensing” is given below.

| | |
|---|---|
| 2002 IEEE International Geoscience and Remote Sensing Symposium [front matter] | Conference proceedings front matter may contain various advertisements, welcome messages, committee or program information, and other miscellaneous conference information. This may in some cases also include the cover art, table of contents, copyright statements, title-page or half title-pages, blank pages, venue maps or other general information relating to the conference that was part of the original conference proceedings. |
| 2002 IEEE International Geoscience and Remote Sensing Symposium [front matter] | Conference proceedings front matter may contain various advertisements, welcome messages, committee or program information, and other miscellaneous conference information. This may in some cases also include the cover art, table of contents, copyright statements, title-page or half title-pages, blank pages, venue maps or other general information relating to the conference that was part of the original conference proceedings. |
| 2002 IEEE International Geoscience and Remote Sensing Symposium [front matter] | Conference proceedings front matter may contain various advertisements, welcome messages, committee or program information, and other miscellaneous conference information. This may in some cases also include the cover art, table of contents, copyright statements, title-page or half title-pages, blank pages, venue maps or other general information relating to the conference that was part of the original conference proceedings. |
| 2002 IEEE International Geoscience and Remote Sensing Symposium [front matter] | Conference proceedings front matter may contain various advertisements, welcome messages, committee or program information, and other miscellaneous conference information. This may in some cases also include the cover art, table of contents, copyright statements, title-page or half title-pages, blank pages, venue maps or other general information relating to the conference that was part of the original conference proceedings. |
| 2002 IEEE International Geoscience and Remote Sensing Symposium [front matter] | Conference proceedings front matter may contain various advertisements, welcome messages, committee or program information, and other miscellaneous conference information. This may in some cases also include the cover art, table of contents, copyright statements, title-page or half title-pages, blank pages, venue maps or other general information relating to the conference that was part of the original conference proceedings. |
| 2013 Index IEEE Transactions on Geoscience and Remote Sensing Vol. 51 | This index covers all technical items - papers, correspondence, reviews, etc. - that appeared in this periodical during the year, and items from previous years that were commented upon or corrected in this year. Departments and other items may also be covered if they have been judged to have archival value. The Author Index contains the primary entry for each item, listed under the first author's name. The primary entry includes the co-authors' names, the title of the paper or other item, and its location, specified by the publication abbreviation, year, month, and inclusive pagination. The Subject Index contains entries describing the item under all appropriate subject headings, plus the first author's name, the publication abbreviation, month, and year, and inclusive pages. Note that the item title is found only under the primary entry in the Author Index. |
| 2007 Index IEEE Transactions on Geoscience and Remote Sensing Vol. 45 | This index covers all technical items - papers, correspondence, reviews, etc. - that appeared in this periodical during the year, and items from previous years that were commented upon or corrected in this year. Departments and other items may also be covered if they have been judged to have archival value. The Author Index contains the primary entry for each item, listed under the first author's name. The primary entry includes the coauthors' names, the title of the paper or other item, and its location, specified by the publication abbreviation, year, month, and inclusive pagination. The Subject Index contains entries describing the item under all appropriate subject headings, plus the first author's name, the publication abbreviation, month, and year, and inclusive pages. Note that the item title is found only under the primary entry in the Author Index. |
| 2013 Index IEEE Journal of Selected Topics in Applied Earth Observations and Remote Sensing Vol. 6 | This index covers all technical items - papers, correspondence, reviews, etc. - that appeared in this periodical during the year, and items from previous years that were commented upon or corrected in this year. Departments and other items may also be covered if they have been judged to have archival value. The Author Index contains the primary entry for each item, listed under the first author's name. The primary entry includes the co-authors' names, the title of the paper or other item, and its location, specified by the publication abbreviation, year, month, and inclusive pagination. The Subject Index contains entries describing the item under all appropriate subject headings, plus the first author's name, the publication abbreviation, month, and year, and inclusive pages. Note that the item title is found only under the primary entry in the Author Index. |
| 2009 Index IEEE Transactions on Geoscience and Remote Sensing Vol. 47 | This index covers all technical items - papers, correspondence, reviews, etc. - that appeared in this periodical during the year, and items from previous years that were commented upon or corrected in this year. Departments and other items may also be covered if they have been judged to have archival value. The Author Index contains the primary entry for each item, listed under the first author's name. The primary entry includes the coauthors' names, the title of the paper or other item, and its location, specified by the publication abbreviation, year, month, and inclusive pagination. The Subject Index contains entries describing the item under all appropriate subject headings, plus the first author's name, the publication abbreviation, month, and year, and inclusive pages. Note that the item title is found only under the primary entry in the Author Index. |

The result shown by the query “Software Quality Assurance” is given below.

| Title | Abstract |
|---|---|
| Software Quality Engineering: Making It Happen | Software quality engineering calls for a formal management of quality throughout the full lifecycle of software or a system. Several quality models were developed in the course of past three decades, some of them recognized mostly by the scientific community, others also gaining recognition within the industry. This chapter presents the most widely known models of McCall, Boehm, Dromey, ISO/IEC 9126, and ISO/IEC 25010. In software quality engineering, measurement is a pivotal concept. The evaluation of software product quality is important to both the acquisition and development of software. Quality design depends on requirements and their completeness, feasibility, and quality. The chapter analyzes conflicts that can appear within the basic change control process of a software project. The chapter presents four diagrams that help the potential user of Software Quality Implementation Model (SQIM) to map this model to a development process of the most popular choice. |
| IEEE Approved Draft Standard for Software Quality Assurance Processes | This standard establishes requirements for initiating, planning, controlling, and executing the Software Quality Assurance processes of a software development or maintenance project. This standard is harmonized with the software life cycle process of ISO/IEC 12207:2008 and the information content requirements of ISO/IEC/IEEE 15289:2011 |
| IEEE Draft Standard for Software Quality Assurance Processes | This standard establishes requirements for initiating, planning, controlling, and executing the Software Quality Assurance processes of a software development or maintenance project. This standard is harmonized with the software life cycle process of ISO/IEC 12207:2008 and the information content requirements of ISO/IEC/IEEE 15289:2011 |
| IEEE Standard for Software Quality Assurance Processes | Requirements for initiating, planning, controlling, and executing the Software Quality Assurance processes of a software development or maintenance project are established in this standard. This standard is harmonized with the software life cycle process of ISO/IEC/IEEE 12207:2008 and the information content requirements of ISO/IEC/IEEE 15289:2011. |
| IEEE Trial-Use Standard--Adoption of ISO/IEC TR 15026-1:2010 Systems and Software Engineering--Systems and Software Assurance--Part 1: Concepts and Vocabulary | This trial-use standard adopts ISO/IEC TR 15026-1:2010, which defines terms and establishes an extensive and organized set of concepts and their relationships for software and systems assurance, thereby establishing a basis for shared understanding of the concepts and principles central to ISO/I EC 15026 across its user communities. It provides information to users of the subsequent parts of ISO/IEC 15026, including the use of each part and the combined use of multiple parts. Coverage of assurance for a service being operated and managed on an ongoing basis is not covered in ISO/I EC 15026. |
| IEEE Draft Trial-Use Standard for Adoption of ISO/IEC TR 15026-1:2010 -- Systems and Software Engineering -- Systems and Software Assurance -- Part 1: Concepts and Vocabulary | This Technical Report defines terms and establishes an extensive and organized set of concepts and their relationships thereby establishing a basis for shared understanding of the concepts and principles central to ISO/IEC 15026 across its user communities. It provides information to users of the other parts of this International Standard including the use of each part and the combined use of multiple parts. Coverage of assurance for a service being operated and managed on an ongoing basis is not covered in this International Standard. |
| Supply Chain Quality Integration: Antecedents and Consequences | This study extends quality management from an individual company perspective to a supply chain perspective. We propose a concept of supply chain quality integration (SCQI) that consists of internal, supplier, and customer integration for quality improvement, and develop a model that specifies the relationships among competitive hostility, the organization-wide approach to quality, three types of SCQI, and quality-related performance. We test the model using data collected from 291 high-performance manufacturing plants from ten countries. The results indicate that competitive hostility has a positive effect on the organization-wide approach to quality, and that both have positive effects on SCQI. In addition, internal quality integration significantly enhances external quality integration with both suppliers and customers. Further, internal quality integration significantly improves all quality-related performance (i.e., product quality, cost, delivery, and flexibility), and both supplier and customer quality integration significantly improve cost performance. Whereas customer quality integration significantly improves delivery performance and supplier quality integration significantly improves quality performance, only internal quality integration can improve flexibility performance. The findings reveal how different types of SCQI are related to quality-related performance and highlight internal quality integration as a core strategic resource for quality improvement. As such, they provide important managerial insights for supply chain quality managers to improve quality-related performance. |
| Towards a new telecommunications industry quality standard | The quality technologies developed by Bell Communications Research (Bellcore) and adopted by its owner/clients are described. At the core of the philosophy is the concept of supplier accountability for product and service quality. A brief history of quality in the former Bell System is presented. A supplier career path for today's telecommunications suppliers is shown, and the various characteristics that accompany each phase of the buyer/supplier relationship are described. Programs that enable buyers and suppliers to achieve a long-term cooperative relationship are presented chronologically according to the phases of the product life cycle. The customer/supplier quality process, which is the newest and most comprehensive of the quality technologies available in the telecommunications field, is summarized |

The result shown by the query "Neural Networks" are given below.

| Title | Abstract |
|---|---|
| Solving the interconnection problem with a linked assembly of neural networks | To reduce the complexity of training a single, large neural network, partitioning of the problem is introduced to facilitate the identification of smaller and, where possible, replicable networks which are more readily trained. A system for the linked assembly of these neural networks (ASLANN) has been developed and is used to generate the final neural system. To demonstrate this approach the authors discuss the application of backpropagation neural networks to the routing of an integrated circuit |
| Neural network models as an alternative to regression | Neural networks can provide several advantages over conventional regression models. They are claimed to possess the property to learn from a set of data without the need for a full specification of the decision model; they are believed to automatically provide any needed data transformations. They are also claimed to be able to see through noise and distortion. An empirical study evaluating the performance of neural network models on data generated from three known regression models is presented. The results of this study indicate that neural network models perform best under conditions of high noise and low sample size. With less noise or larger sample sizes, they become less competitive. However, in two of the three cases, the neural network models were able to maintain mean absolute percentage errors (MAPE) within 2% of those of the true model |
| Bayesian Neural Network with and without compensation for competing risks | This paper addresses the problem of compensation mechanisms which can be used by Bayesian Neural Networks (BNNs) when dealing with skewed training data. The compensation mechanisms are used to balance the training data towards a mean value so that to be able to calculate the marginalized neural network predictions. There are presented 2 compensation mechanisms and each of them is applied to a BNN: a local compensation mechanism and a global mechanism. There is presented a third BNN model which does not use a compensation mechanism. It is shown that in the absence of a compensation mechanism, the marginalized network outputs can still be calculated through a scaling of the Jacobian and Hessian matrixes involved in the respective calculations. The standard BNN is a Partial Logistic Artificial Neural Network with Automatic Relevance Determination, which has multiple competing network outputs which corresponds to the Competing Risks (CRs) type of analysis specific to the medical domain of survival analysis. The resulted model is entitled the PLANN-CR-ARD model. The three versions of the PLANN-CR-ARD model are tested on a very demanding medical dataset taken from the survival analysis. The ARD framework implements the calculation of the network outputs, the marginalization of the network outputs and the model selection. The numerical results show that the neural network model based on the global compensation is very effective. |
| An air quality forecast model based on the BP neural network of the samples self-organization clustering | In practice, the training samples of the neural network usually have intrinsic characteristics and regularity. The paper presents a BP neural network (BPNN) forecast model based on the samples self-organizing clustering. Using the clustering feature of the self-organizing competitive neural network(SOCNN), it improves the effect of the training sample to the performance of BPNN. The momentum - adaptive learning rate adjustment algorithm that makes the convergence speed faster with the higher error precision is used for the BPNN in this model. The experiments of the air quality forecast with this model showed that BPNN forecast model based on the samples self-organizing clustering will improve the convergence rate first and reduce the possibility of falling into the local minimum also and improve the forecast accuracy. |
| Hyperspectral analysis of leaf copper accumulation in agronomic crop based on artificial neural network | Copper is one kind of trace element in soil which is necessary for the growth and development of plants. Much more copper over the needed amount of agronomic crop is harmful to crop growth and becomes pollutants in soil. At present, there are few studies concerning the quantitative impact of heavy metal contamination on crops. This research investigates an alternative approach. Red edge parameters of rice canopy will be obtained based on the first order and second order derivative spectra, and its relationship with agricultural parameters will be analyzed. It is found that there is strong correlation between red edge position and leaf chlorophyll a / leaf chlorophyll b, red edge amplitude and carotenoid, red edge peak area and the leaf area index, margin and fresh leaves quality. There is no obvious correlation between moisture and red edge parameters. BP artificial neural network method is used to study quantitatively the inherent relation between the chlorophyll content of rice and copper contents in soil. Taking red edge parameters mentioned above which have strong correlation with agricultural parameters, as well as ph value as input, copper content as output, four layers BP neural network with five inputs, one output and two hidden layers will be established. It is tested that the network fitting accuracy reaches 98% and the model has a high fitting degree, which prediction accuracy also receives 85.4%. This study is helpful to improve the ability of monitoring the heavy metal contamination of soil and environment in agricultural region. |
| Research on the Mechanical Property of Insect Cuticle with Neural Network | In this work, a neural network material model is built for the simulation of the inelastic behavior of biocomposite insect cuticle. Radial basis function neural network is adopted in the simulation for that the neural network has the characteristic of fast and exactly completing the simulation. In the construction of the neural network, the network is trained based on the experimental data of the load-displacement relationship of a chafer cuticle. A strain-controlled mode and the iterative method of data are adopted in the training process of the neural network. The obtained neural network model is used for the simulation of the inelastic behavior of another kind of insect cuticle. It is shown that the obtained material model of the radial basis function neural network can satisfactorily simulate the inelastic behavior of insect cuticle |
| Interpretation Artificial Neural Network in Remote Sensing Image Classification | The responses of neural networks for uniform and normal distribution are studied, especially the BP and RBF neural networks and the question of combination between neural networks and fuzzy logical is answered by experiments. Linear relationship among sample feature components which impact the time consumption and convergence accuracy of networks has been discussed also. In the condition of feature vector included original bands and good separating degree components, BP and RBF neural networks combined with Fuzzy Reasoning have been used for TM image classification. Overall classification accuracy and Kappa coefficients reached 0.915 and 94.33% in RBF network which is higher than 0.845 and 89.67% in BP network. |

The result shown by the query "Networking" are given below.

| Title | Abstract |
|---|---|
| IEEE Standard for Information Technology--Telecommunications and information exchange between systems--Local and metropolitan area networks--Specific requirements Part 11: Wireless LAN Medium Access Control (MAC) and Physical Layer (PHY) specifications Amendment 10: Mesh Networking | This amendment describes protocols for IEEE 802.11 stations to form self-configuring multi-hop networks that support both broadcast/multicast and unicast data delivery. |
| IEEE Standard for Wireless Access in Vehicular Environments (WAVE) - Networking Services | Wireless Access in Vehicular Environments (WAVE) Networking Services provides services to WAVE devices and systems. Layers 3 and 4 of the open system interconnect (OSI) model and the Internet Protocol (IP), User Datagram Protocol (UDP), and Transmission Control Protocol (TCP) elements of the Internet model are represented. Management and data services within WAVE devices are provided. |
| Networking Capability and New Product Development | Current research on network theory remains largely focused on structures and outcomes without exploring the capability that firms need to build efficient and effective networks to their advantage. In this paper, we take a networking capability view in studying inter-firm relationships. We assume that firms create their networks strategically. We show a more compelling picture of how networking behavior influences new product development performance by developing a research design and statistical approach that addresses the endogeneity problem in network research. We argue theoretically and demonstrate empirically that networking capability is a reliable predictor of new product development performance. We find that interaction cost reduction, opportunity discovery, resource acquisition, and market knowledge generation and technology knowledge generation, respectively, mediate the positive relationship between networking capability and new product development performance. Third, we find that environmental dynamism acts as a moderating factor in the effect of networking capability on new product development performance, and the effect of networking capability on interaction cost reduction, opportunity discovery, resource acquisition, and market knowledge generation and technology knowledge generation are more salient when environmental dynamism is high than when it is low. The implication of the findings is that firms can develop strong networking capability to enhance product and technology innovation. |
| A Survey on Service-Oriented Network Virtualization Toward Convergence of Networking and Cloud Computing | The crucial role that networking plays in Cloud computing calls for a holistic vision that allows combined control, management, and optimization of both networking and computing resources in a Cloud environment, which leads to a convergence of networking and Cloud computing. Network virtualization is being adopted in both telecommunications and the Internet as a key attribute for the next generation networking. Virtualization, as a potential enabler of profound changes in both communications and computing domains, is expected to bridge the gap between these two fields. Service-Oriented Architecture (SOA), when applied in network virtualization, enables a Network-as-a-Service (NaaS) paradigm that may greatly facilitate the convergence of networking and Cloud computing. Recently the application of SOA in network virtualization has attracted extensive interest from both academia and industry. Although numerous relevant research works have been published, they are currently scattered across multiple fields in the literature, including telecommunications, computer networking, Web services, and Cloud computing. In this article we present a comprehensive survey on the latest developments in service-oriented network virtualization for supporting Cloud computing, particularly from a perspective of network and Cloud convergence through NaaS. Specifically, we first introduce the SOA principle and review recent research progress on applying SOA to support network virtualization in both telecommunications and the Internet. Then we present a framework of network-Cloud convergence based on service-oriented network virtualization and give a survey on key technologies for realizing NaaS, mainly focusing on state of the art of network service description, discovery, and composition. We also discuss the challenges brought in by network-Cloud convergence to these technologies and research opportunities available in these areas, with a hope to arouse the research community's interest - n this emerging interdisciplinary field. |

The result shown by the query "Modeling" are given below.

| Title | Abstract |
|---|---|
| IEC/IEEE Behavioural Languages - Part 5: VITAL ASIC (Application Specific Integrated Circuit) Modeling Specification (Adoption of IEEE Std 1076.4-2000) | The VITAL (VHDL Initiative Towards ASIC Libraries) ASIC Modeling Specication is defined in this standard. This modeling specication defines a methodology which promotes the development of highly accurate, efficient simulation models for ASIC (Application-Specific Integrated Circuit) components in VHDL. |
| IEEE Standard for VITAL ASIC (Application Specific Integrated Circuit) Modeling Specification | The VITAL (VHDL Initiative Towards ASIC Libraries)ASIC Modeling Specification is defined in this standard.This modeling specification defines a methodology which promotes the development of highly accurate, efficient simulation models for ASIC (Application-Specific Integrated Circuit)components in VHDL. |
| IEEE Standard for Conceptual Modeling Language Syntax and Semantics for IDEF1X/Sub 97/ (IDEF/Sub Object/) | IDEF1X/sub 97/ consists of two conceptual modeling languages. The key-style language supports data/information modeling and is downward compatible with the US government's 1993 standard, FIPS PUB 184. The identity-style language is based on the object model with declarative rules and constraints. IDEF1X/sub 97/ identity style includes constructs for the distinct but related components of object abstraction: interface, requests, and realization; utilizes graphics to state the interface; and defines a declarative, directly executable Rule and Constraint Language for requests and realizations. IDEF1X/sub 97/ conceptual modeling supports implementation by relational databases, extended relational databases, object databases, and object programming languages. IDEF1X/sub 97/ is formally defined in terms of first order logic. A procedure is given whereby any valid IDEF1X/sub 97/ model can be transformed into an equivalent theory in first order logic. That procedure is then applied to a meta model of IDEF1X/sub 97/ to define the valid set of IDEF1X/sub 97/ models. |
| IEEE Recommended Practice for Validation of Computational Electromagnetics Computer Modeling and Simulations | This recommended practice is a companion document for IEEE Std 1597.1 \™-2008. Examples and problem sets to be used in the validation of computational electromagnetics (CEM) computer modeling and simulation techniques, codes, and models are provided. It is applicable to a wide variety of electromagnetic (EM) applications including but not limited to the fields of antennas, signal integrity (SI), radar cross section (RCS), and electromagnetic compatibility (EMC). This document shows how to validate a particular solution data set by comparing it to the data set obtained by measurements, alternate codes, canonical, or analytic methods. |
| IEEE Standard for Modeling and Simulation [M and S] High Level Architecture [HLA] - Federate Interface Specification | The high level architecture [HLA] has been developed to provide a common architecture for distributed modeling and simulation. The HLA defines an integrated approach that provides a common framework for the interconnection of interacting simulations. This document, the second in a family of three related HLA documents, defines the standard services of and interfaces to the HLA Runtime Infrastructure [RTI]. These services are used by the interacting simulations to achieve a coordinated exchange of information when they participate in a distributed federation. The standards contained in this architecture are interrelated and need to be considered as a product set, when changes are made. They each have value independently. |
| The Modeling Process with System Dynamics | This chapter contains sections titled:<br>System Dynamics Background<br>General System Behaviors<br>Modeling Overview<br>Problem Definition<br>Model Conceptualization<br>Model Formulation and Construction<br>Simulation<br>Model Assessment<br>Policy Analysis<br>Continuous Model Improvement<br>Software Metrics Considerations<br>Project Management Considerations<br>Modeling Tools<br>Major References<br>Chapter 2 Summary<br>Exercises |

The result shown by the query "Neural Networks" are given below.

| Title | Abstract |
|---|---|
| A genetic algorithm as the learning procedure for neural networks | A way in which a neural network can implement a genetic algorithm as its learning algorithm is shown. This model is called GLANN (genetic learning algorithm for neural networks). The components of GLANN can be shown to be biologically plausible. The algorithm itself can be classified as a reinforcement learning algorithm. The neural network has a fixed architecture and processes binary strings using genetic operators. Learning is stored in the form of newly created patterns, which can then be stored in some kind of associative memory. The benefits of GLANN reside in the proven optimizing capabilities of genetic algorithms, and in its parallel implementation. The shallow two-level architecture translates into system scalability, an issue that has not been successfully resolved in the case of other neural network algorithms. |
| Learning, Goals, and Learning Goals | This chapter contains sections titled: Why Goals?, An Everyday Example, Toward a Planful Model of Learning, A Framework for Goal-Driven Learning, Major Issues in Goal-Driven Learning, What is a Goal?, Types of Goals, Role of Goals in Learning, Pragmatic Implications of Goal-Driven Learning, Summary, Acknowledgments, Notes, References |
| Discovery of Action Patterns and User Correlations in Task-Oriented Processes for Goal-Driven Learning Recommendation | With the high development of social networks, collaborations in a socialized web-based learning environment has become increasing important, which means people can learn through interactions and collaborations in communities across social networks. In this study, in order to support the enhanced collaborative learning, two important factors, user behavior patterns and user correlations, are taken into account to facilitate the information and knowledge sharing in a task-oriented learning process. Following a hierarchical graph model for enhanced collaborative learning within a task-oriented learning process, which describes relations of learning actions, activities, sub-tasks and tasks in communities, the learning action pattern and Goal-driven Learning Group, as well as their formal definitions and related algorithms, are introduced to extract and analyze users' learning behaviors in both personal and cooperative ways. In addition, a User Networking Model, which is used to represent the dynamical user relationships, is proposed to calculate user correlations in accordance with their interactions in a social community. Based on these, an integrated mechanism is developed to utilize both user behavior patterns and user correlations for the recommendation of individualized learning actions. The system architecture is described finally, and the experiment results are presented and discussed to demonstrate the practicability and usefulness of our methods. |
| Machine Learning Paradigms for Speech Recognition: An Overview | Automatic Speech Recognition (ASR) has historically been a driving force behind many machine learning (ML) techniques, including the ubiquitously used hidden Markov model, discriminative learning, structured sequence learning, Bayesian learning, and adaptive learning. Moreover, ML can and occasionally does use ASR as a large-scale, realistic application to rigorously test the effectiveness of a given technique, and to inspire new problems arising from the inherently sequential and dynamic nature of speech. On the other hand, even though ASR is available commercially for some applications, it is largely an unsolved problem - for almost all applications, the performance of ASR is not on par with human performance. New insight from modern ML methodology shows great promise to advance the state-of-the-art in ASR technology. This overview article provides readers with an overview of modern ML techniques as utilized in the current and as relevant to future ASR research and systems. The intent is to foster further cross-pollination between the ML and ASR communities than has occurred in the past. The article is organized according to the major ML paradigms that are either popular already or have potential for making significant contributions to ASR technology. The paradigms presented and elaborated in this overview include: generative and discriminative learning; supervised, unsupervised, semi-supervised, and active learning; adaptive and multi-task learning; and Bayesian learning. These learning paradigms are motivated and discussed in the context of ASR technology and applications. We finally present and analyze recent developments of deep learning and learning with sparse representations, focusing on their direct relevance to advancing ASR technology. |
| Introduction to Reinforcement and Systemic Machine Learning | This chapter contains sections titled:<br>Introduction<br>Supervised, Unsupervised, and Semisupervised Machine Learning<br>Traditional Learning Methods and History of Machine Learning<br>What Is Machine Learning?<br>Machine-Learning Problem<br>Learning Paradigms<br>Machine-Learning Techniques and Paradigms<br>What Is Reinforcement Learning?<br>Reinforcement Function and Environment Function<br>Need of Reinforcement Learning<br>Reinforcement Learning and Machine Intelligence<br>What Is Systemic Learning?<br>What Is Systemic Machine Learning?<br>Challenges in Systemic Machine Learning<br>Reinforcement Machine Learning and Systemic Machine Learning<br>Case Study Problem Detection in a Vehicle<br>Summary<br>Reference |

The result shown by the query “Data Mining” is given below.

| Title | Abstract |
| --- | --- |
| Toward Scalable Systems for Big Data Analytics: A Technology Tutorial | Recent technological advancements have led to a deluge of data from distinctive domains (e.g., health care and scientific sensors, user-generated data, Internet and financial companies, and supply chain systems) over the past two decades. The term big data was coined to capture the meaning of this emerging trend. In addition to its sheer volume, big data also exhibits other unique characteristics as compared with traditional data. For instance, big data is commonly unstructured and require more real-time analysis. This development calls for new system architectures for data acquisition, transmission, storage, and large-scale data processing mechanisms. In this paper, we present a literature survey and system tutorial for big data analytics platforms, aiming to provide an overall picture for nonexpert readers and instill a do-it-yourself spirit for advanced audiences to customize their own big-data solutions. First, we present the definition of big data and discuss big data challenges. Next, we present a systematic framework to decompose big data systems into four sequential modules, namely data generation, data acquisition, data storage, and data analytics. These four modules form a big data value chain. Following that, we present a detailed survey of numerous approaches and mechanisms from research and industry communities. In addition, we present the prevalent Hadoop framework for addressing big data challenges. Finally, we outline several evaluation benchmarks and potential research directions for big data systems. |
| Data-Mining Concepts | This chapter contains sections titled:<br>Introduction<br>Data-Mining Roots<br>Data-Mining Process<br>Large Data Sets<br>Data Warehouses for Data Mining<br>Business Aspects of Data Mining: Why a Data-Mining Project Fails<br>Organization of This Book<br>Review Questions and Problems<br>References for Further Study |
| Advances in Data Mining | This chapter contains sections titled:<br>Graph Mining<br>Temporal Data Mining<br>Spatial Data Mining (SDM)<br>Distributed Data Mining (DDM)<br>Correlation Does Not Imply Causality<br>Privacy, Security, and Legal Aspects of Data Mining<br>Review Questions and Problems<br>References for Further Study |
| Data mining: an overview from a database perspective | Mining information and knowledge from large databases has been recognized by many researchers as a key research topic in database systems and machine learning, and by many industrial companies as an important area with an opportunity of major revenues. Researchers in many different fields have shown great interest in data mining. Several emerging applications in information-providing services, such as data warehousing and online services over the Internet, also call for various data mining techniques to better understand user behavior, to improve the service provided and to increase business opportunities. In response to such a demand, this article provides a survey, from a database researcher's point of view, on the data mining techniques developed recently. A classification of the available data mining techniques is provided and a comparative study of such techniques is presented |
| Introduction to Scientific Data Mining: Direct Kernel Methods and Applications | This chapter contains sections titled:<br>Introduction<br>What Is Data Mining?<br>Basic Definitions for Data Mining<br>Introduction to Direct Kernel Methods<br>Direct Kernel Ridge Regression<br>Case Study #1: Predicting the Binding Energy for Amino Acids<br>Case Study #2: Predicting the Region of Origin for Italian Olive Oils<br>Case Study #3: Predicting Ischemia from Magnetocardiography<br>Fusion of Soft Computing and Hard Computing<br>Conclusions |
| Data Extraction for Deep Web Using WordNet | Our survey shows that the techniques used in data extraction from deep webs need to be improved to achieve the efficiency and accuracy of automatic wrappers. Further investigations indicate that the development of a lightweight ontological technique using existing lexical database for English (WordNet) is able to check the similarity of data records and detect the correct data region with higher precision using the semantic properties of these data records. The advantages of this method are that it can extract three types of data records, namely, single-section data records, multiple-section data records, and loosely structured data records, and it also provides options for aligning iterative and disjunctive data items. Experimental results show that our technique is robust and performs better than the existing state-of-the-art wrappers. Tests also show that our wrapper is able to extract data records from multilingual web pages and that it is domain independent. |

The result shown by the query “Computer Graphics” are given below.

| Title | Abstract |
|---|---|
| Computer-Aided Geometric Design: A bibliography with Keywords and Classified Index | Computer-aided geometric design is a subject which involves the representation, specification, design, manipulation, display, and analysis of free-form curves and surfaces. This eclectic field draws on techniques and principles from numerical analysis, approximation theory, computer graphics, interactive computer systems, mechanical and geometric design, and many other areas. Nevertheless, computer-aided geometric design is an endeavor with a character, emphasis, and purpose of its own, combining techniques and principles from other fields with unique approaches that are often somewhat enigmatic and confusing to the novice. Many things sound and seem familiar, yet the directions and motivations are distinct to this discipline. This decade is proving to be an era of intense interest in computer-aided geometric design, an activity attracting people from many sectors including computing and manufacturing. As the discipline more firmly establishes valid principles and standard techniques, it will undoubtedly be of increasing importance in computer-aided manufacturing and a wide range of other applications. |
| IEEE Standard Computer Dictionary: A Compilation of IEEE Standard Computer Glossaries | Identifies terms currently in use in the computer field. Standard definitions for thoseterms are established. Compilation of IEEE Stds IEEE Std 1084, IEEE Std 610.2, IEEE Std 610.3, IEEE Std 610.4, IEEE Std 610.5 and IEEE Std 610.12 |
| Computer Graphics, Interactive Techniques, and Image Processing 1970-1975: A Bibliography | Computer graphics, interactive techniques, and image processing are among the developments in the constantly evolving computer science field that impact the potential user ever more rapidly. This bibliography attempts to compile all articles, books, conference papers, and technical reports about computer graphics and man-machine interaction that have been published in English from 1970 to 1975. Because the literature pertaining to computer graphics and man-machine interaction is immense, this bibliography will no doubt be incomplete. Suggestions and contributions for future supplements to the bibliography should be sent to the compiler. |
| IEEE Standard Glossary of Computer Hardware Terminology | Terms pertaining to computer hardware are defined. Terms falling under the categories of computer architecture, computer storage, general hardware concepts, peripherals, and processors and components are included. |
| Review of computer vision education | Computer vision is becoming a mainstream subject of study in computer science and engineering. With the rapid explosion of multimedia and the extensive use of video and image-based communications over the World Wide Web, there is a strong demand for educating students to become knowledgeable in computer imaging and vision. The purpose of this paper is to review the status of computer vision education today. |
| Background and Source Information About Computer Graphics | This up-to-date summary* contains a selection of representative books, articles, magazines, newsletters, and journals in the computer graphics field. It singles out some specific articles and also includes professional society names and addresses; titles and dates of conferences, seminars, national meetings, and short courses; names and addresses of commercial exhibitions with technical programs; product reviews and directories; and market data sources. |
| An Updated Guide to Sources of Information About Computer Graphics | This source guide* contains a selection of representiative books, magazines, and journals from the computer graphics field. It singles out some specific magazine and journal articles and also includes professional society names and addresses; titles and dates of conferences, symposia, meetings, and short courses; and names and addresses of product review and market data sources. |
| IEEE Standard Glossary of Computer Applications Terminology | This glossary defines terms in the field of computer applications. Topics covered include automated language processing, automatic indexing, business data processing, character recognition, computer-aided design and manufacturing, computer-assisted instruction, control systems, critical path method, library automation, medical applications, micrographics, office automation, operations research, personal computing, scientific and engineering applications, telecommunication applications, and word processing.<> |
| An Indexed Bibliography on Computer Animation | Although the computer plays an ever increasing role in animation, the term Â¿computer animationÂ¿ is imprecise and sometimes can be misleading, since the computer can play a variety of different roles. A popular and simple way of classifying animation systems is to distinguish between computer-assisted and modeled animation. |
| Graphics Software-from Techniques to Principles | The history of graphics software mirrors that of computing in general: we work out basic techniques; we develop algorithms; we begin to search for standards. |
| Hardware Design for 3D Graphics | Three-dimensional graphics has played an important role in multimedia systems and virtual reality systems. Since its applications emerge rapidly from technical areas to nontechnical areas, state-of-the-art 3-D graphics hardware design focuses not only on high performance and quality, but also on low cost and system integration. In this chapter, we discuss the methodology and bottlenecks in hardware design. The techniques for high- performance 3-D graphics synthesis and various considerations are discussed, especially parallelism and advanced memory I/O architecture. Then, some existing architectures for various design considerations are illustrated. Because most programmers rely on application program interfaces (APIs) to develop applications on hardware, the API becomes important and affects the 3-D hardware design in integrated system. |
| A Guide to Sources of Information about Computer Graphics | Computer graphics is now a mature discipline, supported by an extensive literature, a number of professional societies, and a full schedule of conferences, workshops, and short courses. |

The result shown by the query "Clustering" are given below.

| Title | Abstract |
|---|---|
| DSCLU: A New Data Stream Clustring Algorithm for Multi Density Environments | Recently, data stream has become popular in many contexts of data mining. Due to the high amount of incoming data, traditional clustering algorithms are not suitable for this family of problems. Many data stream clustering algorithms proposed in recent years considered the scalability of data, but most of them did not attend the following issues: (1) The quality of clustering can be dramatically low over the time. (2) Some of the algorithms cannot handle arbitrary shapes of data stream and consequently the results are limited to specific regions. (3) Most of the algorithms have not been evaluated in multi-density environments. Identifying appropriate clusters for data stream by handling the arbitrary shapes of clusters is the aim of this paper. The gist of the overall approach in this paper can be stated in two phases. In online phase, data manipulate with specific data structure called micro cluster. This phase is activated by incoming of data. The offline phase is manually activated by coming a request from user. The algorithm handles clusters by considering with micro clusters created by the online phase. The experimental evaluation showed that proposed algorithm has suitable quality and also returns appropriate results even in multi-density environments. |
| GDCLU: A New Grid-Density Based ClustrIng Algorithm | This paper addresses the density based clustering problem in data mining where clusters are established based on density of regions. The most well-known algorithm proposed in this area is DBSCAN [1] which employs two parameters influencing the shape of resulted clusters. Therefore, one of the major weaknesses of this algorithm is lack of ability to handle clusters in multi-density environments. In this paper, a new density based grid clustering algorithm, GDCLU, is proposed which uses a new definition for dense regions. It determines dense grids based on densities of their neighbors. This new definition enables GDCLU to handle different shaped clusters in multi-density environments. Also this algorithm benefits from scale independency feature. The time complexity of the algorithm is O(n) in which n is number of points in dataset. Several examples are presented showing promising improvement in performance over other basic algorithms like optics in multi-density environments. |
| Detection of QRS complex in ECG signal based on classification approach | Electrocardiogram (ECG) signals are used to analyze the cardiovascular activity in the human body and have a primary role in the diagnosis of several heart diseases. The QRS complex is the most important and distinguishable component in the ECG because of its spiked nature and high amplitude. Automatic detection and delineation of the QRS complex in ECG is of extreme importance for computer aided diagnosis of cardiac disorder. Therefore, the accurate detection of this component is crucial to the performance of subsequent machine learning algorithms for cardiac disease classification. The aim of the present work is to detect the QRS wave from electrocardiogram (ECG) signals. Initially the baseline drift has been removed from the signal followed by the decomposition using continuous wavelet transform. Modulus maxima approach proposed by Mallat has been used to compute the Lipschitz exponent of the components. By using the property of R peak, having highest and prominent amplitude and Lipschitz exponents, we have applied the K means clustering technique to classify QRS complex. In order to evaluate the algorithm, the analysis has been done on MIT-BIH Arrhythmia database. |
| Improving threshold assignment for cluster head selection in hierarchical wireless sensor networks | WSNs (wireless sensor networks) are networks that hundreds or thousands of nodes poured in a field and nodes try to send sensed event to base station (BS). In many cases, the BS isn't in the field and is so far away from nodes. Energy efficiency is one of the major concerns in wireless sensor networks since it impacts the network lifetime. So instead of transmitting directly to BS, in hierarchical network, we choose a cluster head (CH) to send aggregated data from neighbors to BS. In this paper we investigate a new threshold assignment for LEACH and xLEACH that improves energy consumption. We insert the distances of nodes from BS in threshold assignment in order to unbalance the CH selection to reduce energy consumption in the network. |
| A Mathematics Morphology Based Algorithm of Obstacles Clustering | As a large amount of data stored in spatial databases, people may like to find groups of data which share similar features. Thus cluster analysis becomes an important area of research in data mining. In the real world, there exist many physical obstacles such as rivers, lakes and highways, and their presence may affect the result of clustering substantially. However, most of clustering algorithms can not deal with obstacles. In this paper, a new clustering algorithm MMO is proposed for the problem of clustering in the presence of obstacles. The main contributions are: two new mathematics morphological operators are introduced to discover clusters in the presence of obstacles. Our new operators are more accurate than the ordinary operators: open and close. The performance tests show that: MMO is effective in discovering clusters of arbitrary shape in the presence of obstacles; it is very efficient with a complexity of O(N+M) , where N is the number of data points, and M is the number of obstacles; it is not sensitive to noise. |
| Fast Nonparametric Image Segmentation with Dirichlet Processes | Among nonparametric clustering methods Dirichlet processes mixture models have proven to be very effective for unsupervised clustering. Image segmentation is an area where clustering has become a frequently used method. Many existing cluster type segmentation algorithms face problems such as slowness or parametric nature. We propose an effective method based on variational Dirichlet processes to achieve a great speed. In our approach we apply kd-tree to partition images and Dirichlet processes to cluster pixel color values in those partitions. Our experiments have shown that our method of clustering is fast compared to other methods of clustering using Dirichlet processes and also well performing compared spectral clustering. |
| Clustering Algorithms in Mobile Ad Hoc Networks | Clustering of nodes provides an efficient means of establishing a hierarchical structure in mobile ad hoc networks. In mobile ad hoc networks, the movement of the network nodes may quickly change the topology resulting in the increase of the overhead message in topology maintenance; the clustering schemes for mobile ad hoc networks therefore aim at handling topology maintenance, managing node movement or reducing overhead. This paper presents the reasons for clustering algorithms in ad hoc networks, as well as a short survey of the basic ideas and priorities of existing clustering algorithms. |

The result shown by the query “Cloud Computing” is given below.

Show Springer Open Response

| Title | Abstract |
|---|---|
| Cloud-Based Augmentation for Mobile Devices: Motivation, Taxonomies, and Open Challenges | Recently, Cloud-based Mobile Augmentation (CMA) approaches have gained remarkable ground from academia and industry. CMA is the state-of-the-art mobile augmentation model that employs resource-rich clouds to increase, enhance, and optimize computing capabilities of mobile devices aiming at execution of resource-intensive mobile applications. Augmented mobile devices envision to perform extensive computations and to store big data beyond their intrinsic capabilities with least footprint and vulnerability. Researchers utilize varied cloud-based computing resources (e.g., distant clouds and nearby mobile nodes) to meet various computing requirements of mobile users. However, employing cloud-based computing resources is not a straightforward panacea. Comprehending critical factors (e.g., current state of mobile client and remote resources) that impact on augmentation process and optimum selection of cloud-based resource types are some challenges that hinder CMA adaptability. This paper comprehensively surveys the mobile augmentation domain and presents taxonomy of CMA approaches. The objectives of this study is to highlight the effects of remote resources on the quality and reliability of augmentation processes and discuss the challenges and opportunities of employing varied cloud-based resources in augmenting mobile devices. We present augmentation definition, motivation, and taxonomy of augmentation types, including traditional and cloud-based. We critically analyze the state-of-the-art CMA approaches and classify them into four groups of distant fixed, proximate fixed, proximate mobile, and hybrid to present a taxonomy. Vital decision making and performance limitation factors that influence on the adoption of CMA approaches are introduced and an exemplary decision making flowchart for future CMA approaches are presented. Impacts of CMA approaches on mobile computing is discussed and open challenges are presented as the future research directions. |
| Heterogeneity in Mobile Cloud Computing: Taxonomy and Open Challenges | The unabated flurry of research activities to augment various mobile devices by leveraging heterogeneous cloud resources has created a new research domain called Mobile Cloud Computing (MCC). In the core of such a non-uniform environment, facilitating interoperability, portability, and integration among heterogeneous platforms is nontrivial. Building such facilitators in MCC requires investigations to understand heterogeneity and its challenges over the roots. Although there are many research studies in mobile computing and cloud computing, convergence of these two areas grants further academic efforts towards flourishing MCC. In this paper, we define MCC, explain its major challenges, discuss heterogeneity in convergent computing (i.e. mobile computing and cloud computing) and networking (wired and wireless networks), and divide it into two dimensions, namely vertical and horizontal. Heterogeneity roots are analyzed and taxonomized as hardware, platform, feature, API, and network. Multidimensional heterogeneity in MCC results in application and code fragmentation problems that impede development of cross-platform mobile applications which is mathematically described. The impacts of heterogeneity in MCC are investigated, related opportunities and challenges are identified, and predominant heterogeneity handling approaches like virtualization, middleware, and service oriented architecture (SOA) are discussed. We outline open issues that help in identifying new research directions in MCC. |
| Market-Oriented Cloud Computing and The Cloudbus Toolkit | This chapter introduces the fundamental concepts of market-oriented Cloud computing systems and presents a reference model. The model, together with the state-of-the-art technologies presented in the chapter, contribute significantly towards the mainstream adoption of Cloud computing technology. However, any technology brings with it new challenges and breakthroughs. The chapter focuses on the major challenges faced by the industry when adopting Cloud computing as a mainstream technology as part of the distributed computing paradigm. It presents a utility-oriented Cloud vision that is a generic model for realizing market-oriented Cloud computing vision. Cloudbus realized this by developing various tools and platforms that can be used individually or together as an integrated solution. The author's demonstrate through experiments that their toolkit can provide applications based on deadline, optimize cost and time of applications, and manage real-world problems through an integrated solution. |
| A Survey on Service-Oriented Network Virtualization Toward Convergence of Networking and Cloud Computing | The crucial role that networking plays in Cloud computing calls for a holistic vision that allows combined control, management, and optimization of both networking and computing resources in a Cloud environment, which leads to a convergence of networking and Cloud computing. Network virtualization is being adopted in both telecommunications and the Internet as a key attribute for the next generation networking. Virtualization, as a potential enabler of profound changes in both communications and computing domains, is expected to bridge the gap between these two fields. Service-Oriented Architecture (SOA), when applied in network virtualization, enables a Network-as-a-Service (NaaS) paradigm that may greatly facilitate the convergence of networking and Cloud computing. Recently the application of SOA in network virtualization has attracted extensive interest from both academia and industry. Although numerous relevant research works have been published, they are currently scattered across multiple fields in the literature, including telecommunications, computer networking, Web services, and Cloud computing. In this article we present a comprehensive survey on the latest developments in service-oriented network virtualization for supporting Cloud computing, particularly from a perspective of network and Cloud convergence through NaaS. Specifically, we first introduce the SOA principle and review recent research progress on applying SOA to support network virtualization in both telecommunications and the Internet. Then we present a framework of network-Cloud convergence based on service-oriented network virtualization and give a survey on key technologies for realizing NaaS, mainly focusing on state of the art of network service description, discovery, and composition. We also discuss the challenges brought in by network-Cloud convergence to these technologies and research opportunities available in these areas, with a hope to arouse the research community's interest - n this emerging interdisciplinary field. |
| Cloud Mobile Media: Reflections and Outlook | This paper surveys the emerging paradigm of cloud mobile media. We start with two alternative perspectives for cloud mobile media networks: an end-to-end view and a layered view. Summaries of existing research in this area are organized according to the layered service framework: i) cloud resource management and control in infrastructure-as-a-service (IaaS), ii) cloud-based media services in platform-as-a-service (PaaS), and iii) novel cloud-based systems and applications in software-as-a-service (SaaS). We further substantiate our proposed design principles for cloud-based mobile media using a concrete case study: a cloud-centric media platform (CCMP) developed at Nanyang Technological University. Finally, this paper concludes with an outlook of open research problems for realizing the vision of cloud-based mobile media. |

The result shown by the query "Big data" are given below.

Show Springer Open Response

| Title | Abstract |
|---|---|
| Toward Scalable Systems for Big Data Analytics: A Technology Tutorial | Recent technological advancements have led to a deluge of data from distinctive domains (e.g., health care and scientific sensors, user-generated data, Internet and financial companies, and supply chain systems) over the past two decades. The term big data was coined to capture the meaning of this emerging trend. In addition to its sheer volume, big data also exhibits other unique characteristics as compared with traditional data. For instance, big data is commonly unstructured and require more real-time analysis. This development calls for new system architectures for data acquisition, transmission, storage, and large-scale data processing mechanisms. In this paper, we present a literature survey and system tutorial for big data analytics platforms, aiming to provide an overall picture for nonexpert readers and instill a do-it-yourself spirit for advanced audiences to customize their own big-data solutions. First, we present the definition of big data and discuss big data challenges. Next, we present a systematic framework to decompose big data systems into four sequential modules, namely data generation, data acquisition, data storage, and data analytics. These four modules form a big data value chain. Following that, we present a detailed survey of numerous approaches and mechanisms from research and industry communities. In addition, we present the prevalent Hadoop framework for addressing big data challenges. Finally, we outline several evaluation benchmarks and potential research directions for big data systems. |
| Data mining with big data | Big Data concern large-volume, complex, growing data sets with multiple, autonomous sources. With the fast development of networking, data storage, and the data collection capacity, Big Data are now rapidly expanding in all science and engineering domains, including physical, biological and biomedical sciences. This paper presents a HACE theorem that characterizes the features of the Big Data revolution, and proposes a Big Data processing model, from the data mining perspective. This data-driven model involves demand-driven aggregation of information sources, mining and analysis, user interest modeling, and security and privacy considerations. We analyze the challenging issues in the data-driven model and also in the Big Data revolution. |
| Governing Big Data: Principles and practices | As data-intensive decision making is being increasingly adopted by businesses, governments, and other agencies around the world, most organizations encountering a very large amount and variety of data are still contemplating and assessing their readiness to embrace "Big Data." While these organizations devise various ways to deal with the challenges such data brings, the impact and importance of Big Data to information quality and governance programs should not be underestimated. Drawing upon implementation experiences of early adopters of Big Data technologies across multiple industries, this paper explores the issues and challenges involved in the management of Big Data, highlighting the principles and best practices for effective Big Data governance. |
| BigDataBench: A big data benchmark suite from internet services | As architecture, systems, and data management communities pay greater attention to innovative big data systems and architecture, the pressure of benchmarking and evaluating these systems rises. However, the complexity, diversity, frequently changed workloads, and rapid evolution of big data systems raise great challenges in big data benchmarking. Considering the broad use of big data systems, for the sake of fairness, big data benchmarks must include diversity of data and workloads, which is the prerequisite for evaluating big data systems and architecture. Most of the state-of-the-art big data benchmarking efforts target evaluating specific types of applications or system software stacks, and hence they are not qualified for serving the purposes mentioned above. This paper presents our joint research efforts on this issue with several industrial partners. Our big data benchmark suite-BigDataBench not only covers broad application scenarios, but also includes diverse and representative data sets. Currently, we choose 19 big data benchmarks from dimensions of application scenarios, operations/ algorithms, data types, data sources, software stacks, and application types, and they are comprehensive for fairly measuring and evaluating big data systems and architecture. BigDataBench is publicly available from the project home page http://prof.ict.ac.cn /BigDataBench. Also, we comprehensively characterize 19 big data workloads included in BigDataBench with varying data inputs. On a typical state-of-practice processor, Intel Xeon E5645, we have the following observations: First, in comparison with the traditional benchmarks: including PARSEC, HPCC, and SPECCPU, big data applications have very low operation intensity, which measures the ratio of the total number of instructions divided by the total byte number of memory accesses; Second, the volume of data input has non-negligible impact on micro-architecture characteristics, which may impose challenges for simulation-based- big data architecture research; Last but not least, corroborating the observations in CloudSuite and DCBench (which use smaller data inputs), we find that the numbers of L1 instruction cache (L1I) misses per 1000 instructions (in short, MPKI) of the big data applications are higher than in the traditional benchmarks; also, we find that L3 caches are effective for the big data applications, corroborating the observation in DCBench. |
| Requirements of an Open Data Based Business Ecosystem | Emerging opportunities for open data based business have been recognized around the world. Open data can provide new business opportunities for actors that provide data, for actors that consume data, and for actors that develop innovative services and applications around the data. Open data based business requires business models and a collaborative environment-called an ecosystem-to support businesses based on open data, services, and applications. This paper outlines the open data ecosystem (ODE) from the business viewpoint and then defines the requirements of such an ecosystem. The outline and requirements are based on the state-of-the-art knowledge explored from the literature and the state of the practice on data-based business in the industry collected through interviews. The interviews revealed several motives and advantages of the ODE. However, there are also obstacles that should be carefully considered and solved. This paper defines the actors of the ODE and their roles in the ecosystem as well as the business model elements and services that are needed in open data based business. According to the interviews, the interest in open data and open data ecosystems is high at this moment. However, further research work is required to establish and validate the ODE in the near future. |

The result shown by the query “Application Programming Interface” is given below.

| Title | Abstract |
|---|---|
| Multi-dimensional exploration of API usage | This paper is concerned with understanding API usage in a systematic, explorative manner for the benefit of both API developers and API users. There exist complementary, less explorative methods, e.g., based on code search, code completion, or API documentation. In contrast, our approach is highly interactive and can be seen as an extension of what IDEs readily provide today. Exploration is based on multiple dimensions: i) the hierarchically organized scopes of projects and APIs; ii) metrics of API usage (e.g., number of project classes extending API classes); iii) metadata for APIs; iv) project- versus API-centric views. We also provide the QUAATLAS corpus of Java projects which enhances the existing QUALITAS corpus to enable API-usage analysis. We implemented the exploration approach in an open-source, IDE-like, Web-enabled tool EXAPUS. |
| Mining API mapping for language migration | To address business requirements and to survive in competing markets, companies or open source organizations often have to release different versions of their projects in different languages. Manually migrating projects from one language to another (such as from Java to C#) is a tedious and error-prone task. To reduce manual effort or human errors, tools can be developed for automatic migration of projects from one language to another. However, these tools require the knowledge of how Application Programming Interfaces (APIs) of one language are mapped to APIs of the other language, referred to as API mapping relations. In this paper, we propose a novel approach, called MAM (Mining API Mapping), that mines API mapping relations from one language to another using API client code. MAM accepts a set of projects each with two versions in two languages and mines API mapping relations between those two languages based on how APIs are used by the two versions. These mined API mapping relations assist in migration of projects from one language to another. We implemented a tool and conducted two evaluations to show the effectiveness of MAM. The results show that our tool mines 25,805 unique mapping relations of APIs between Java and C# with more than 80% accuracy. The results also show that mined API mapping relations help reduce 54.4% compilation errors and 43.0% defects during migration of projects with an existing migration tool, called Java2CSharp. The reduction in compilation errors and defects is due to our new mined mapping relations that are not available with the existing migration tool. |
| Automated API Property Inference Techniques | Frameworks and libraries offer reusable and customizable functionality through Application Programming Interfaces (APIs). Correctly using large and sophisticated APIs can represent a challenge due to hidden assumptions and requirements. Numerous approaches have been developed to infer properties of APIs, intended to guide their use by developers. With each approach come new definitions of API properties, new techniques for inferring these properties, and new ways to assess their correctness and usefulness. This paper provides a comprehensive survey of over a decade of research on automated property inference for APIs. Our survey provides a synthesis of this complex technical field along different dimensions of analysis: properties inferred, mining techniques, and empirical results. In particular, we derive a classification and organization of over 60 techniques into five different categories based on the type of API property inferred: unordered usage patterns, sequential usage patterns, behavioral specifications, migration mappings, and general information. |
| Web API growing pains: Stories from client developers and their code | Web APIs provide a systematic and extensible approach for application-to-application interaction. Developers using web APIs are forced to accompany the API providers in their software evolution tasks. In order to understand the distress caused by this imposition on web API client developers we perform a semi-structured interview with six such developers. We also investigate how major web API providers organize their API evolution, and we explore how this affects source code changes of their clients. Our exploratory study of the Twitter, Google Maps, Facebook and Netflix web APIs analyzes the state of web API evolution practices and provides insight into the impact of service evolution on client software. Our study is complemented with a set of observations regarding best practices for web API evolution. |
| Recommending Proper API Code Examples for Documentation Purpose | Code examples are important resources for expressing correct application programming interface (API) usages. However, many framework and library APIs fail in offering sufficient code examples in corresponding API documentations. This is because constructing proper code examples for documentation purpose takes significant developers' efforts. To reduce such effort, this work proposes a methodology, PropER-Doc, that recommends proper code examples for documentation purpose. PropER-Doc accepts queries from API developers and utilizes code search engines (CSEs) to collect corresponding code example candidates. The structural and conceptual links between API elements are captured from the API implementation and available API documents to guide candidate recommendation. During recommendation, PropER-Doc groups collected candidates based on involved API types for distinguishing different API usages. To assist API developers in selecting proper candidates, a diagrammatic presentation and three code example appropriateness metrics are also developed in {PropER-Doc}. Two case studies on Eclipse JDT framework are conducted to confirm the effectiveness of PropER-Doc. |
| How Does Web Service API Evolution Affect Clients? | Like traditional local APIs, web service APIs (web APIs for short) evolve, bringing new and improved functionality as well as incompatibilities. Client programs have to be modified according to these changes in order to use the new APIs. Unlike client programs of a local API, which could continue to use the old API, clients of a web API often do not have the option not to upgrade, since the old version of the API may not be provided as a service anymore. Therefore, migrating clients of web APIs is a more critical task. Research has been done in the evolution of local APIs and different approaches have been proposed to support the migration of client applications. However, in practice, we seldom observe that web API providers release automated tools or services to assist the migration of client applications. In this paper, we report an empirical study on web API evolution to address this issue. We analyzed the evolution of five popular web APIs, in total 256 hanged API elements, and carefully compared our results with existing empirical study on API evolution. Our findings are threefold: 1) We summarize the API changes into 16 change patterns, which provide grounded supports for future research, 2) We identify 6 completely new challenges in migrating web API clients, which do not exist in the migration of local API clients, 3) We also identify several unique characteristics in web API evolution. |

The result shown by the query “Artificial Intelligence” is given below.

| | |
|---|---|
| Editorial Board | |
| Dynamic Programming Used to Provide Artifical Intelligence Batch Processes | The automation of the batch chemical plant generally begins and ends with providing the real time control and recipe handling. This level of automation replaces manpower in some basic functions and generates incentives in the capacity area. However, the true incentive in the automation of a chemical operation is the decision analysis made by the selection of products to be run in specific equipment. Currently, even with automated systems, critical decisions which impacting product cost such as scheduling and equipment allocation, are left to operators. Secondly, process supervision makes allocations based on intuition rather than true cost calculations. Scheduling techniques to soft out campaigns, schedule recipe implementation, allocate process equipment and to enforce these schedules have been implemented with success. These specialty chemical and resin plants have reaped benefit in energy, raw material and capacity areas, plus the product quality area. This provides a compounding beneficial effect. |
| Integrating multimedia and artifical intelligence for pest prediction and aeration control of stored grain bins | In the present study, an intelligent system with human-machine interface of knowledge acquisition is established to implement the pest prediction and the aeration control of stored grain bins. In the system, recurrent neuro-fuzzy network models are proposed to predict temperature evolvement in the grain bins. 3D multimedia displays of node sensor-measured temperatures, and its gradient distributions of the given grain layer and their variations in the interval of the given time are used to extract the system knowledge in the current and future. The results of the experiment in the two grain depots in northeastern China have verified the effectiveness of the system. |
| An Overview of Publications on Artificial Intelligence Research: A Quantitative Analysis on Recent Papers | The study on artificial intelligence (AI) is highly interdisplinary, which involves increasing number of researchers from different academic fields. In order to provide an overview for the researchers on recent publications in related fields, we conducted a statistic analysis using bibliometric methods. Using the data source from Web of Science (ISI), we have studied the articles published in the SCI and SSCI journals on the the subject of AI between the year of 2000 and 2011. The data were analyzed from six aspects, including article distribution by years, journals, languages, countries/regions, research fields, and authors. The result revealed the most influential journals, language and authors in this field, which could help researchers from the related area to find useful in formations and direct their researches. |
| An electronic advisor/companion: Computers able to function as personal consultants require breakthroughs in the field of artificial intelligence | Discusses the future development of computers as personal consultants. A great deal of research, however, is required in the field of artificial intelligence. |
| Technical papers sought on Artificial Intelligence | Prospective authors are invited to submit papers on Artificial Intelligence for presentation at sessions during the AIEE Winter General Meeting in New York City on January 27-February 1 and for publication as AIEE Transactions Papers. |
| On Methods of Transfer as a Learning Based on Artifical Neural Networks | This paper studies learning transfer based on the adaptive learning theory, intelligence cognitive model theory and knowledge input theory under the large framework of updated cognitive theory of psychology. It mainly focuses on the method and experiment of positive transfer in production system self-adaptability model with the help of updated Artificial Neural Networks communication technology. These methods can help provide students with strategies in coping with their learning so that they could become self-directed learners through the guilded adaptive learning and intelligence teaching. |
| Artificial Intelligence Planning Problems in a Petri Net Framework | Artificial Intelligence planning systems determine a sequence of actions to be taken to solve a problem. This is accomplished by generating and evaluating alternative courses of action. A special type of Petri net is first defined and then used to model a class of Artificial Intelligence planning problems. A planning strategy is developed using results from the theory of heuristic search. In particular, the A* algorithm is utilized. From the Petri net framework it is shown how to develop an admissible and consistent A* algorithm. As an illustration of the results three Artificial Intelligence planning problems are modelled and solved. |

# 5.SUMMARY

## Conclusion

In the era of science and technology semantic web is playing a very important rule in many frameworks to organized data in a meaningful way. User is also capable to communicate with each other data over different sources at the same time using these web semantic approaches. This technology gives an organized data to facilitate their researcher and clients. It is known that the huge amount of data on internet is increasing day to day, year to year. To deal this huge amount of data and to fetch desired information, web semantics technologies are put into work in this study. Most of the research engines that use semantic web technologies are based on different frameworks in view of their requirements. Over WWW (World Wide Web) semantic web is the future of internet and KDD (Knowledge Base Data Driven) these technologies aim to mine data from a huge database in such a way that one can search his/ her related information from internet. In future searches will become personalized and desired results will be more accurate and précised on the basis of semantic web. The existing model of internet resources were able to defined intelligent services along with talking lot of time to query or result because of their old architectures. Semantic web frameworks can overcome all of these issues by providing better user experience over the world. JSON and XML are the form of data that can be easily understandable and transferable over different devices at the same time. Number of devices can use this format to work properly in a parallel way. There is a need to study on web semantic because of its importance and usage in many different field of science.

In semantic web researchers have together different kinds of information in a meaningful way from different resources. However, more than one research portals are easily accessible to get query results but there is a need of clients / researchers to deal more than one research journals websites to get desired amount and type of data.

In this study we make an attempt to design and implement semantic web framework in such a way that researcher can find their desired data in a meaningful way and gather research related data from IEEE Explore and Springer. The results found in this study are on the basis of API (Application Programming Interface) this framework shows the working of project in a detail way. Some of other Ontology is also proposed together data in a semantically organized way. The main purpose of this study is to collect data from many web portals in semantically way to facilitate client/ users as well as researcher. In future work we can add many other data resources to achieve this different task. This study is API (Application Programming Interface) has features to provide data for different devices at the same time.